\documentclass[usenatbib]{mn2e}

\def\kms{\,km~s$^{-1}$}
\def\Reff{$R_{\rm e}$}
\usepackage{graphicx}
\usepackage{subfigure}
\usepackage{journal_names}
\usepackage{float}
\usepackage{xcolor} 
\usepackage{amsmath}
\usepackage{caption}
\usepackage{hyperref}

\title[SLUGGS kinemetry]{The SLUGGS Survey: stellar kinematics, kinemetry and trends at large radii in 25 early-type galaxies}
\author[C. Foster et al.]{Caroline Foster,$^{1}$\thanks{E-mail: cfoster@aao.gov.au} Nicola Pastorello,$^2$ Joel Roediger,$^{3,4}$ Jean P. Brodie,$^{3,5}$ \newauthor Duncan A. Forbes,$^{2}$ Sreeja S. Kartha,$^2$ Vincenzo Pota,$^6$ Aaron J. Romanowsky,$^{5,7}$ \newauthor Lee R. Spitler,$^{8,1}$ Jay Strader,$^{9}$ Christopher Usher,$^{2,10}$ Jacob A. Arnold$^3$
\\
$^1$Australian Astronomical Observatory, PO Box 915, North Ryde, NSW 1670, Australia\\
$^2$Centre for Astrophysics \& Supercomputing, Swinburne University, Hawthorn, VIC 3122, Australia\\
$^3$Department of Astronomy and Astrophysics, University of California, Santa Cruz, CA 95064, USA\\
$^4$NRC Herzberg Astronomy \& Astrophysics, 5071 W Saanich Rd, Victoria BC, V9E 2E7, Canada\\
$^5$University of California Observatories, 1156 High Street, Santa Cruz, CA 95064, USA\\
$^6$ INAF - Osservatorio Astronomico di Capodimonte, Salita Moiariello, 16, I-80131 Napoli, Italy\\
$^7$Department of Physics and Astronomy, San Jos\'e State University, One Washington Square, San Jos\'e, CA 95192, USA\\
$^8$Department of Physics \& Astronomy, Macquarie University, Sydney, NSW 2109, Australia\\
$^9$Department of Physics and Astronomy, Michigan State University, East Lansing, MI 48824, USA\\
$10$Astrophysics Research Institute, Liverpool John Moores University, 146 Brownlow Hill, Liverpool L3 5RF, UK\\
}

\begin{document}
\date{Accepted 17 December 2015}
\pagerange{\pageref{firstpage}--\pageref{lastpage}} \pubyear{2015}
\maketitle
\label{firstpage}

\begin{abstract}
Due to longer dynamical timescales, the outskirts of early-type galaxies retain the footprint of their formation and assembly. Under the popular two-phase galaxy formation scenario, an initial in-situ phase of star formation is followed by minor merging and accretion of ex-situ stars leading to the expectation of observable transitions in the kinematics and stellar populations on large scales. However, observing the faint galactic outskirts is challenging, often leaving the transition unexplored. The large scale, spatially-resolved stellar kinematic data from the SAGES Legacy Unifying Galaxies and GlobularS (SLUGGS) survey are ideal for detecting kinematic transitions.
We present kinematic maps out to 2.6 effective radii on average, kinemetry profiles, measurement of kinematic twists and misalignments, and the average outer intrinsic shape of 25 SLUGGS galaxies. We find good overall agreement in the kinematic maps and kinemetry radial profiles with literature. We are able to confirm significant radial modulations in rotational versus pressure support of galaxies with radius so that the central and outer rotational properties may be quite different. We also test the suggestion that galaxies may be more triaxial in their outskirts and find that while fast rotating galaxies were already shown to be axisymmetric in their inner regions, we are unable to rule out triaxiality in their outskirts. We compare our derived outer kinematic information to model predictions from a two-phase galaxy formation scenario. We find that the theoretical range of local outer angular momentum agrees well with our observations, but that radial modulations are much smaller than predicted.
\end{abstract}


\section{Introduction}

The development of integral field spectrographs (IFS) has been accompanied by a proportional rise in interest in the spatially-resolved kinematics, stellar populations, interstellar media, etc of galaxies. Examples include the pioneering work of the Spectrographic Areal Unit for Research on Optical Nebulae (SAURON) survey \citep{deZeeuw02}, its successor the ATLAS$^{\rm 3D}$ survey \citep{Cappellari11a}, the ongoing Calar Alto Legacy Integral Field spectroscopy Area (CALIFA) survey \citep{Sanchez12} and the MASSIVE survey \citep{Ma14}. A major focus of these is the study of the stellar and gas spatially-resolved line-of-sight velocity distribution (LOSVD) or 2D kinematics. 

An important finding of the SAURON and ATLAS$^{\rm 3D}$ surveys is the apparent dichotomy in the central kinematics of early-type galaxies. Building systematically on earlier hints from \citet{Davies83} and \citet{Kormendy96}, \citet{Emsellem07} and \citet{Cappellari07} have found that the specific angular momentum of stars in the centres of early-type galaxies covers a range of values, with galaxies showing a very high degree of pressure support labelled as ``slow rotators'', and galaxies with significant rotational support and high central specific angular momentum labelled ``fast rotators''.

The comparison of the central LOSVD in 2D to that expected from simple rotation or purely pressure-supported systems, and their correspondence (or otherwise) with photometrically identified substructures, have already greatly informed our understanding of the structures of galaxies \citep[e.g.][]{Krajnovic08,Emsellem11, BarreraBallesteros14,Emsellem14}. Indeed, while photometric studies reveal the projected 2D distribution of stars on the sky, the added dimension of spectroscopy affords the possibility of disentangling the true 3D shapes of galaxies \citep{Weijmans14}. Moreover, important galaxy formation events may leave obvious imprints on the LOSVD that are undetectable, ambiguous or unexpected from photometry alone \citep[e.g.][]{Krajnovic11,Krajnovic12,GarciaLorenzo15}.

These studies have led to the development of analysis tools used for sifting through 2D spectroscopy. In particular, with regards to the analysis of kinematic maps, a number of techniques have been employed. A notable example is kinemetry \citep{Krajnovic06}. Similar in principle to photometry of extended objects, kinemetry fits trigonometric functions and ellipses to the maps of the odd (recession velocity, $V_{\rm obs}$, and $h_3$, related to the skewness) and even (velocity dispersion, $\sigma$; and $h_4$, related to the kurtosis) velocity moments of the LOSVD. This allows for the easy identification of asymmetries and determination of basic parameters such as e.g. the kinematic position angle ($PA_{\rm kin}$) or the kinematic axis ratio ($q_{\rm kin}$), analogous to their photometric cousins $PA_{\rm phot}$ and $q_{\rm phot}$, respectively.

Hand-in-hand with the IFS observational campaigns is the ongoing efforts to understand the theoretical aspects of galaxy formation. Many recent models have aimed to explain the diversity and complexity of these findings. The photometric and kinematic properties of fast rotators are easily explained by binary major merger encounters \citep[e.g.][]{Cox06,Hoffman09,Hoffman10,Moody14}. The photometric and kinematic properties of slow rotators are however more difficult to reproduce via binary major mergers of disc galaxies \citep[e.g.][]{Jesseit09,Bois11}. Possible formation pathways for slow rotators include multiple minor mergers or cosmologically meaningful assembly histories \citep[e.g.][]{Moody14,Naab14,Wu14}. 

Looking forward, a new generation of state-of-the-art instruments is now seeing first light at observatories around the world. Multi-object IFSs such as the Sydney-AAO (Australian Astronomical Observatory) Multi-object IFS \citep[SAMI,][]{Croom12} and the K-band Multi-Object Spectrograph \citep[KMOS,][]{Sharples13} were developed to address limitations associated with the observational cost of obtaining a large sample of galaxies using single IFSs by allowing for multiple targets to be observed simultaneously. These instruments will greatly improve our ability to make statistically meaningful inferences about the spatially-resolved properties of galaxies. Examples of planned and ongoing surveys using multi-object IFSs include the SAMI Galaxy Survey \citep{Allen15} and Mapping Nearby Galaxies at APO (MaNGA, PI Kevin Bundy, http://www.sdss3.org/future/manga.php).

The number of targets observed with IFSs is increasing exponentially as a result. Unfortunately, a clear shortcoming of most current IFS studies, preventing important advances in our understanding of galaxy formation, is the lack of radial coverage beyond roughly one effective radius ($R_{e}$, the radius within which half the stellar light is enclosed, see figure 2 and table 1 of \citealt{Brodie14}). This is a direct result of the limited field-of-view of most IFSs \citep[but see][for a notable exception]{Raskutti14} and implies that only half of the stellar LOSVD and less than 5 or 10 per cent of the total mass or angular momentum is typically probed, respectively \citep[see][]{Brodie14}. Moreover, important theoretically expected large scale kinematic transitions may be missed altogether. This is a liability because stellar kinematics at large radii are essential for delineating between various galaxy formation scenarios.

One such scenario was developed in order to explain recent findings from deep and high resolution imaging of high-redshift massive galaxies revealing that they have undergone intense size growth since $z\sim2$ when compared with their local counterparts \citep{Daddi05,Trujillo07,Cimatti08,vanDokkum08}. Their extreme density is confirmed by their correspondingly high velocity dispersions \citep{vanDokkum09}. The main driver of this size growth has been a matter of debate in the literature, however a consensus is slowly emerging. This consensus focuses on the two-phase galaxy formation scenario, wherein massive galaxies first form a substantial bulge via quick successive multiple major mergers \citep[e.g.][]{Cook09} or turbulent discs / inflows \citep[e.g.][]{Dekel09,Ceverino15}. This in-situ bulge is then augmented by the continuing accretion via minor mergers that deposit stars preferentially in the outskirts, hence contributing to the size growth without implying significant stellar mass growth \citep[e.g.][]{Naab09,Bezanson09,Oser10,vanDokkum10}. A qualitative prediction of this scenario is a kinematic transition between the central in-situ and outer accreted stars of galaxies at some yet to be determined radius. Even galaxies formed via equal mass gas-rich merging are expected to show such a kinematic transition \citep{Novak06,Hoffman09} between 1 and 3 $R_e$.

The SAGES Legacy Unifying Galaxies and GlobularS (SLUGGS, \citealt{Brodie14}) survey was designed to address the radial coverage issue. The SLUGGS survey targets both the stars and globular clusters around early-type (lenticular and elliptical) galaxies at distances up to $\sim30$ Mpc. The spectroscopic follow-up aspect of the survey is carried out using the DEep Imaging Multi-Object Spectrograph instrument on the Keck telescope in multiple-slit mode. \citet{Proctor09} showed that spatially-resolved galaxy light spectra of the host could be extracted from the very same slits as the target globular clusters, hereby affording the possibility of probing the 2D LOSVD out to $3R_e$. Combined with the use of other spatial smoothing and extrapolation techniques such as kriging \citep{Matheron63,Cressie90,Foster13,Pastorello14} and kinemetry adapted to sparsely sampled datasets \citep{Proctor09,Foster11,Foster13}, the SLUGGS survey enables the study of the outer kinematics of nearby early-type galaxies and reveals the predicted kinematic transition.

In their proof-of-concept study, \citet{Proctor09} showed that the degree of rotational support of galaxies changes dramatically with galactocentric radius. The reality of the kinematic transition at large galactocentric radii is also confirmed using planetary nebulae as kinematic tracers \citep{Coccato09} and in other SLUGGS galaxies \citep{Foster11,Foster13}. However, just how common these transitions are has yet to be assessed using a well-defined galaxy sample and their implication for galaxy formation still needs to be addressed in a statistically meaningful sense. Thanks to the generous radial extent probed and its representative sample selection, the SLUGGS survey is particularly well-suited for conducting such a systematic study of early-type galaxies.

In this paper, we present the systematic kinemetry fitting of our targets. In a companion paper, \citet{Arnold14} have already introduced the kinematic maps for the majority of galaxies in the SLUGGS sample. Using the large scale kinematic maps, they were able to confirm that galaxies commonly showed strong radial variations in their local angular momentum. This finding was however recently put into question by \citet{Raskutti14}. This question will be readdressed here. Moreover, we analyse the kinematic position angles and their variations for our sample galaxies. \citet{Schauer14} recently suggested rings of enhanced velocity dispersion values (so-called `sigma bump') should be common in the remnants of major mergers. This provides an interesting avenue for understanding the assembly history of galaxies in SLUGGS via a search for this tell-tale signature in large scale kinematic maps.

This paper is divided as follows. Section \ref{sec:data} describes the SLUGGS sample and our data reduction steps while Section \ref{sec:method} describes kinemetry adapted to sparse sampling and our analysis of the results. We discuss and provide an interpretation of our results in Section \ref{sec:discussion}. A summary and our conclusions can be found in Section \ref{sec:conclusions}.

\section{Data}\label{sec:data}

We here give a description of the imaging and spectroscopic data used in this work. A summary of salient properties of the 25 galaxies in our sample can be found in Table \ref{table:sample}.

\subsection{Spectra}
This work mainly uses data from the SLUGGS\footnote{http://sluggs.swin.edu.au/} survey \citep{Brodie14}. The survey includes a total sample of 25 main and 3 ``bonus'' galaxies of representative stellar masses, environments, and early-type galaxy sub-types. The survey combines wide-field imaging of the host galaxy and its globular cluster systems, as well as follow-up spectroscopy using the DEep Imaging Multi-Object Spectrograph (DEIMOS, \citealt{Faber03}) on the Keck telescope in Hawaii. Observations spanned multiple nights over 2009-2014.  To the dataset presented in \citet{Arnold14}, we include additional data from a total of ten recently observed slit-masks taken on the nights of 2013 September 29, 2013 November 2 and 2014 January 27 for galaxies NGC~720 (2 masks), NGC~821 (1 mask), NGC~1023 (1 mask), NGC~3608 (2 masks), NGC~4564 (2 masks) and NGC~7457 (2 masks). 

{ In all cases, the 1200 lines grating centred around 7800 $\AA$ with slits of 1 arcsec width and $>5$ arcsec length is used. This yields a spectral resolution of FWHM$\sim1.5\AA$ around the Calcium Triplet ($\sim8600\AA$). See \citet{Brodie14} for more detail and a full description of the survey data, aims and main results to date.}

In this work, we use the stellar spectroscopic component of the survey for 23 out of 25 main and 2 out of 3 bonus galaxies. This subsample is selected based on the individual dataset size and overall signal-to-noise. In other words, 2 main and 1 bonus galaxies have either not been observed yet, or did not have sufficient sampling and/or signal-to-noise to enable us to perform the kinemetry analysis. The DEIMOS spectra are reduced using the {\sc idl spec2d} data reduction pipeline \citep{Cooper12}. The pipeline performs flat-fielding using internal flats, wavelength calibration using ArKrNeXe arcs, as well as local background (sky $+$ galaxy light) subtraction. The outputs include the globular cluster spectra with their corresponding fully propagated variance arrays as well as the subtracted background or ‘sky’ spectra for each slit.

The galaxy light spectra are extracted out of the same slits as the globular clusters themselves with the Stellar Kinematics using Multiple Slits (SKiMS) technique developed by \citet{Proctor09} and \citet{Norris08}. This technique applies non-local sky subtraction for each slit. The pure galaxy spectrum is obtained from the background spectrum in each slit by subtracting a linear combination of ``pure'' sky spectra taken from selected slits at large galactocentric radii. At these large radii the contribution from the galaxy light is insignificant \citep[see][for more detail]{Arnold14}. This technique has been shown to yield a continuum level accurate to within 0.7 per cent of the noise in the skyline residuals \citep{Foster09}.

Our outer (reaching out to $2.6R_e$ on average) spatially-resolved kinematics are supplemented with central or inner (reaching out to $\sim0.9R_e$ on average) data from the SAURON spectrograph as published by the ATLAS$^{\rm 3D}$ survey \citep{Cappellari11a,Emsellem11} for the 21 galaxies in common. In what follows, we refer to parameters derived using the ATLAS$^{\rm 3D}$ data as ``inner'' or ``central'' properties, while measurements from the SLUGGS data are referred to as ``outer''.

\begin{table*} 
\begin{center}
\begin{tabular}{rcrrclrcrccccc}
\hline
Galaxy & Dist. & $\sigma_{\rm kpc}$ &  \Reff & Morph. & $PA_{\rm phot}$ & $\epsilon_{\rm phot}$ &  $V_{\rm sys}$ & Env. & F/S & KinStruct & $\Psi_0$& $\Psi_{\rm outer}$ \\
{NGC} & (Mpc) & (km\,s$^{-1}$) & (arcsec) &  & (deg) &  & (km\,s$^{-1}$) &  & & & (deg) & (deg)\\
(1)&(2)&(3)&(4)&(5)&(6)&(7)&(8)&(9)&(10)&(11)&(12)&(13)\\
\hline
{\bf Main:}&    &      &     &      &         &        &        &         &          \\
 720 & 26.9 & 227 &  35 & E5      & 140.0 & 0.49      & 1745 & F & F & RR/NF&  -- &14$\pm$28\\
 821 & 23.4 & 193 &  40 & E6      &  31.2 & 0.35      & 1718 & F & F & RR/NF&1.3&8$\pm$30\\
1023  & 11.1 & 183 &  48 & S0      &  83.3 & 0.63      &  602 & G & F & RR/NF &5.2&1$\pm$4\\
1400  & 26.8 & 236 &  28 & E1/S0   &  40.0 & 0.13      &  558 & G & F &  RR/NF&  --&4$\pm$15\\
1407  & 26.8 & 252 &  63 & E0      &  35.0 & 0.07     & 1779 & G & S & NRR/LV &  --&42$\pm$46\\
2768 & 21.8 & 206 &  63 & E6/S0  &  91.6 & 0.57      & 1353 & G & F & RR/NF &0.9&1$\pm$5\\
2974 & 20.9 & 231 &  38 & E4/S0      &  44.2 & 0.37     & 1887 & F & F & RR/NF&1.2&2$\pm$4\\
3115 &  9.4 & 248 & 35 & S0      &  40.0 & 0.66     &  663 & F & F & RR/NF &  --& $6\pm$6\\
3377 & 10.9 & 135 &  36 & E5--6    &  46.3 & 0.33   &  690 & G & F & RR/NF&2.3&$15\pm36$\\
3608 & 22.3 & 179 &  30 & E1--2    &  82.0 & 0.20      & 1226 & G & S & NRR/CRC&3.5&$16\pm19$\\
4111 & 14.6 & 161 &  12 & S0       & 150.3 & 0.79  &  792 & G & F & RR/2m&0.8&$0\pm3$\\
4278 & 15.6 & 228 &  32 & E1--2     &  39.5 & 0.09     &  620 & G & F &  RR/NF&29.5&$55\pm56$\\
4365 & 23.1 & 253 &  53 & E3      &  40.9 & 0.24      & 1243 & G & S & NRR/KDC&75.9&62$\pm41$\\
4374 & 18.5 & 284 &  53 & E1      & 128.8 & 0.05      & 1017 & C & S &  NRR/LV&42.7&--\\
4473 & 15.2 & 189  & 27 & E5      &  92.2 & 0.43     & 2260 & C & F & NRR/2s&0.2&$36\pm100$\\
4486 & 16.7 & 307 &  81 & E0/cD    & 151.3 & 0.16     & 1284 & C & S &  NRR/LV&46.2&--\\
4494 & 16.6 & 157 &  49 & E1--2     & 176.3 & 0.14     & 1342 & G & F &  RR/2m&8.7&$3\pm12$\\
4526 & 16.4 & 233 & 45 & S0       & 113.7 & 0.76       &  617 & C & F & RR/2m&5.2&1$\pm$3\\
4564 & 15.9 & 153 &  20 & E6       &  48.5 & 0.53     & 1155 & C & F & RR/NF&0.5&$0\pm5$\\
4649 & 16.5 & 308 &  66 & E2/S0   &  91.3 & 0.16     & 1110 & C & F &  RR/NF&0.2&$4\pm15$\\
4697 & 12.5 & 180 &  62 & E6       &  67.2 & 0.32      & 1252 & G & F & RR/NF&0.3&$6\pm17$\\
5846  & 24.2 & 231 &  59 & E0--1/S0  &  53.3 & 0.08   & 1712 & G & S & NRR/LV&79.2&--\\
7457 & 12.9 &  74 &  36 & S0       & 124.8 & 0.47   &  844 & F & F & RR/NF&0.8&$0\pm3$\\
\hline                                                                                         
{\bf Bonus:}&        &     &      &    &         &        &            &   &      &       &         &\\
3607 & 22.2 & 229  & 39 & S0       & 124.8 & 0.13      &  942 & G & F & RR/NF&3.3&$11\pm64$\\
5866 & 14.9 & 163  & 36 & S0       & 125.0 & 0.58        &  755 & G & F & RR/NF&1.5&$9\pm12$\\
\hline 
\end{tabular} 
\end{center}
\caption{Sample photometric and kinematic parameters of the 23 main SLUGGS + 2 bonus (at the bottom of the table) galaxies included in this work.
Unless otherwise specified, details of the parameter derivations can be found in \citet{Brodie14}.
Column descriptions: (1) galaxy NGC number;
(2) distance;
(3) central stellar velocity dispersion within 1\,kpc;
(4) effective (half-light) radius; 
(5) morphological type;
(6) position angle, for the outer isophotes and (7) ``global'' ellipticity (i.e. $\epsilon=1-q$, with q the axis ratio);  
(8) systemic velocity;
(9) local environment type: field (F), group (G), or cluster (C);
(10) fast (F) or slow (S) rotator classification within $1R_{e}$ as per \citet{Emsellem11} and \citet{Arnold14};
(11) central kinematic classification as per \citet[][RR: Regular rotator, NRR: non-regular rotator / NF: no kinematic features, CRC: counter-rotating core, 2m: double velocity maximum, KDC: kinematically decoupled core, LV: low level velocity, 2s: $2\sigma$ galaxy]{Krajnovic11}. For galaxies not in common with ATLAS$^{\rm 3D}$, we kinematically classify using the analysis presented in this work;
(12) and (13), the central and large scale kinematic misalignment, $\Psi_0$ \citep[as reported in][]{Krajnovic11}) and $\Psi_{\rm outer}$ (this work), respectively.}
\label{table:sample}
\end{table*}

\subsection{Isophotal analysis}\label{sec:iso}

{ In this section, we explore the isophotal shape and luminosity of our sample galaxies in order to enable a consistent comparison of photometric and kinematic properties.}

Of the 25 SLUGGS + 3 ``bonus'' galaxies, we measure surface brightness profiles for 17 galaxies { (i.e. all but NGC~720, NGC~820, NGC~1023, NGC~1400, NGC~1407, NGC~2974, NGC~3115 and NGC~3607)} with optical imaging available from the Sloan Digital Sky Survey (SDSS).  Calibrated mosaics of 0.25 deg$^2$ size were obtained in the $ugriz$ bandpasses through the SDSS DR10 Science Archive Server mosaic tool\footnote{\tt http://data.sdss3.org/mosaics}.  The size of the mosaics was chosen to approximately match the field of view of our complimentary Subaru data and allows us to measure the light distributions of our galaxies out to extremely large radii (32 kpc, in the median), where the stellar halo should be the sole emitter.  All mosaics correspond to median stacks of all SDSS fields (not just primaries) which lie within the specified area about the galaxy center and are generated via SWarp.  Since DR8, the background subtraction performed by the {\tt photo} pipeline on SDSS images has been upgraded to better treat cases of bright, nearby galaxies with extended halos \citep{Blanton11}, which many SLUGGS galaxies fall under.  Inspection of the delivered mosaics confirms that the majority of them have background residuals which are quite uniform and low over their full extents.  Nevertheless, we still take steps (described below) to quantify the accuracy of the {\tt photo} background subtraction algorithm as part of our isophotal analysis.

To extract radial profiles of isophotal surface brightness, ellipticity, and position angle for SLUGGS galaxies, we fit elliptical contours to our $i$-band images using the {\sc XVista} software package\footnote{http://astronomy.nmsu.edu/holtz/xvista/index.html}.  The procedures in {\sc XVista} which facilitate isophotal fitting are described in detail in \citet{Courteau96}.  More recent applications of the {\sc XVista} package for similar purposes as ours may be found in \citet{MacArthur03}, \citet{McDonald09}, \citet{McDonald11}, and \citet{Hall12}.  We focus on the $i$-band since it generally has the best combination of signal-to-noise ratio and residual background variations of the five SDSS bands, and is less susceptible to the perturbing effects of hot stars and dust, which are occasionally found in early-type galaxies.

The basic order of operations for our isophotal analysis is as follows.  We begin with certain preparations that must precede the fitting process.  First, the residual background is measured by calculating the modes of pixel values within five boxes, free of sources, placed randomly and at large distances from the galaxy center.  For SDSS data, the mean and rms dispersion of these five modes indicate the fidelity of the background subtraction by the {\tt photo} pipeline across the mosaic.  We generally find that the pipeline removes backgrounds to an accuracy of $\pm$0.01 nanomaggies.  Second, we locate the centre of the galaxy by computing the centroid of pixels lying within a box, roughly $50\times50$ pixels in size, placed over the image center.  We fix the centers of the isophotes to this location during fitting.  Third, foreground stars are automatically masked with circular apertures having sizes equal to four times their full width at half-maxima.  Afterwards, the mask is visually inspected and manually updated (if necessary) to blank out pixels containing background/satellite galaxies and artifacts caused by bright stars.

Following the above preparations, we may fit for the light distribution of the galaxy.  A preliminary isophotal solution is calculated using a set of elliptical contours, spaced at one-pixel increments in semi-major axis radius, starting from the galaxy center and ending approximately halfway to the image perimeter.  During this step, each contour is perturbed in ellipticity and position angle (plus higher-order Fourier moments, up to $n=4$) over many iterations until the best-fit values are found, that is, those which minimise the deviations in intensity around the circumference.  In regions of low signal-to-noise ratio or strong non-axisymmetries (e.g. bars), this preliminary solution can exhibit large variations from one isophote to the next.  To correct for this, a single isophote is chosen to represent the outermost regions of the galaxy and strong/high-frequency variations in the underlying structure are smoothed out.  SLUGGS galaxies being early-types, and thus largely devoid of non-axisymmetries, this correction is mostly used to force a smooth solution in the regime of intermediate-to-large radii.  At this point, the solution can also be extended in radial coverage as far as desired; we choose the edge of the image.  Afterwards, isophotal surface brightnesses are re-measured and curves of growth are constructed for both the fixed and freely-varying isophotal parameters.  Finally, the measured surface brightnesses are transformed into units of mag arcsec$^{-2}$ using known photometric zeropoints (i.e. the nanomaggie system) and pixel scales for SDSS DR10.

We generate isophotal isolutions for two cases of spatial resolution: one at the native SDSS value (pixel scale $=0.396127$ arcsec) and the other with pixels 10$\times$ larger.  The latter case is treated in order to achieve greater depths in our solutions and has proven a successful strategy in other studies of the (faint) halo structure around massive galaxies \citep{Tal11}.  Degraded images were obtained by using {\sc SWarp} to bin pixels, where the binning factor (10) was chosen for providing the best compromise between the subsequent spatial resolution and signal-to-noise ratio.  By and large, the agreement between the two sets of solutions for our galaxies is excellent and degrading the resolution affords us gains in limiting surface brightness between $\sim$0.3-2.5 mag arcsec$^{-2}$.

\section{Method}\label{sec:method}

\begin{figure}
\begin{center}
\includegraphics[width=0.5\textwidth]{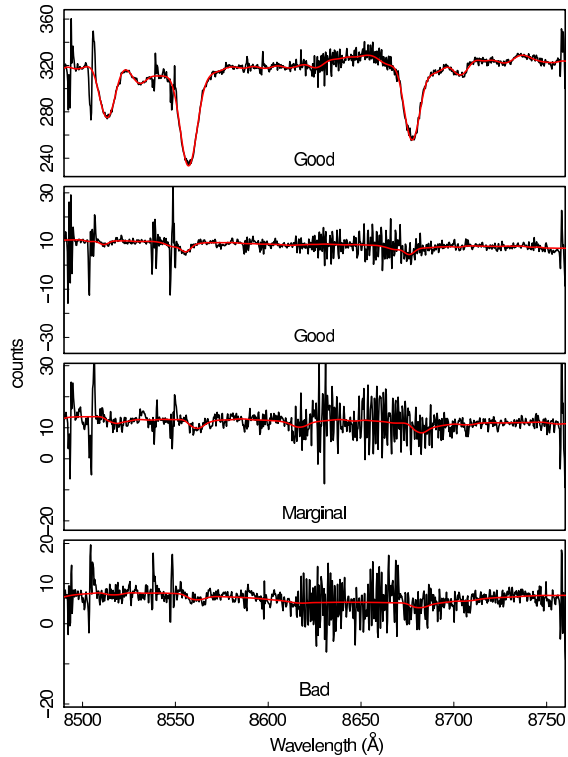}
\caption{{ Example spectra (black) and fits (red) of declining quality from top to bottom as labelled. Good spectra show reliable fits to at least 2 clear CaT lines, while marginal spectra are either too noisy for a reliable fit or only show one clear absorption feature. Bad spectra do not show any hint of the CaT. This work uses data only from spectra with good fits.}}
\label{fig:qcexamples}
\end{center}
\end{figure}

\subsection{Kinematics}

As the kinematic measurements have already been presented in \citet{Arnold14}. We simply give a brief summary here.

We use the penalised pixel cross-correlation fitting ({\sc pPXF}) code of \citet{Cappellari04} to extract the kinematics from the reduced galaxy spectra. The {\sc pPXF} code uses the corresponding fully propagated variance array to weight or ``penalise'' pixels in the reduced spectrum. Each spectrum is fitted with a linear combination of template spectra convolved with Gauss-Hermite polynomials. { The template library contains 42 stars from the \citet{Cenarro01} library ranging from late-type K giants, which are the main contributors in the wavelength range that we analyse, to main sequence and early-type stars to account for the possible presence of Paschen lines \citep{Foster10}. Spectra of M dwarf stars are also included to allow for possible molecular absorption features. This broad set ensures minimal template mismatch.}  The best set of kinematic parameters and templates is identified by minimising the residuals between the observed and fitted spectra. The first four velocity moments (i.e., recession velocity, velocity dispersion, $h_{3}$, and $h_{4}$) are simultaneously fitted and uncertainties are measured using Monte Carlo methods.

\begin{figure}
\hspace{-0.2in}\includegraphics[width=98mm]{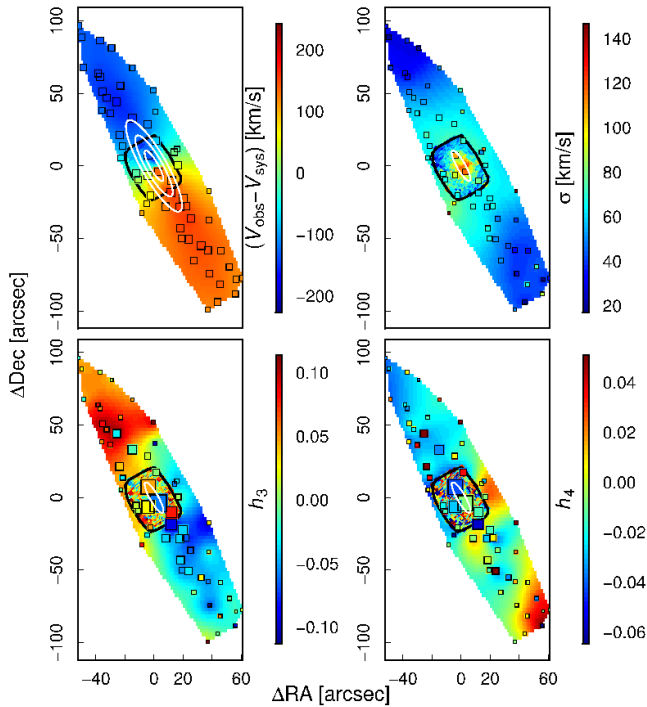}
\caption{Velocity (top left), velocity dispersion (top right), $h_3$ (lower left) and $h_4$ (lower right) kriging maps for NGC~4111. The central maps from ATLAS$^{\rm 3D}$ are shown within the black polygon. White ellipses show 1, (2 and 3) $R_e$ in each map (the velocity map) as per Table \ref{table:sample}. Individual SKiMS data are shown as coloured squares with sizes inversely proportional to the uncertainties. The ATLAS$^{\rm 3D}$ velocity dispersions have been offset empirically to match the SKiMS data (see text).}
\label{fig:krig4111}
\end{figure}

The uncertainties do not completely account for all systematics, so we add 5 and 8 \kms\, in quadrature to the Monte Carlo uncertainties on the measured recession velocity and velocity dispersion, respectively \citep[see][for a quantitative justification]{Foster11}.

Apart from the new data included here, our dataset differs from that of \citet{Arnold14} on one more important account. That is, we perform a stringent visual inspection of every single spectral fit in order to carefully discard bad fits and outliers. { Examples of good, marginal and bad fits are shown in Figure \ref{fig:qcexamples}. Only spectra with ``good'' quality flags are used in this work (i.e. marginal and bad fits are discarded). In other words, the spectra used here have all been reliably fitted and exhibit at least 2 clear CaT lines.}

\subsection{Kriging maps}

Following \citet{Foster13}, here we use kriging to produce continuous maps of the various kinematic moments based on our sparsely sampled datasets. Kriging is a well established geostatistical technique \citep[e.g.][]{Matheron63,Cressie90,Furrer02} that uses a simple function (polynomial) to fit randomly and sparsely sampled datasets. { Provided that regions of localised large changes in the field are well sampled, kriging is a reliable visualisation and interpolation tool. As is true of any interpolation tool, the final map depends strongly on the distribution and quality of the input datapoints. For example, undersampled substructured can be smoothed out.  For a more thorough description of kriging, including other potential shortcomings and another example of an application to astronomy, see \citet{Pastorello14}.} In practice, we use the {\sc Krig} function in {\sc r} and weight each data point inversely with its uncertainty to create the kriging maps. As with \citet{Pastorello14}, we first reject slits associated with nearby galaxy NGC~5846A for the analysis of NGC~5846 in what follows.

{ In general, we find that the overall standard error on the kriging maps is less than 5 and 7 km s$^{-1}$ for the recession velocity and velocity dispersion fields, respectively, and less than 0.02 for both the h$_3$ and h$_4$ fields.}

We find a systematic offset in the measured velocity dispersion between the SLUGGS and ATLAS$^{\rm 3D}$ data, so we subtract the measured offset from the ATLAS$^{\rm 3D}$ dispersion data for each galaxy. The origin of this offset is as yet unknown, but several possibilities are discussed in both Section \ref{sec:literature} and \citet{Arnold14}.

An example kriging map for NGC~4111 is shown in Fig. \ref{fig:krig4111} while other maps for the rest of the sample can be found in Appendix \ref{section:krigmaps}.

\subsection{Kinemetry fits}

To obtain a physically meaningful fit to the data, we use kinemetry \citep{Krajnovic06} adapted to sparse samples following \citet{Proctor09}. Whenever a galaxy is also part of the ATLAS$^{\rm 3D}$ survey, we apply the same algorithm to the ATLAS$^{\rm 3D}$ data to compare the results self-consistently. { We stress that the kinemetry fits and all derived quantities shown in this section are completely independent from the kriging fits discussed previously.}

We use rolling radial fits \citep{Proctor09,Foster11} by first binning the data in elliptical annuli:
\begin{equation}
R=R_{\rm circ}=\sqrt{X^2\times q_{\rm phot}+Y^2/ q_{\rm phot}},
\end{equation}
where $X$ and $Y$ are the rotated coordinate system such that the semi-major axis of the galaxy lies along the photometric $x$-axis. In other words, data are binned in elliptical annuli according to the photometric parameters given in Table \ref{table:sample}.

When modelling the recession velocity map, we vary all or a subset of the free variables: kinematic position angle ($PA_{\rm kin}$), axis ratio ($q_{\rm kin}$) and/or amplitude of the rotational velocity ($V_{\rm rot}$), to minimise the following equation for the $j^{\rm th}$ bin:
\begin{equation}\label{eq:Vobs1}
\chi^2_{V,j}=\sum^{i=N_j}_{i=1} \frac{1}{(\Delta V_{{\rm obs},i})^2} \left( V_{{\rm obs},i}-V_{{\rm mod},i,j} \right)^2,
\end{equation}
where
\begin{equation}\label{eq:Vobs2}
V_{{\rm mod},i,j}=V_{{\rm sys}} + V_{{\rm rot},j}\cos(\phi_{i,j}),
\end{equation}
where
\begin{equation}\label{eq:Vobs3}
\tan(\phi_{i,j})=\frac{\tan(PA_{i,j}-PA_{{\rm kin},j})}{q_{{\rm kin},j}}.
\end{equation}

In Equations \ref {eq:Vobs1}, \ref {eq:Vobs2} and \ref {eq:Vobs3}, $PA_i$ and $\Delta V_{{\rm obs},i}$ are the position angle and the uncertainty on the recession velocity measurement of the $i^{{\rm th}}$ data point, respectively. The systemic velocity ($V_{\rm sys}$) is determined first as a free parameter by fitting only the inner highest signal-to-noise data for each galaxy.

For the $j^{\rm th}$ bin we fit the even moment velocity dispersion ($\sigma_j$) using $\chi^2$-minimisation where 
\begin{equation}\label{eq:vd}
\chi^2_{\sigma,j}=\sum^{i=N_j}_{i=1} \left( \frac{\sigma_{{\rm obs},i}-\sigma_j}{\Delta\sigma_{{\rm obs},i}} \right)^2.
\end{equation}
In Eq. \ref{eq:vd}, $\sigma_{{\rm obs},i}$ and $\Delta\sigma_{{\rm obs},i}$ are the measured velocity dispersion of the $i^{\rm th}$ data point and associated random uncertainty, respectively.

Similarly, we minimise
\begin{equation}\label{eq:H3_1}
\chi^2_{H_{3},j}=\sum^{i=N_j}_{i=1} \frac{1}{(\Delta h_{3,{{\rm obs},i}})^2}(h_{3,{\rm obs},i}-h_{3,{{\rm mod},i,j}})^2,
\end{equation}
where
\begin{equation}\label{eq:H3_2}
h_{3,{{\rm mod},i,j}}=H_{3,j}\cos(\phi_{i,j})
\end{equation}
to obtain $H_{3}$ for the $j^{\rm th}$ bin. The symbols $h_{3,{{\rm obs},i}}$ and $\Delta h_{3,{{\rm obs},i}}$ correspond to the measured $h_3$ value and uncertainty of the $i^{\rm th}$ data point, respectively.

Finally, we compute $H_4$ by minimising
\begin{equation}\label{eq:H4}
\chi^2_{H_{4},j}=\sum^{i=N_j}_{i=1} \left( \frac{h_{4,{{\rm obs},i}}-H_{4,j}}{\Delta h_{4,{{\rm obs},i}}} \right)^2,
\end{equation}
where $h_{4,{{\rm obs},i}}$ and $\Delta h_{4,{{\rm obs},i}}$ are the measured $h_4$ value and its uncertainty for the $i^{\rm th}$ slit.

As in \citet{Foster11}, the kinemetry fits for the higher order velocity moments (i.e. $\sigma$, $h_3$ and $h_4$) are carried out simultaneously for each bin using the values of $PA_{{\rm kin},j}$ and $q_{{\rm kin},j}$ from the recession velocity.

In order to obtain confidence contours, we iterate the procedure above 500 times for each galaxy using bootstrapping by sampling each dataset with replacement. This method was introduced in \citet{Foster13}.

\begin{figure} 
\hspace{-0.5cm}\includegraphics[width=90mm]{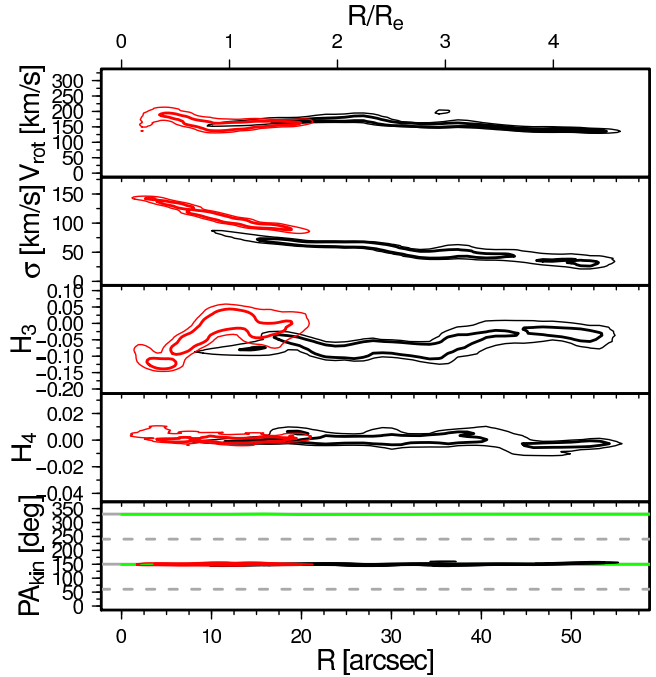}
\caption{Results of the radial kinemetry for NGC~4111. The rotational velocity ($V_{\rm rot}$), velocity dispersion ($\sigma$), amplitudes of $h_3$ and $h_4$ ($H_3$ and $H_4$, respectively) and kinematic position angle ($PA_{\rm kin}$) profiles are shown in their respective panels. Black thick and thin contours are the 1 and 2 $\sigma$ confidence intervals, respectively, for the SLUGGS data. Similarly, thick and thin red contours show the 1 and 2 $\sigma$ confidence intervals, respectively, of the ATLAS$^{\rm 3D}$ kinemetry. Green solid lines show our measured light profile, while grey solid and dashed lines show the literature photometric major axis/axis ratio and minor axis, respectively, as given in Table \ref{table:sample}. While the agreement between the ATLAS$^{\rm 3D}$ and SLUGGS data is usually good, there is a systematic offset in the velocity dispersion between the two surveys (see text). As is common for fast rotators, $PA_{\rm kin}$ is well aligned with the photometric major axis.}\label{fig:kin4111}
\end{figure}

Given the varying quality (number of data points, radial and azimuthal coverage) of our data from galaxy to galaxy, it is difficult to identify a single set of parameter restrictions that suits all galaxies. In general, the more free parameters, the less stable the fits. For galaxies with modest sampling and/or complex velocity fields (NGC~4374 and NGC~5846), we fix $PA_{\rm kin}=PA_{\rm phot}$ and $q_{\rm kin}=q_{\rm phot}$. For most other galaxies in our sample, we apply our \emph{fiducial case} which allows the following variables to vary within the allowed ranges listed: \begin{enumerate}
\item 0 $\le V_{\rm rot}\le$ 300 \kms,
\item 0 $\le PA_{\rm kin}\le$ 360 deg,
\item $q_{\rm kin} = q_{\rm phot}$ (as given in Table \ref{table:sample}),
\item 0 $\le \sigma\le$300 \kms,
\item -0.7 $\le H_{3}, H_{4}\le$ 0.7.
\end{enumerate}
An exception is NGC~4486 for which we have set $PA_{\rm kin}=PA_{\rm phot}\pm90$ deg (outer minor axis rotation). For a variety of cases with exceptionally good quality dataset, we have attempted to fit for $q_{\rm kin}$ also, but we have found that it cannot be reliably recovered for sparsely sampled datasets. It is worth noting that our assumption that $q_{\rm kin}=q_{\rm phot}$ may lead to an underestimation of $V_{\rm rot}$ of up to ~10-20 percent in the case of embedded discs. 

One further consideration of this method has to do with the size of the rolling bins. There is a tradeoff with larger bins ($>30$) yielding more stable fits, but diluting substructures (such as seen in NGC~4473). We opted for a fiducial rolling bin size of 15 data points, but clearly this bin size may be optimised for each galaxy.

An example of the kinemetry results for NGC~4111 as a function of radius can be found in Fig. \ref{fig:kin4111} while the rest of the sample can be found in Appendix \ref{section:kinemetryfigures}. The median galactocentric radius reached by our kinemetry fits is 2.5 $R_e$, with NGC~4111 reaching as far out as 4.7 $R_e$.

\subsection{Comparing SLUGGS and ATLAS$^{\rm 3D}$}\label{sec:literature}

\begin{figure}
\begin{center}
\includegraphics[width=84mm]{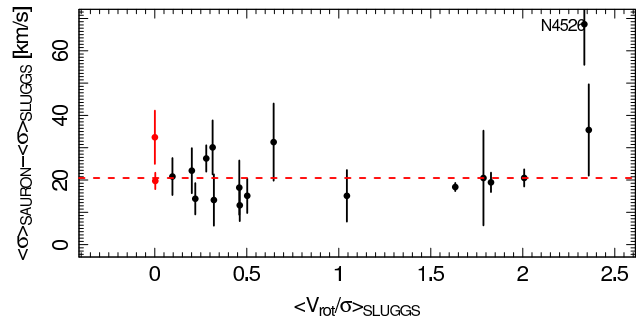}
\caption{Measured velocity dispersion offset in individual slow (red) and fast (black) rotator galaxies for the kinemetry fits in the overlapping radial range between ATLAS$^{\rm 3D}$ and SLUGGS as a function of the ratio of the local measured rotational over dispersion velocity ($V_{\rm rot}/\sigma$). One outlier (NGC~4526) is labelled. The red dashed line shows the mean offset.}
\label{fig:offsets}
\end{center}
\end{figure}

{ We find that there is usually a smooth transition and good overall agreement in the overlap regions between our data and those of ATLAS$^{\rm 3D}$ for three of the velocity moments: $V_{\rm rot}$, $H_3$ and $H_4$, as well as $PA_{\rm kin}$ whenever fitted. We stress that higher order moments are the most difficult to reliably measure and thus are more uncertain, especially at larger galactocentric radii where the signal-to-noise is worse. Hence, while we do find meaningful measurements of both $h_3$ and $h_4$ in the inner regions of our target galaxies, this is not necessarily the case in the outskirts where measurement uncertainties start to dominate, thus yielding uncertain higher order kinemetry profiles ($H_3$ and $H_4$) at large galactocentric radii. Similarly, the ATLAS$^{\rm 3D}$ data are more agressively binned as one moves out in radius in order to have a constant signal to noise. Hence, in the overlap region, the SLUGGS data are most reliable, while the opposite is true for the ATLAS$^{\rm 3D}$ data. Nevertheless, it is reassuring that except for a few cases, most kinematic moments and fitted parameters usually match well between the two surveys. We further note that good agreement is only possible if and when the same set of variable and fixed kinemetry parameters are chosen for both SAURON and SLUGGS, emphasizing the importance of analysing both datasets in a self consistent manner for a meaningful comparison.

However, as can be seen from the kinemetry (Fig. \ref{fig:kinemetry}) and as previously reported in \citet{Proctor09} and \citet{Arnold14}, the velocity dispersion measured in SLUGGS is systematically lower than that published by ATLAS$^{\rm 3D}$. We confirm that the mean offset for co-spatial points is $\sim$20 \kms, while the weighted mean is $\sim15$ \kms. In Fig. \ref{fig:offsets}, we compare the mean offset between ATLAS$^{\rm 3D}$ and SLUGGS as measured from the kinemetry at overlapping radii. The measured median offset for all galaxies with radial overlap is $20$\kms, in agreement with that found by \citet{Arnold14}. There is one clear unexplained outlier (NGC~4526) with an offset of $68$\kms. We emphasise that the kinemetry is fitted in a self-consistent manner in each target for both surveys and with consistent bin shapes (hence the radii correspond). There is no clear systematic trend between the amplitude of the offset and the mean velocity dispersion in the overlap radial region, central velocity dispersion nor $V_{\rm rot}/\sigma$. The same is true for individual SKiMS data points. This appears to rule out systematics to do with spectral resolution (either in the templates or spectra) since if that were the case, the offset should correlate with some measure of $\sigma$.

New data for NGC~1023 presented in Pastorello et al. (2015, submitted) has allowed us to show that the offset vanishes in that galaxy when one omits points that overlap with the outer ATLAS$^{\rm 3D}$ data where the spatial binning is more agressive. Indeed, in order to conserve a minimum signal-to-noise ratio throughout their dataset, the ATLAS$^{\rm 3D}$ low signal-to-noise spectra are co-added in the outskirts before fitting the velocity moments. This leads to an artificial enhancement of the measured velocity dispersion. We test the amplitude of this effect  in our target galaxies by comparing the SLUGGS data with only the unbinned ATLAS$^{\rm 3D}$ data. The mean offset is noticeably reduced: down to 13 km s$^{-1}$. Hence, the agressive binning and/or measurement uncertainties in the outskirts of the ATLAS$^{\rm 3D}$ data are artificially enhancing the measured dispersion by $\sim 7$ km s$^{-1}$.

We suspect that the remaining $\sim 13$ km s$^{-1}$ offset may be due to inevitable systematics. For example, in the ideal case of spatially overlapping high signal-to-noise ratio data, with matched spatial coverage, matching wavelength ranges, and matching kinematic templates applied with the same kinematics extraction routine, \citet{Shapiro06} still found a residual offset in velocity dispersion of 5\kms\, between their SAURON and the OASIS IFS data, which they attributed to systematic uncertainties in homogenisation of the spectral resolution across the IFS fields-of-view, and accounted for with a fixed offset. However, this is still much smaller than the offsets measured here. Using a different template library (\citealt{Jones97} versus MILES; \citealt{SanchezBlazquez06}) and non-matched (but largely overlapping) spectral range for the OASIS data, the offset in dispersion increased to $\sim 28$\kms\, (R. McDermid, private communication). Hence, given the large difference in the wavelength ranges probed by SLUGGS and ATLAS$^{\rm 3D}$, it is perhaps not surprising to find such offsets. In other words, these offsets may be due to unavoidable systematics to do with the vastly different wavelengths probed and the use of different spectral templates. We do not find strong evidence for a radial variation in the offset, so therefore correct for it with a constant offset on a per galaxy basis.

With no intended judgement on the quality of either dataset, we subtract the corresponding offset from the ATLAS$^{\rm 3D}$ data to match our SLUGGS values on a per galaxy basis in Fig. \ref{fig:krig}. We point out that a systematic velocity dispersion offset of 20 \kms\, for galaxies with local dispersions of order 150-200 \kms\, does lead to significant systematic uncertainties in the mass estimates of order $\sim 20$ percent. This possibility needs to be taken into account when comparing mass estimates between different surveys. However, relative masses within a given survey are still meaningful. It also has important implications for the outcomes of detailed dynamical modelling that should not be ignored.}

\subsection{Specific angular momentum profiles}\label{sec:kinvar}

We use the kriging maps to compute the \emph{local} specific angular momentum proxy ($\lambda(R)$) at circularised radius R. In practice, for each elliptical bin of $N$ kriging ``pixels'', we compute:
\begin{equation}
\lambda(R)=\frac{\sum_{n=1}^N{R_n|V_{{\rm krig},n}|}}{\sum_{n=1}^NR_n\sqrt{V_{{\rm krig},n}^2+\sigma_{{\rm krig}, n}^2}},
\end{equation}
where $R_{n}$, $V_{{\rm krig}, n}$, $\sigma_{{\rm krig}, n}$, are the circularised radius, fitted velocity and velocity dispersion at the position of the $n^{\rm th}$ ``pixel''. We emphasise that this local $\lambda(R)$ differs from the cummulative $\lambda_{R}$, which is typically defined within a pre-determined radius (i.e. 1$R_{e}$ in ATLAS$^{\rm 3D}$, e.g. \citealt{Emsellem07,Emsellem11}). In order to take full advantage of the generous radial range probed by our data and as we are interested in the variation in \emph{local} specific angular momentum with radius, we purposely avoid using the more ``traditional'' $\lambda_{R}$ definition in order to have an idea of the specific angular momentum variations with radius. Indeed, \citet[][e.g. their figure 12]{Wu14} showed that the luminosity weighted $\lambda_{R}$ radial profiles are invariably either increasing or flat in the outskirts, while the locally determined $|v|/\sigma$ (a different \emph{local} proxy for specific angular momentum) shows much more variation with radius, including sharp drops. The same was also observed in \citet{Coccato09} and \citet{Raskutti14}. We emphasise that $\lambda(R)$ is a proxy, but not the actual angular momentum, which would be derived differently \citep{RomanowskyFall12}. { The local $\lambda_R$ radial profiles are shown in Fig. \ref{fig:LambdaR}.}

{
\begin{figure*}
\begin{center}
\includegraphics[width=190mm]{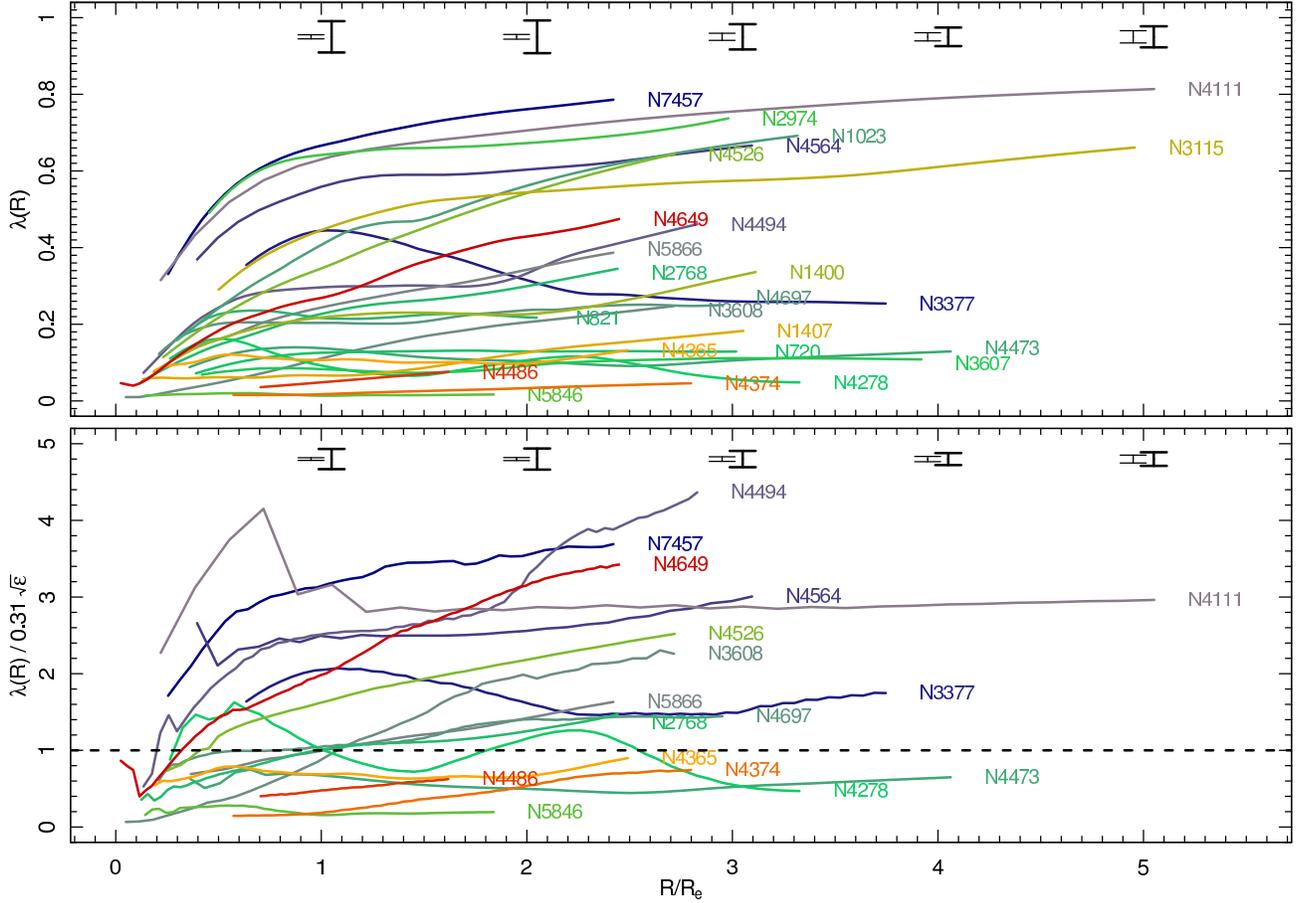}
\caption{The top panel shows the local specific angular momentum proxy $\lambda(R)$ profiles as a function of radius ($R/R_e$) for our sample of 25 galaxies as labelled. The lower panel shows the radial profiles of $\frac{\lambda(R)}{0.31\times \sqrt(\epsilon)}$ for the 17 galaxies with available light profiles. In this lower panel, the dashed line is the transition between fast and slow rotation according to \citet{Emsellem11}. There are large variations in local specific angular momentum with radius. In particular, the galaxy NGC~4278 crosses the dashed line multiple times. In both panels, the minimum (thin errorbars) and maximum (thick errorbars) uncertainties among all galaxies are shown for selected radii. Galaxies are ordered as a function of their central velocity dispersion with purple/red profiles being lower/higher dispersion galaxies.}
\label{fig:LambdaR}
\end{center}
\end{figure*}
}

In order to check whether radial variations in $\lambda(R)$ are associated with variations in the ellipticity with radius, we plot our profiles in { a projection of this} three dimensional space in Fig. \ref{fig:LambdaR}. Variations in $\lambda(R)$ are not always accompanied by variations in $\epsilon$ as one might have naively expected. Moreover, we find that several galaxies show substantial variations of both variables with radius. This is discussed in further detail in Section \ref{sec:mapsvariations}.

\subsection{Kinematic twist and misalignment}

\begin{figure}
\hspace{-0.5cm}\includegraphics[width=90mm]{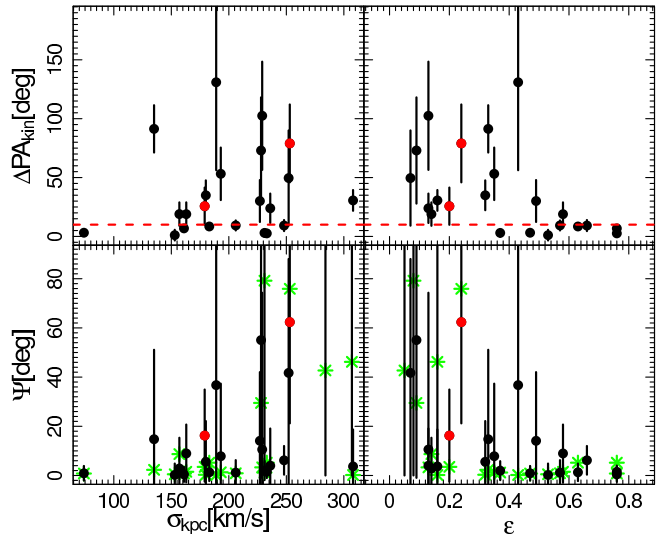}
\caption{Amplitude of inner (measured by ATLAS$^{\rm 3D}$ green asterisks) and outer (filled circles with errorbars) kinematic { twists (top) and misalignments (bottom)} as a function of central velocity dispersion (left) and ellipticity (right). Known central slow rotators are highlighted with red symbols. Galaxies with amplitudes above the red dashed threshold line of 10 degrees present a kinematic twist. { There is a tendency for galaxies with higher velocity dispersions and lower ellipticities to have a variety of kinematic misalignments. An equivalent trend in kinematic twist is only evident with ellipticity.}}
\label{fig:twists}
\end{figure}

To quantify to what measure the galaxies in our sample exhibit shape changes in the outskirts, we look for large scale kinematic twists \citep[KT;][]{Krajnovic11} within the SLUGGS data, which are defined as a smooth variation in $PA_{{\rm kin}}$ of at least 10 degrees throughout the probed radial range. In Fig. \ref{fig:twists}, we show the maximum amplitude of this kinematic twist ($\Delta PA_{\rm kin}$) { over the entire radial range} in degrees as a function of the galaxy central velocity dispersion and ellipticity. Rounder galaxies show a higher fraction of KTs.

\begin{figure}
\begin{center}
\includegraphics[width=84mm]{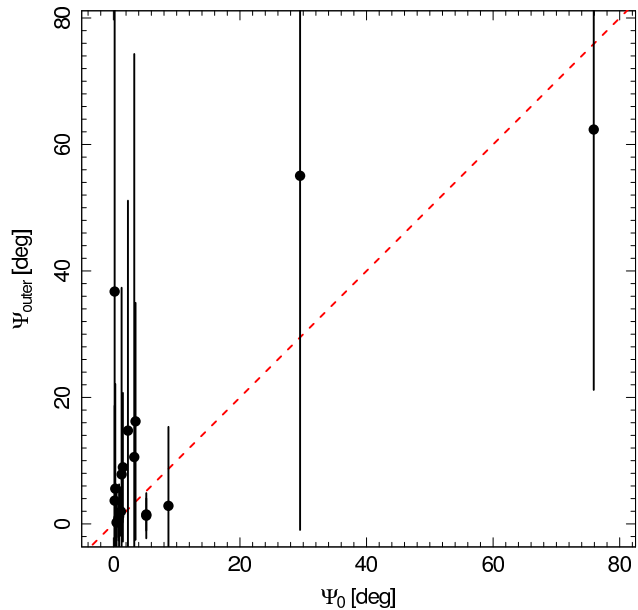}
\caption{Outer kinematic misalignment measured in SLUGGS compared to the central value reported in \citet{Krajnovic11}. The dashed red line shows the one-to-one relation.}
\label{fig:psi}
\end{center}
\end{figure}

Similarly, we also compute the mean kinematic misalignment angle ($\Psi$) { within the measured radial range for each sample} according to 
\begin{equation}
\sin{\Psi}=|\sin(PA_{\rm phot}-PA_{\rm kin})|,
\end{equation}
as in \citet{Krajnovic11}. The $\Psi$ value is a measure of the departure of the kinematic position angle from its photometric value and is a good diagnostic test for complex kinematics. A comparison of the central and outer measured values for $\Psi$ can be found in Table \ref{table:sample} and visualised in Fig. \ref{fig:psi}. While the uncertainties are substantial as a result of sparse sampling, we find good qualitative agreement between the inner and outer values of $\Psi$. 

Kinematic twists and misalignments are discussed in further detail in Section \ref{sec:kinemetryvariations}.

\subsection{Intrinsic (3D) outer shape}\label{sec:Weijmans}

We also attempt to invert the distribution of ellipticities ($\epsilon$) and kinematic misalignments ($\Psi$) to infer the mean intrinsic (three-dimensional) shape of both the inner and outer regions of our sample galaxies. We follow the method outlined in section 4.4 and appendix A of \citet{Weijmans14}. This method aims to recover the mean value of the 3D axis ratios ($p$ and $q$ such that $1\ge p\ge q$) for a sample of galaxies based on their measured $\Psi$ and $\epsilon$ distributions \citep[see also][]{Binney85,Franx91}. If galaxies are on average axisymmetric, we expect their mean $p\sim1$, while a triaxial system would have $p\ne1$. For this analysis to be meaningful and possible, we must make two important assumptions: 1) that the galaxies in our sample are inclined randomly on the sky, and 2) that the intrinsic kinematic misalignment (in 3D) coincides with the viewing direction that generates a null projected ellipticity\footnote{We have also tried setting the intrinsic kinematic misalignment to zero, essentially assuming that any projected misalignment is purely due to projection effects. This only had minimal impact on the output intrinsic shape.}:

\begin{equation} 
\tan\theta_{\rm int}=\sqrt{\frac{T}{1-T}},
\end{equation}
where $T$ is the triaxiality parameter defined in the usual way:
\begin{equation} 
T=\frac{1-p^2}{1-q^2}.
\end{equation}

 As mentioned in \citet{Arnold14}, our sample is biased toward the more rotationally dominated lower luminosity systems. In other words, we have preferentially selected more edge-on inclinations (see figure 19 of \citealt{Brodie14} for a comparison of our sample properties with those of ATLAS$^{\rm 3D}$). Hence, an important caveat is that the first assumption above is not entirely satisfied.

\begin{figure}
\hspace{-0.5cm}\includegraphics[width=94mm]{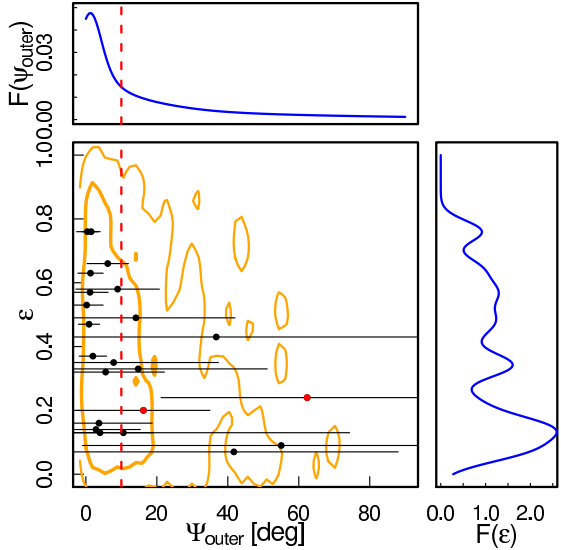}
\caption{Distribution of kinematic misalignment versus global ellipticity of SLUGGS centrally fast (black) and slow (red) rotators. The 1 and 2 $\sigma$ predicted distribution for fast rotators derived from the intrinsic shape analysis is shown in orange. The smoothed density functions $F(\Psi)$ and $F(\epsilon)$ are shown in blue at the top and right, respectively. The dashed red line represents a kinematic misalignment of 10 degrees to guide the eye. There is more scatter in the outer kinematic misalignment for rounder galaxies (i.e. whenever $\epsilon\sim0$).}
\label{fig:psi_eps}
\end{figure}

In practice, we fit for the Gaussian distribution of $q$ with free parameters mean $\mu_q$ and dispersion $\sigma_q$ and a log normal distribution of $p$ such that $Y=\ln(1-p)$ with free parameters mean $\mu_Y$ and dispersion $\sigma_Y$ using optimisation. We first estimate the observed probability density distributions of kinematic misalignments and ellipticities ($F(\Psi)$ and $F(\epsilon)$, respectively) by averaging the Gaussians arising from all data points taking into account their uncertainties. We note that in so doing, the area under the $F(\Psi)$ and $F(\epsilon)$ curves is automatically normalised to one and measurement uncertainties are explicitly taken into account. Since the ellipticity measures have very low uncertainties, we smooth the observed distribution with a constant Gaussian kernel of standard deviation $\sigma_{\epsilon}=0.035$. These smoothed distributions can be visualised in Fig. \ref{fig:psi_eps}.

We use the equations and method described in appendix A1 of \citet{Weijmans14} to translate the intrinsic shape parameters $p$ and $q$ into observed/projected distributions of $\epsilon$ and $\Psi$. In contrast to \citet{Weijmans14}, we simultaneously fit all four parameters. To perform the fit, we minimise the sum of the squares of the differences between the observed ($F_{\rm obs}$) and modelled ($F_{\rm mod}$) projected $i^{\rm th}$ and $j^{\rm th}$ binned distributions of $\Psi$ and $\epsilon$, respectively, as follows:
\begin{equation}\label{eq:3Dchisq}
{\sum}^{2}_{\rm 3D}= A \sum_{i} (F_{\rm obs}(\Psi_{i})-F_{\rm mod}(\Psi_{i}))^{2} + \sum_{j} (F_{\rm obs}(\epsilon_{j})-F_{\rm mod}(\epsilon_{j}))^{2}
\end{equation}
where $A=\frac{\max F_{\rm obs}(\epsilon)}{\max F_{\rm obs}(\Psi)}$ is a scaling factor ensuring that both variables receive similar weights in the optimisation process. We have tried a variety of functional forms for ${\sum}^{2}_{\rm 3D}$ and note that our conclusions are not affected by either the exact form or varying the scaling factor A within reasonable values. We use the {\sc DEoptim r} package to efficiently find the global minimum of Equation \ref{eq:3Dchisq}. This routine uses the principles of differential evolution to rapidly find and converge to the true global minimum \citep[see][for more detail]{Mullen11}. While we allow for up to 500, we find that the solution has converged within less than 100 iterations. Variables may range within the following bounds: $-5\le\mu_Y\le0$, $0.001\le\sigma_Y\le2$ and $0\le\mu_q\le1$, $0.001\le\sigma_q\le1$. The modelled distributions $F_{\rm mod}(\epsilon)$ and $F_{\rm mod}(\Psi)$ are derived in the same manner as the observed ones by assuming corresponding uncertainties for each parameter. We refer the reader to the comprehensive work of \citet{Weijmans14} and references therein for further detail.

We perform this analysis on our entire sample (slow and fast rotators included), on fast rotators only and on the central ATLAS$^{\rm 3D}$ values for fast rotators in common (see Table \ref{table:sample}). The resulting best-fit predicted distribution of $\Psi$ and $\epsilon$ is shown in Fig. \ref{fig:psi_eps}. The best fit models for the outer $\Psi_{\rm outer}$--$\epsilon$ and central $\Psi_{\rm 0}$--$\epsilon$ distributions of (centrally) fast rotator galaxies are further discussed in Section \ref{sec:3Dshape}. As a result of larger uncertainties and smaller sample size, the model contours are not as tight as for the full ATLAS$^{\rm 3D}$ sample presented in \citet{Weijmans14}. Hence, outer triaxiality cannot be ruled out confidently.

\section{Discussion}\label{sec:discussion}

In this section, we provide an interpretation of the results presented in Section \ref{sec:method}. We first look at radial kinematic variations based on the kriging maps and kinemetry fits. Also, our findings regarding the intrinsic shape of our galaxies are discussed. Moreover, we compare and contrast our results with theoretical galaxy formation models.

\subsection{Large scale kinematic variations from stellar kinematic maps}\label{sec:mapsvariations}

A particular advantage of the SLUGGS sample is its ability to probe the outer angular momentum. Using cumulative $\lambda_R$ as defined in ATLAS$^{\rm 3D}$ would emphasise the centre of the stellar component (i.e. $<10$ percent of the total angular momentum), leaving little influence for the outer regions. As we are interested in global angular momentum, we have redefined a ``local'' $\lambda(R)$. Hence, our values are not directly comparable to those published in ATLAS$^{\rm 3D}$.

In a preface to this work, \citet{Arnold14} have analysed the SKiMS data for 22 SLUGGS galaxies. In contrast to this work, they used smoothed, weighted and binned maps instead of kriging. Based on their maps, they also measured the local $\lambda(R)$ parameter along with its variation between 1 and 2.5 $R_{e}$ ($\Delta\lambda$). A clear highlight of their analysis is the realisation that a significant fraction of galaxies centrally classified as fast rotators show declining $\lambda(R)$ profiles. The proposed interpretation of this result is that fast rotators (whether they be ellipticals or lenticulars) may contain multiple kinematic components and that a radial decline in $\lambda(R)$ signals the presence of embedded discs. This is consistent with what was found at large radii using kinematic tracers such as PNe \citep[e.g.][]{Coccato09}. \citet{Arnold14} reported cases of declining angular momentum with radius in the galaxies NGC~821, NGC~3115, NGC~3377 and NGC~4278. This was corroborated by the anti-correlation of $V_{\rm rot}/\sigma$ and $h_{3}$ that usually accompany the $\lambda(R)$ decline, as well as the possible presence of multiple photometric components. While our SKiMS data reach out to much larger galactocentric radii than typical IFS surveys, it still does not probe the full extent of the total angular momentum, which has a characteristic scale-length of $\sim5R_{e}$. \citet{Emsellem11} and \citet{Cappellari11b} have suggested that the widely used morphological classifications are misleading based on their studies of the central kinematics of ATLAS$^{\rm 3D}$ galaxies. However, \citet{Arnold14} postulated that the standard morphological classifications based on photometry alone may still be relevant with centrally fast and slow rotators exhibiting divergent kinematic signatures at larger radii. 

In the same vein, \citet{Foster13} used kinemetry adapted to sparsely sampled data \citep[also see][]{Proctor09,Foster11} to confirm the presence of two counter-rotating disc components in the inner regions of the fast rotator NGC~4473. This galaxy is a so-called 2$\sigma$ galaxy as it exhibits two clear maxima in its velocity dispersion map. They also found a statistically significant minor axis rotating component in its outskirts. It is not clear whether the inner disc and minor axis rotation in the outskirts arise from two separate components or whether the inner disc twists into minor axis rotation in the outskirts. These possibilities remain to be explored. Regardless of the number of sub-components in NGC~4473, it is clear that the combination of both minor and major axis rotation in the outskirts is an unstable configuration and would be short lived in a flattened system. Hence, it is a tell-tale sign of triaxiality and \citet{Foster13} concluded that NGC~4473 is a centrally fast rotating triaxial elliptical galaxy.

\begin{figure}
\hspace{-0.5cm}\includegraphics[width=94mm]{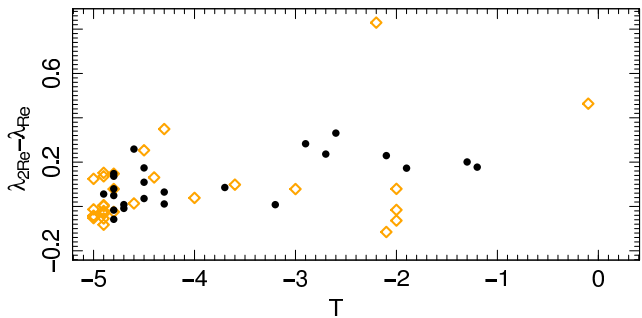}
\caption{{ Measured gradient in local specific angular momentum between 2 and 0.5 $R_e$ for the SLUGGS (black filled circles) and \citet[][orange diamonds]{Raskutti14} samples as a function of morphological type proxy T. Elliptical galaxies are toward the left and lenticulars to the right. While a trend is visible in the SLUGGS data, the \citet{Raskutti14} data show large scatter around T $\sim-2$. This possible discrepancy is discussed in the text.}}
\label{fig:Raskutti}
\end{figure}

On the other hand, \citet{Raskutti14} reported that they found no evidence for drastic radial changes in the local specific angular momentum proxy (called $\Lambda$ in their work) in their sample. { Figure \ref{fig:Raskutti} shows the measured change in local $\lambda_R$ between 2 and 0.5 $R_e$ as a function of morphological type for both SLUGGS and the \citet{Raskutti14} sample. { There seems to be a trend, albeit with some scatter, confirming} the result of \citet{Arnold14} that later type SLUGGS systems exhibit larger gradients. However, this trend is not visible in the \citet{Raskutti14} data alone. Perhaps the large scatter in $\lambda(R)$ gradients around $T=-2$ for the \citet{Raskutti14} sample is a result of selecting the most massive S0s.} They did however report the presence of large scale changes in the kinematic position angle reminescent of the kinematically distrinct haloes (KDHs) first reported in \citet{Foster13}.

In this work, we expand on the above analyses and further look for signs of kinematic changes with galactocentric radius. While \citet{Arnold14} used the SAURON definition of slow and fast rotators ($\lambda_{R}=0.1$), we here use the new definition from ATLAS$^{\rm 3D}$ which also takes into account a galaxy's ellipticity ($\epsilon$) in order to mitigate projection effects. Following \citet{Emsellem11}, fast rotators are now defined as those galaxies which lie above the $\lambda_{R_{e}}=0.31\sqrt{\epsilon}$ line in the $(\lambda_{R_{e}}$--$\epsilon)$ plane. This refinement of the fast versus slow rotator definition requires two dimensions, so in order to visualise variations in rotational support with radius, we plot our radial profiles in the { projected} three-dimensional $(\lambda(R)$--$\epsilon$--$R/R_{e})$-space (Fig. \ref{fig:LambdaR}). We do not see a clear dichotomy in slow versus fast rotators in this space, hence we refrain from labelling individual galaxies as either slow or fast rotators based on their outer kinematics. It is clear that galaxies move around in that space varying between more or less pressure/rotational support with radius. In fact, NGC~4278, moves back and forth through the dividing plane. As in \citet{Arnold14}, we find a significant fraction of centrally fast rotating galaxies show declining angular momentum in their outskirts. Other noteworthy cases are those of NGC~2768 and the massive elliptical galaxy NGC~4365, which interestingly move from the regime of slow into that of fast rotators. In the latter case, this is a result of the decreasing contribution of the kinematically decoupled core with radius. A more pronounced example of this behaviour is the case of  NGC~4649, which transitions from the high pressure support within $R\sim0.5R_e$ to strongly rotationally dominated for radii beyond $R\sim2.5R_e$. { These vatiations are often above and beyond the uncertainties and hence are significant.}

\subsection{Large scale kinematic variations from kinemetry}\label{sec:kinemetryvariations}

The $\lambda(R)$ profiles discussed so far rely solely on the kriging maps. We next show that information about multiple kinematic components can also be gleaned from the kinemetry fits. Indeed, kinemetry allows the precise quantification of the amplitude of KTs ($\Delta PA_{\rm kin}$) and kinematic misalignments ($\Psi$) as well as their variations with galactocentric radius.

The outer values of $\Delta PA_{\rm kin}$ are more uncertain as a result of our sparse SKiMS sampling. For many galaxies, the nominal 10 degrees separation is smaller than the uncertainty on the mean $PA_{\rm kin}$. Fig. \ref{fig:twists} shows the range of measured $\Delta PA_{\rm kin}$ values as a function of central velocity dispersion. Several galaxies (NGC~821, NGC~3377, NGC~4365, NGC~4649 and NGC~4697) have outer monotonic KTs larger than 10 degrees within the uncertainties. Other galaxies show strong variations in $PA_{\rm kin}$ (NGC~3607, NGC~4278, NGC~4473) either because the kinematic solution is noisy due to slow rotation (NGC~4278) or as a result of multiple components such as kinematically distinct haloes  (KDH, e.g. NGC~3607 and NGC~4473) as can be seen in Fig. \ref{fig:kinemetry}.

Fig. \ref{fig:twists} shows the measured maximum amplitude of the outer kinematic twists as a function of central velocity dispersion and ellipticity for all galaxies in our sample. Given that slow rotators tend to be the most massive galaxies (or have the highest $\sigma_{\rm kpc}$) and that slow rotators were shown to have the largest KTs in \citet{Krajnovic11}, one might expect a positive trend between the amplitude of the KT and central velocity dispersion or a higher fraction of KTs in higher mass galaxies. We do not see evidence for such a trend in the outer KT data. Moreover, since flattened galaxies (i.e. $\epsilon\sim1$) are more likely to be fast rotators, one may expect smaller KTs in these galaxies. We confirm that this is the case as can be seein in Fig. \ref{fig:twists}.

We also study the kinematic misalignment between $PA_{\rm kin}$ and $PA_{\rm phot}$ ($\Psi$) as described in Section \ref{sec:kinvar}. A comparison of the inner and outer kinematic misalignment is shown in Fig. \ref{fig:psi}. Once again, uncertainties on this value are quite large, reflecting the limitations of kinemetry applied to sparse data. We find no measurable differences between our outer $\Psi_{\rm outer}$ values and the inner values ($\Psi_{0}$) quoted in \citet{Krajnovic11} within these large uncertainties. This is consistent with results presented in \citet{BarreraBallesteros14} for the CALIFA sample who found that the inner and outer kinematic position angles of gas and stars are usually well aligned.

We do find however that there is more scatter in the measured inner and outer $\Psi$ for rounder or galaxies with higher central velocity dispersions (see Fig. \ref{fig:twists}). Hence, more massive and/or galaxies that appear rounder on the sky are more likely to have their $PA_{\rm kin}$ offset from their $PA_{\rm phot}$ as expected.

On the other hand, kinematic position angle variations are not always mirrored in the $PA_{\rm phot}$ profiles (when available). Again, a case in point is NGC~4473, which shows a very constant photometric position angle at all radii despite its complex distribution of observed kinematic position angles. Once again, this shows that the kinematics can help unveil structures in galaxies that are otherwise not visible photometrically.

\subsection{Large scale intrinsic shape variations}\label{sec:3Dshape}

\begin{figure}
\hspace{-0.35in}\includegraphics[width=104mm]{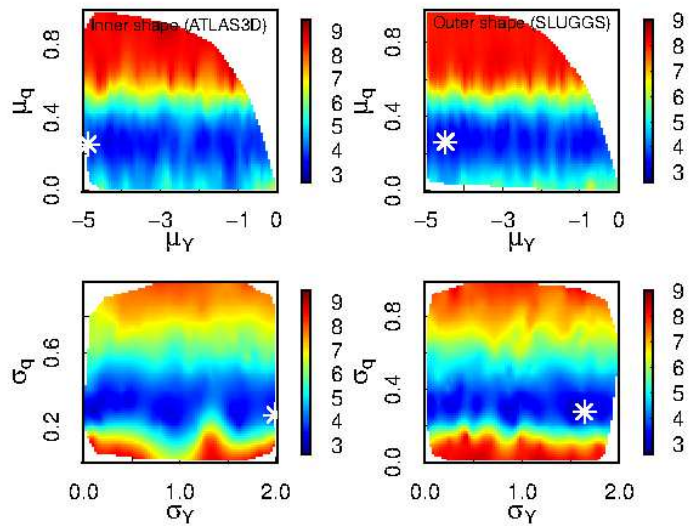}
\caption{Maps of projected measured pseudo-$\chi^2$ (${\sum}^{2}_{\rm 3D}$, see Equation \ref{eq:3Dchisq}) for the intrinsic central shape (mean $\mu$ and dispersion $\sigma$ of the 3D axis ratios $q$ and $Y=\log(1-p)$) of ATLAS$^{\rm 3D}$ fast rotator galaxies in common with SLUGGS (left) and all outer SLUGGS fast rotator galaxies in our sample (right). Lowest values of ${\sum}^{2}_{\rm 3D}$ correspond to best fits. Top panels show $\mu_{Y}$ and $\mu_{q}$ projection maps, while lower panels compare $\sigma_{Y}$ and $\sigma_{q}$. The position of the lowest ${\sum}^{2}_{\rm3D}$ found is shown as a white asterisk. In agreement with \citet{Weijmans14}, the average central and outer intrinsic shape of fast rotators suggest axisymmetry (i.e. $Y=\log(1-p)<<0$, albeit with large $\sigma_{Y}$). However, $\mu_Y$ and $\sigma_Y$ are not well constrained, hence triaxiality cannot be confidently ruled out.}
\label{fig:Chisqmaps}
\end{figure}

{ The kinematic radial transitions seen here in the kriging maps and $\lambda_R$ profiles and in the previous literature may point to multiple stellar components and/or the radially varying intrinsic shapes of galaxies (in 3D) as projected on the sky. One interesting test of this conjecture is afforded by the readily available central (through the ATLAS$^{\rm 3D}$ results) and outer kinematic misalignment measurements in combination with global ellipticities as described in Section \ref{sec:Weijmans}. Fig. \ref{fig:Chisqmaps} shows the output of this exercise. In agreement with \citet{Weijmans14}, we find that galaxies in our sample that are also in the ATLAS$^{\rm 3D}$ sample are consistent with, on average, being centrally axisymmetric (i.e. $Y=\log(1-p)<<0$), albeit with large intrinsic scatter (i.e. large $\sigma_{Y}$). For the outer intrinsic shapes, the best fit model distribution of $\Psi$ and $\epsilon$ can be seen in Fig. \ref{fig:psi_eps}. Again, the output intrinsic shapes can be visualised in Fig. \ref{fig:Chisqmaps}. The same result is obtained for the outer intrinsic shape of fast rotators.

In both cases, however, we note that both $\mu_{Y}$ and $\sigma_Y$ are not well constrained. This may indicate that either 1) the data are limited due to either measurement uncertainties and/or the small sample, thereby prohibiting a constraint on the outer shapes of fast rotators, or 2) there is a range in the intrinsic shape of the outskirts of fast rotators with both triaxial and axisymmetric systems being present. The latter scenario will need to be confirmed with a reliably measured large $\sigma_Y$ value using a larger sample.

The intrinsic flattening ($q$) itself is much better constrained at $q=0.26$, with $\sigma_q=0.27$ for fast rotators, or $q=0.29$ with $\sigma_q=0.26$ for all SLUGGS galaxies. This indicates that the average SLUGGS galaxy is not spherical, but rather quite flattened with one axis being about three to four times larger than another on average. The outer intrinsic flattening is also consistent with the inner value measured using the ATLAS$^{\rm 3D}$ data, suggesting that on average there is no change in the intrinsic flattening of galaxies with galactocentric radius. This result echoes the findings of \citet{Cappellari15}, whose Jeans anisotropic models (JAM) fit well out to $\sim4 R_e$, thus implying that the simple axisymmetric structure extends that far for the fast rotators.

Hence, this analysis cannot detect or rule out outer triaxiality of the average SLUGGS galaxy. It does however show that galaxies on average do not become intrinsically rounder with galactocentric radius, as one might naively assume if a significant fraction of galaxies (including fast rotators) contain embedded discs.}

\subsection{Compiled radial profiles}

\begin{figure}
\hspace{-0.25in}\includegraphics[width=90mm]{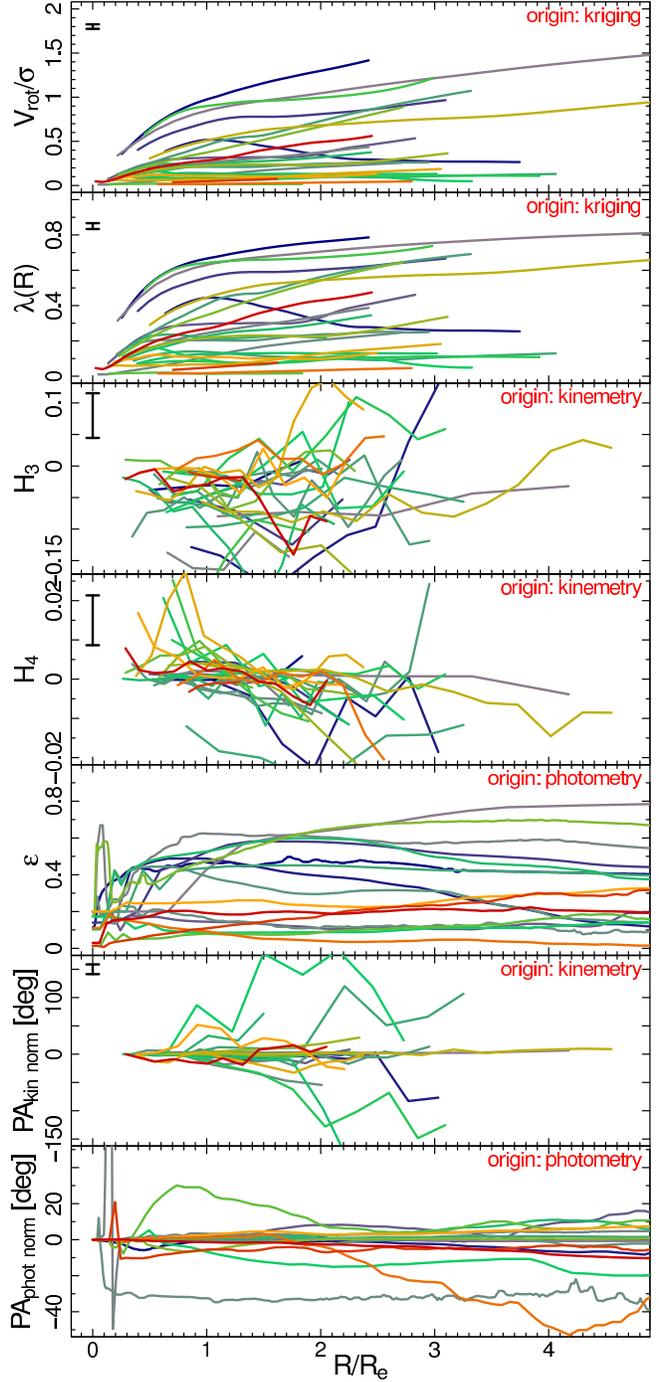}
\caption{{ Compilation of kinematic and photometric radial profiles with colours as per Fig. \ref{fig:LambdaR}.} The top six panels show the compilation of $|V|/\sigma$, $\sigma$, $H_3$, $H_4$, $\lambda(R)$ and ellipticity ($\epsilon$) radial profiles. The lowest two panels show the normalised photometric ($PA_{\rm phot norm}$) and kinematic ($PA_{\rm kin norm}$) position angles profiles where we have subtracted the innermost values from each profile to ease comparison. The origin of the data is labelled in red for each panel { and median uncertainties on kinematic parameters are shown on the top right. The kinemetry and kriging analyses are performed independently.}}
\label{fig:comp_profiles}
\end{figure}

In Fig. \ref{fig:comp_profiles}, we show a compilation of the various radial profiles (both kinematic and photometric, when available). As expected, there is a trend for the most/least massive galaxies to exhibit the lowest/highest local angular momentum (i.e. $\lambda(R)$ and $|V|/\sigma\sim0$ / $\lambda(R)$ and $|V|/\sigma>>0$) and roundest/flattest isophotes (i.e. $\epsilon\sim0$ / $\epsilon>>1$) at all radii. Interestingly, while both $|V|/\sigma$ and $\lambda(R)$ are proxies for the local angular momentum, there is not an exact one-to-one correspondence between the shape and relative position of various galaxies in these spaces. Moreover, there is no clear correspondence between the shape of the $PA_{\rm phot}$ and $PA_{\rm kin}$ profiles for a given galaxy. This again confirms that not all interesting features of a galaxy can be gleaned based on its phometry alone.

\subsection{Assembly and formation histories}

\citet{Naab14} used detailed cosmological hydrodynamical simulations of 44 central galaxies evolved within the two-phase galaxy formation context \citep{Oser10, Oser12} to explain various possible origins for fast and slow rotators as well as inner kinematic features. They  followed the assembly and formation histories of their simulated galaxies within a cosmological context and found that their remnants could be separated into six different classes of assembly and formation histories.  Of these classes, only two did not involve a recent major merger.

{ The data derived in this work have been used in a companion paper (Forbes et al. 2015, submitted) to compare the predicted kinematic properties of the six classes of remnants with those of SLUGGS galaxies. In particular, Forbes et al. examines 2D kinematic maps, 1D local $\lambda_R$ profiles and the distribution of $h_3$ and $h_4$ versus $V/\sigma$ at each spatial position. In Forbes et al. we find, following the classification scheme of \citet[][their table 2]{Naab14}, 14 of our sample galaxies are consistent with having formed through gas-rich minor mergers and gradual dissipation, while 3 galaxies exhibit characteristics of formation via late gas-rich major mergers. One elliptical galaxy is consistent with having formed through a late gas-poor major merger. The remaining 6 galaxies are consistent with having formed through gas-poor mergers (minor and/or major).}

In a companion paper to \citet{Naab14}, \citet{Wu14} have { examined} profiles of $|V|/\sigma$ (an angular momentum proxy) as a function of galactocentric radius out to 5$R_e$ for edge-on galaxies (see e.g. their figure 12). A broad comparison of these theoretical profiles to our observed profiles (see top panel of Fig. \ref{fig:comp_profiles}) shows that the range of values we measure agrees well with theoretical expectations. We do not find, however, the extreme dips in angular momentum beyond $\sim 2 R_e$ for galaxies with $|V|/\sigma>1$ described by \citet{Wu14}. The modulations in our observed profiles are much more gradual.

We have also looked for signs of the velocity dispersion-bump described by \citet{Schauer14}. Based on the results presented in Fig. \ref{fig:kin4111} and Appendix \ref{fig:kinemetry}, we indeed find signs of a sigma bump in NGC~720, NGC~1407 \citep[as first identified by][]{Proctor09} and a possible drop in the velocity dispersion profile of NGC~4365. { In every case, these variations occur around $R\sim60$ arcsec.} 
We note however that the detection of velocity dispersion-bumps in sparsely sampled data can be challenging, especially if the amplitude of the bump is comparable to our uncertainties or if the region of the bump is not particularly well sampled.

\section{Summary and conclusions}\label{sec:conclusions}

Recent years have seen a flury of 2D kinematic studies of galaxies \citep[e.g.][]{Bacon01,Cappellari11a,Sanchez12,Ma14}. Due to the faint nature of galaxy outskirts, most of these studies have typically been limited to the inner effective radius \citep[but see][]{Raskutti14}, thereby only probing half the total stellar mass and less than 10 percent of the total mass and angular momentum. Using the SLUGGS survey \citep{Brodie14} and the SKiMS technique \citep{Norris08,Proctor09}, we are able to extract the stellar kinematics of 25 nearby early-type galaxies out to unprecentedly large galactocentric radii, thereby allowing for the study of the resolved stellar kinematics of the outskirts of early type galaxies.

The following is a summary of our methods and most salient results.

\begin{itemize}
\item We use kriging to model and visualise the spatial distribution of the various velocity moments. Combining with data from the ATLAS$^{\rm 3D}$ survey \citep{Cappellari11a}, we produce 2D maps from the very inner regions out to the faint outskirts. We find generally good agreement between the measured ATLAS$^{\rm 3D}$ and SLUGGS maps, modulo a systematic offset of $\sim20$ \kms\,, such that the velocity dispersion published by ATLAS$^{\rm 3D}$ is systematically higher than that of SLUGGS \citep[also discussed in][]{Arnold14}. This systematic offset is non-negligible and should be a caveat for any modelling and/or mass estimate study. The amplitude of the offset is consistent with previously observed systematics and it is unclear which study (if either) has measured the correct value. 

\item The {interpolated} kriging maps of the recession velocity and velocity dispersion are also used to measure a proxy for the local specific angular momentum. We avoid the traditional, luminosity weighted and cumulative $\lambda_{R}$ parameter in favour of a local version ($\lambda(R)$) as this allows us to better probe local variations in angular momentum with radius. The choice of local $\lambda(R)$ versus cumulative and luminosity weighted $\lambda_{R}$ depends on the specific science case, however we caution that one must use a consistent set of parameters when comparing various studies. We find a diversity of local angular momentum radial profiles ranging from monotonically increasing, flattening or decreasing profiles at large radii, in agreement with previous literature (e.g. \citealt{Proctor09}; \citealt{Arnold14}; \citealt{Coccato09}, but see \citealt{Raskutti14}). In general, we find that the most massive galaxies tend to have the lowest rotation support and local specific angular momentum.

\item We use the kinemetry \citep{Krajnovic06} adapted to sparsely sampled datasets \citep[e.g.][]{Proctor09,Foster13} to fit our data with a simple rotation model. This yields parameters such as the amplitude of the various velocity moments, kinematic position angle and (in principle) the kinematic axis ratio or ellipticity. We find however that fitting for the kinematic axis ratio is notably difficult, even in our best dataset and yields unstable fits. We hence fix it to the photometric axis ratio. We use the kinemetry output to quantify variations in the kinematic position angle greater than 10 degrees (i.e. kinematic twists) and its departure from the photometric value (kinematic misalignment). In agreement with \citet{Krajnovic11}, we find a higher fraction of kinematic twists as well as a larger range of kinematic misalignments in rounder galaxies. We do not find a similar trend with galaxy central velocity dispersion (a proxy for mass), in contrast with previous studies. This is possibly due to the large uncertainties associated with precisely determining the kinematic position angle in sparsely sampled data. A close examination of the kinematic position angle radial profiles reveals that two galaxies in our sample may harbour a kinematically decoupled halo (NGC 3607 and NGC 4473, \citealt{Foster13}). Moreover, we find that kinematic position angle variations are not always accompanied by corresponding variations in the photometry.

\item Dark matter only simulations \citep[e.g.][]{Novak06,VeraCiro11} predict a higher degree of triaxiality in the outskirts of galaxies than in the inner parts. On the other hand, the two-phase galaxy formation scenario \citep[e.g.][]{Wu14} predicts galaxies should have correlated outer and inner shapes. We repeat the analysis performed by \citet{Weijmans14} for our sample of galaxies in order to see whether SLUGGS galaxies are on average more triaxial in the outskirts. \citet{Weijmans14} were able to rule out triaxiality of fast rotators and we are unable to do so in our sample. This may indicate that either 1) there is on average a higher degree of triaxiality in the outskirts of early-type galaxies than in the inner parts or 2) the intrinsic shape of galaxies is not measurable due to our limited sample size and/or large uncertainties. Hence, we cannot rule out the result that early-type galaxies may be more triaxial in their outskirts on average { until a larger sample is available.}


\item The theoretical range in angular momentum profiles reported in \citet{Wu14} agrees well with that observed here. We do not find however the extreme drops in the angular momentum of the fastest rotating galaxies beyond $2R_e$. The modulations in local angular momentum observed in the fastest rotating SLUGGS galaxies are either much more gradual or monotonically increasing.

\end{itemize}

Our work highlights the importance of combining photometry with spatially-resolved inner and outer kinematic studies in order to obtain a complete picture of the properties of galaxies and infer their most likely formation pathways. The added information from the globular cluster component of the SLUGGS survey has the additional potential of probing even further in the haloes of galaxies \citep[e.g.][]{Pota13}.  Moreover, the richness and generous radial extent of the SLUGGS dataset combined with readily available literature data allow for the dynamical bulge-disc decomposition of S0 galaxies \citep[e.g.,][Cortesi et al. 2015 submitted]{Cortesi13}, dynamical decomposition of kinematically decoupled haloes (Mowla et al. in preparation) and dynamical mass modelling \citep[e.g.][]{Napolitano13}.


The stellar kinematic data presented in this work are now publicly available through the SLUGGS website: http://sluggs.swin.edu.au.

\section*{Acknowledgments}
The team would like to acknowledge Gregory Wirth from the W.M. Keck Observatory for his valuable support through many nights on DEIMOS. JPB acknowledges support from NSF grant AST-1211995. DAF thanks the ARC for financial support via DP130100388. 
The data presented herein were obtained at the W.M. Keck Observatory, which is operated as a scientific partnership among the California Institute of Technology, the University of California and the National Aeronautics and Space Administration. The Observatory was made possible by the generous financial support of the W.M. Keck Foundation. 
The analysis pipeline used to reduce the DEIMOS data was developed at UC Berkeley with support from NSF grant AST-0071048. 
Funding for SDSS-III has been provided by the Alfred P. Sloan Foundation, the Participating Institutions, the National Science Foundation, and the U.S. Department of Energy Office of Science. The SDSS-III web site is http://www.sdss3.org/. SDSS-III is managed by the Astrophysical Research Consortium for the Participating Institutions of the SDSS-III Collaboration including the University of Arizona, the Brazilian Participation Group, Brookhaven National Laboratory, Carnegie Mellon University, University of Florida, the French Participation Group, the German Participation Group, Harvard University, the Instituto de Astrofisica de Canarias, the Michigan State/Notre Dame/JINA Participation Group, Johns Hopkins University, Lawrence Berkeley National Laboratory, Max Planck Institute for Astrophysics, Max Planck Institute for Extraterrestrial Physics, New Mexico State University, New York University, Ohio State University, Pennsylvania State University, University of Portsmouth, Princeton University, the Spanish Participation Group, University of Tokyo, University of Utah, Vanderbilt University, University of Virginia, University of Washington, and Yale University.
We acknowledge the usage of the NASA/IPAC Extragalactic Database (NED), which is operated by the Jet Propulsion Laboratory, California Institute of Technology, under contract with the National Aeronautics and Space Administration.

\bibliographystyle{mn2e}
\bibliography{biblio}

\begin{appendix}

\section{Kriging maps}\label{section:krigmaps}

\begin{figure*}
     \begin{center}
        \subfigure{
            \hspace{-0.2in}\includegraphics[width=130mm]{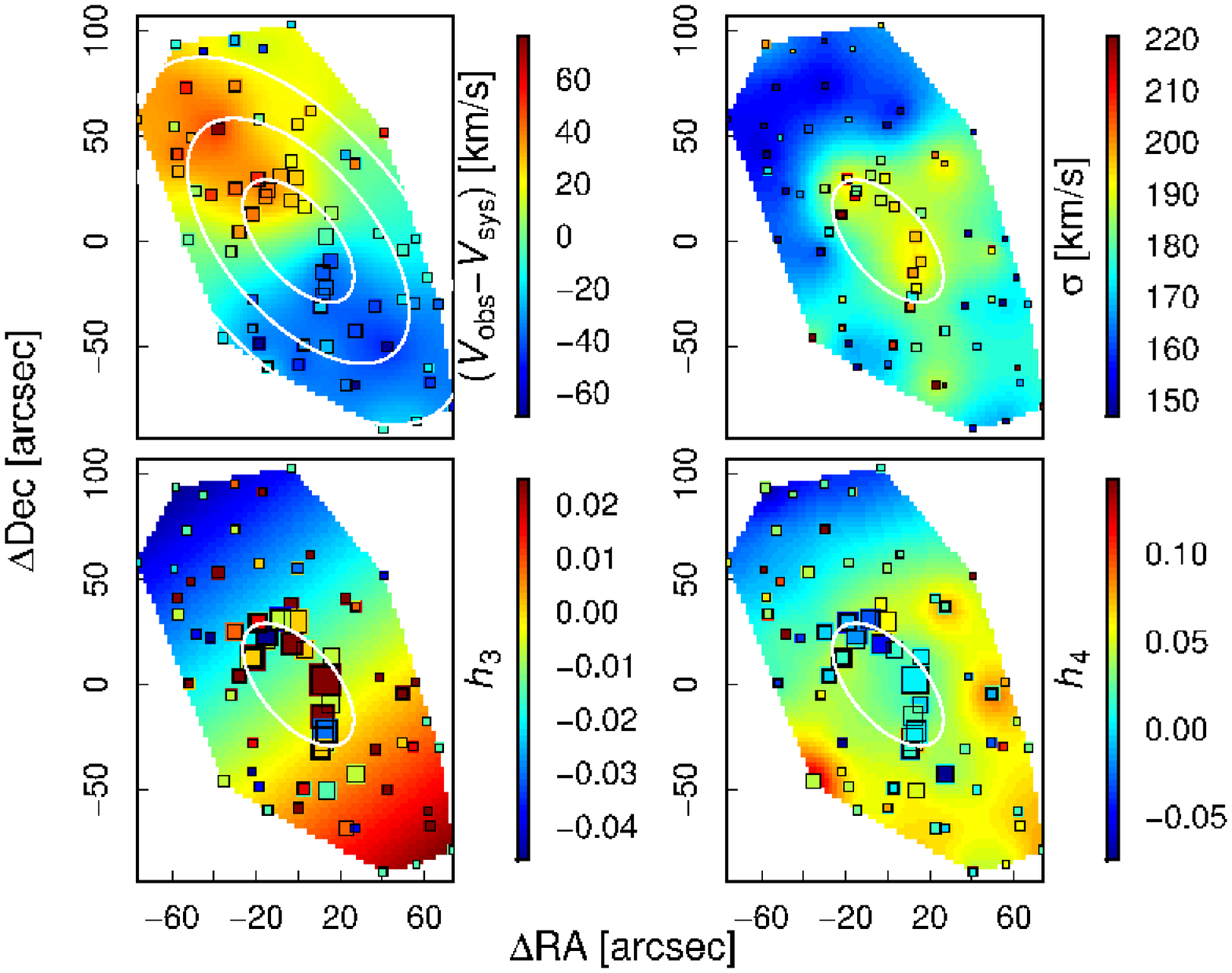}
        }
    \end{center}
    \caption{Velocity (top left), velocity dispersion (top right), $h_3$ (lower left) and $h_4$ (lower right) kriging maps for NGC~720. The central maps from ATLAS$^{\rm 3D}$ are shown within the black polygon whenever available. White ellipses show 1, (2 and 3) $R_e$ in each map (the velocity map) as per Table \ref{table:sample}. Individual SKiMS data are shown as coloured squares with sizes inversely proportional to the uncertainties. The ATLAS$^{\rm 3D}$ velocity dispersions have been offset empirically to match the SKiMS data when possible (see text).}
   \label{fig:krig}
\end{figure*}

\begin{figure*} 
 \ContinuedFloat
     \begin{center}
        \subfigure{
           \hspace{-0.2in}\includegraphics[width=130mm]{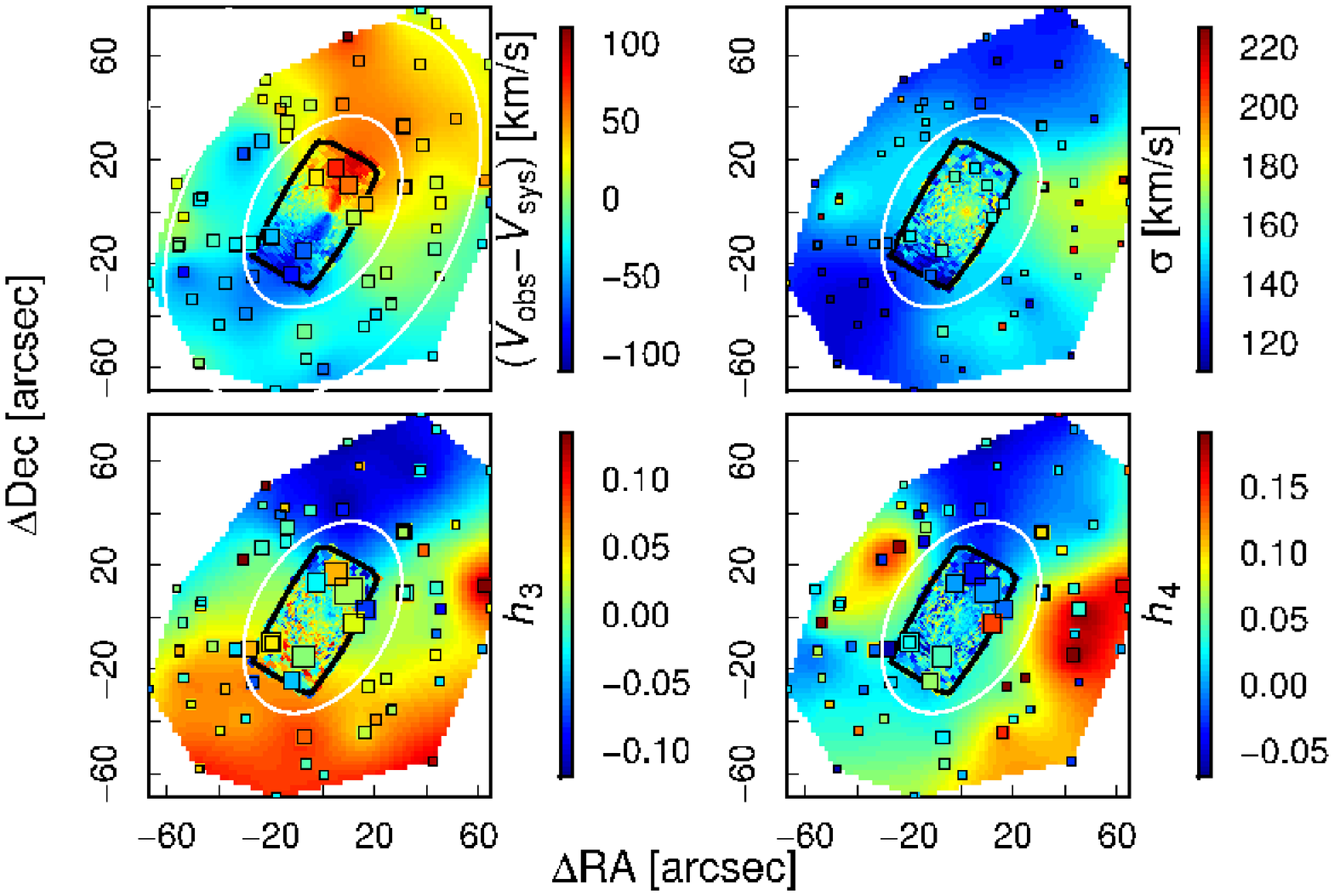}
        }
    \end{center}
    \caption{Continued. Kriging maps for NGC~821. }
\end{figure*}

\begin{figure*} 
 \ContinuedFloat
     \begin{center}
        \subfigure{
            \hspace{-0.3in}\includegraphics[width=130mm]{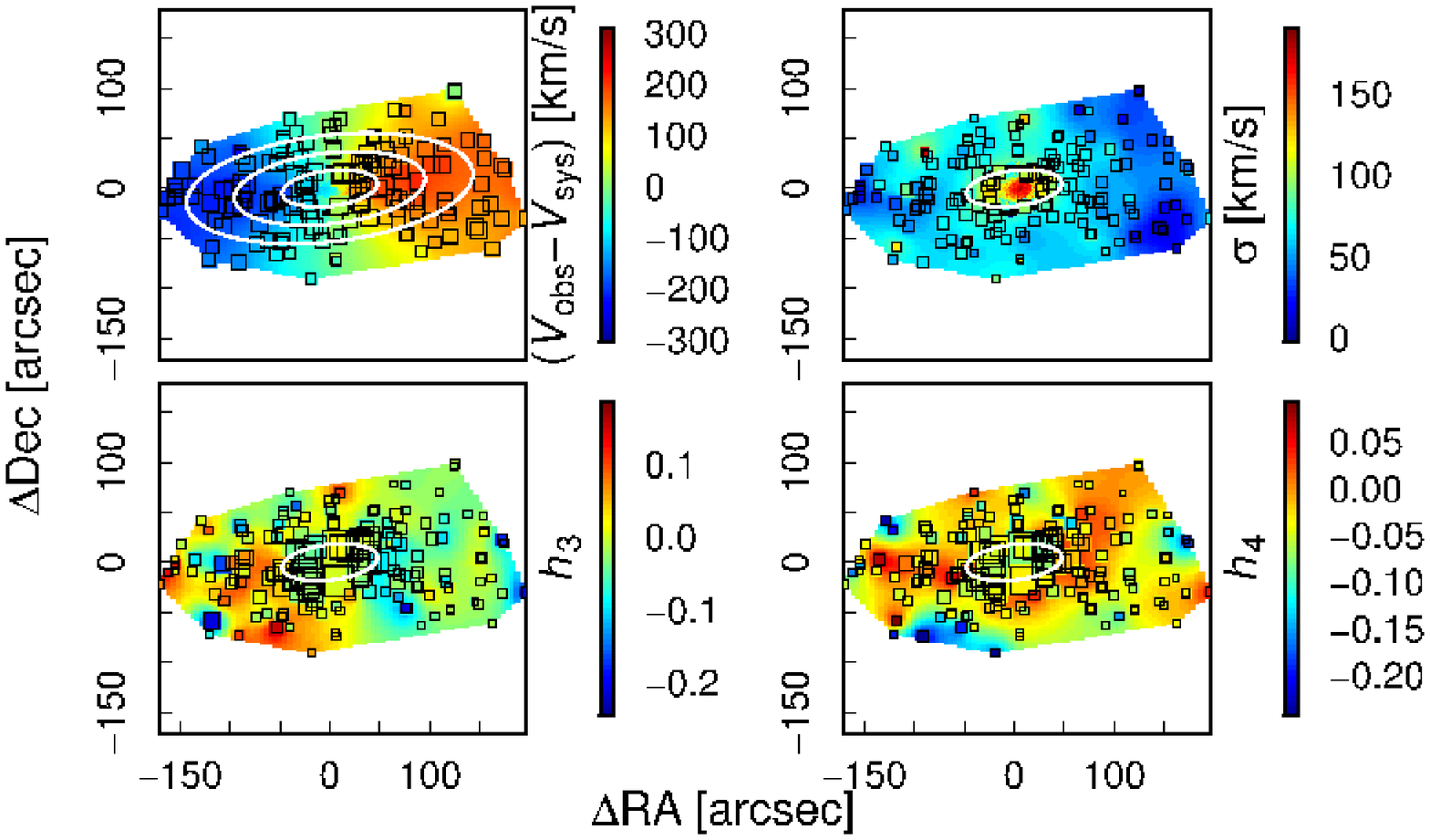}
        }
    \end{center}
    \caption{Continued. Kriging maps for NGC~1023.}
\end{figure*}

\begin{figure*} 
 \ContinuedFloat
     \begin{center}
       \subfigure{
            \hspace{-0.2in}\includegraphics[width=130mm]{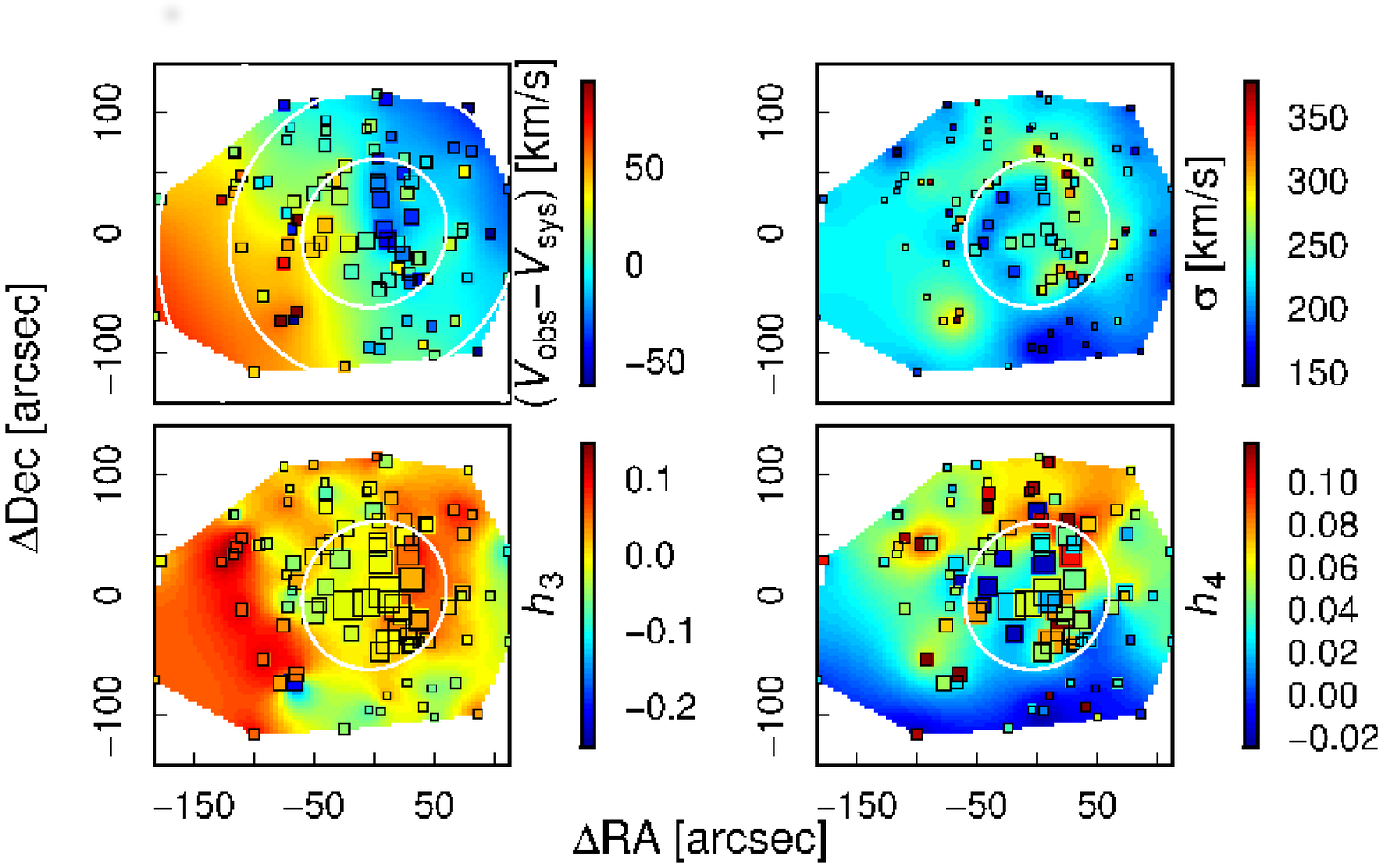}
        }
    \end{center}
    \caption{Continued. Kriging maps for NGC~1407.}
\end{figure*}

\begin{figure*} 
 \ContinuedFloat
     \begin{center}
        \subfigure{
            \hspace{-0.2in}\includegraphics[width=130mm]{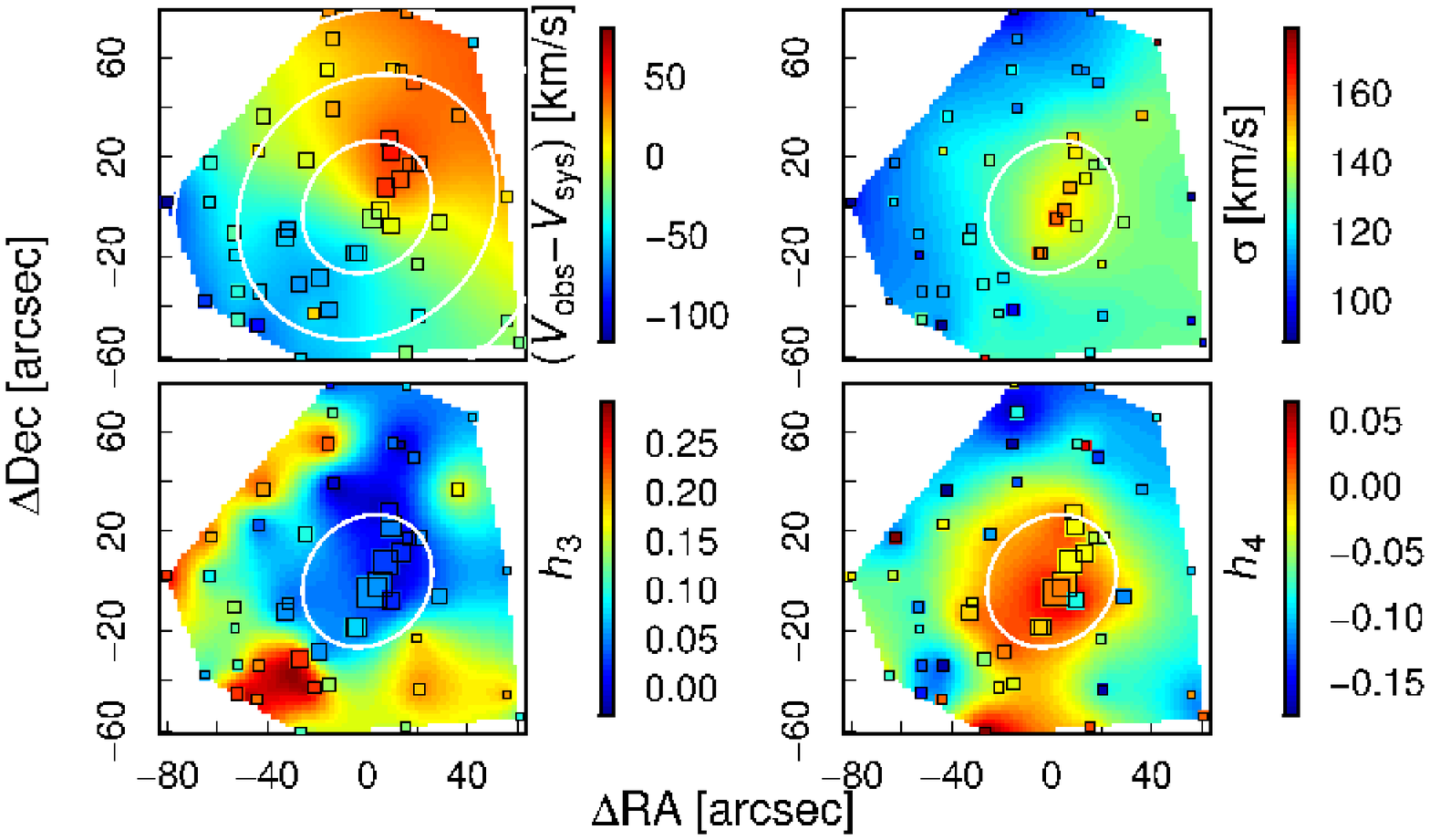}
        }
    \end{center}
    \caption{Continued. Kriging maps for NGC~1400.}
\end{figure*}

\begin{figure*} 
 \ContinuedFloat
     \begin{center}
        \subfigure{
           \hspace{-0.2in}\includegraphics[width=130mm]{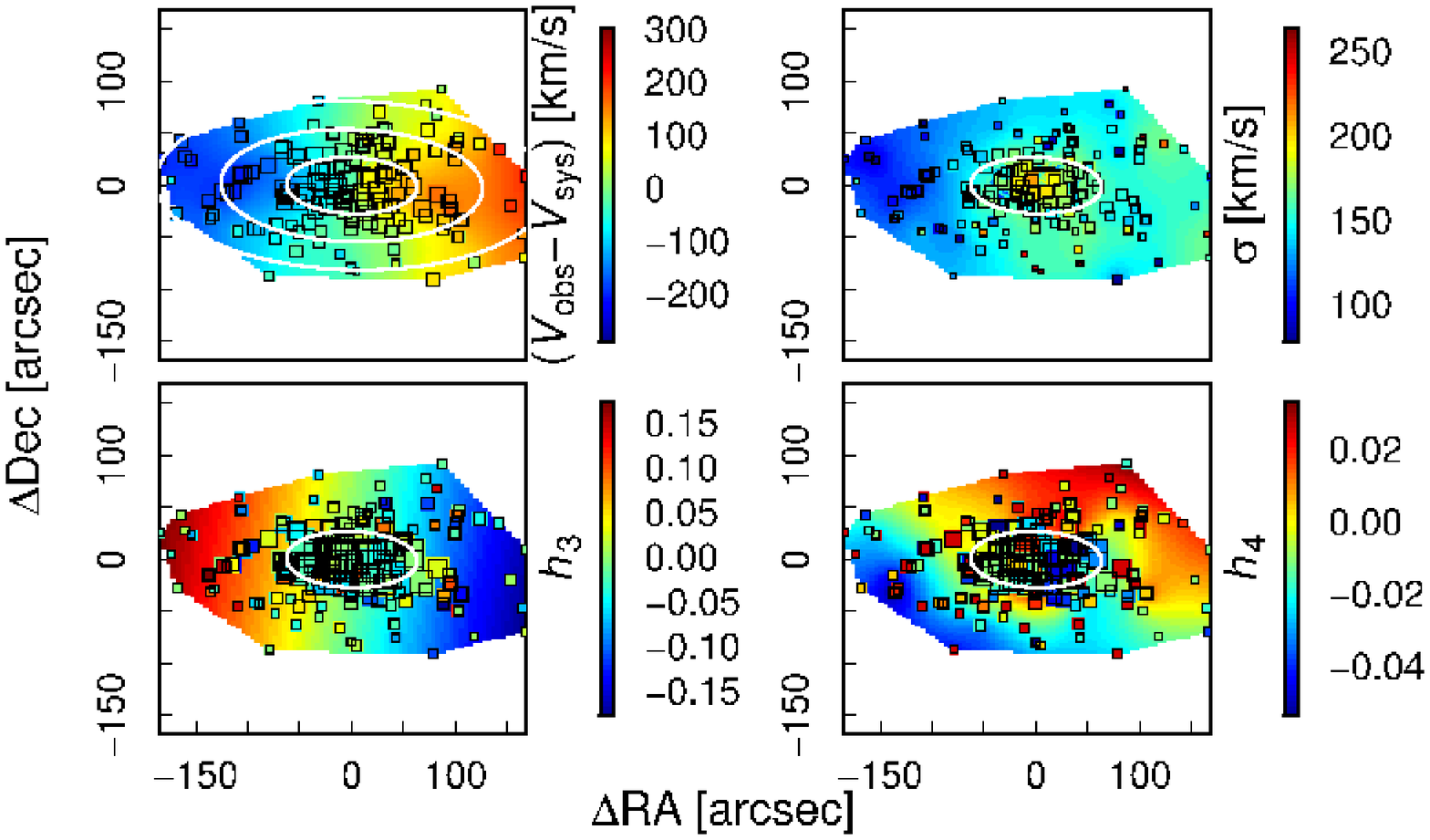}
        }
    \end{center}
    \caption{Continued. Kriging maps for NGC~2768.}
\end{figure*}

\begin{figure*} 
 \ContinuedFloat
     \begin{center}
        \subfigure{
            \label{fig:third}
            \hspace{-0.2in}\includegraphics[width=130mm]{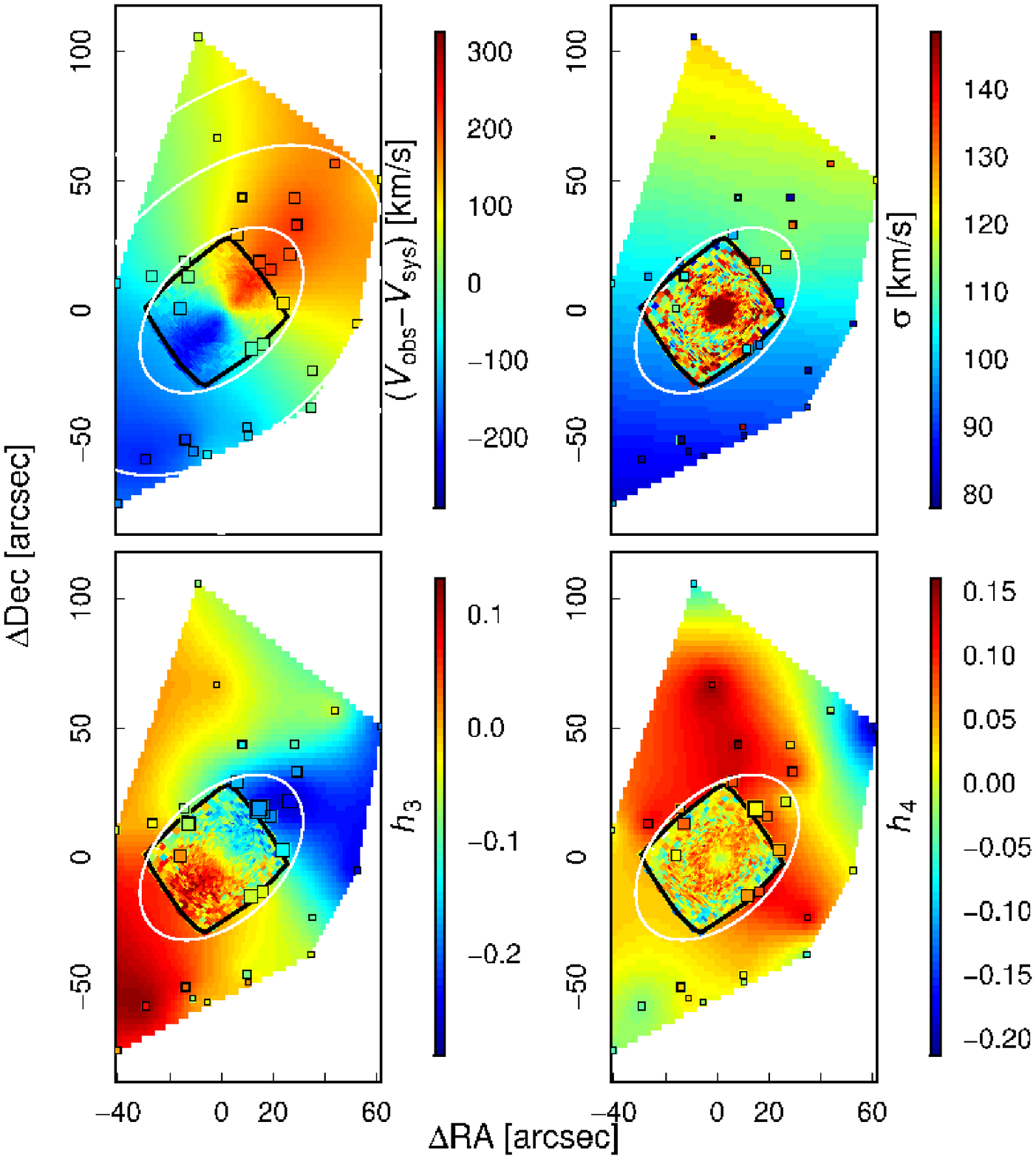}
        }
    \end{center}
    \caption{Continued. Kriging maps for NGC~2974.}
\end{figure*}

\begin{figure*} 
 \ContinuedFloat
     \begin{center}
        \subfigure{
            \hspace{-0.2in}\includegraphics[width=130mm]{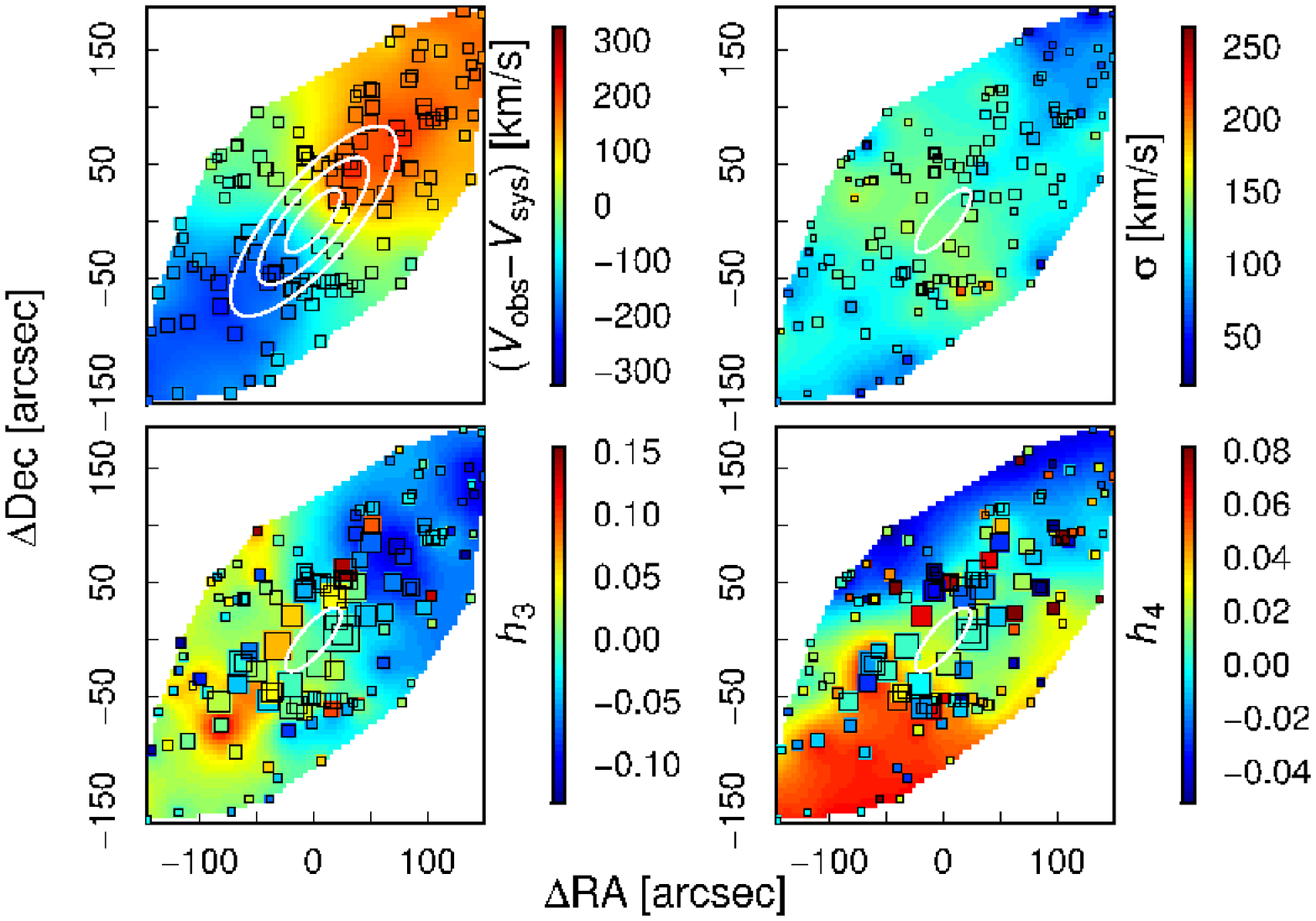}
        }    
    \end{center}
    \caption{Continued. Kriging maps for NGC~3115.}
\end{figure*}

\begin{figure*} 
 \ContinuedFloat
     \begin{center}
        \subfigure{
            \hspace{-0.2in}\includegraphics[width=130mm]{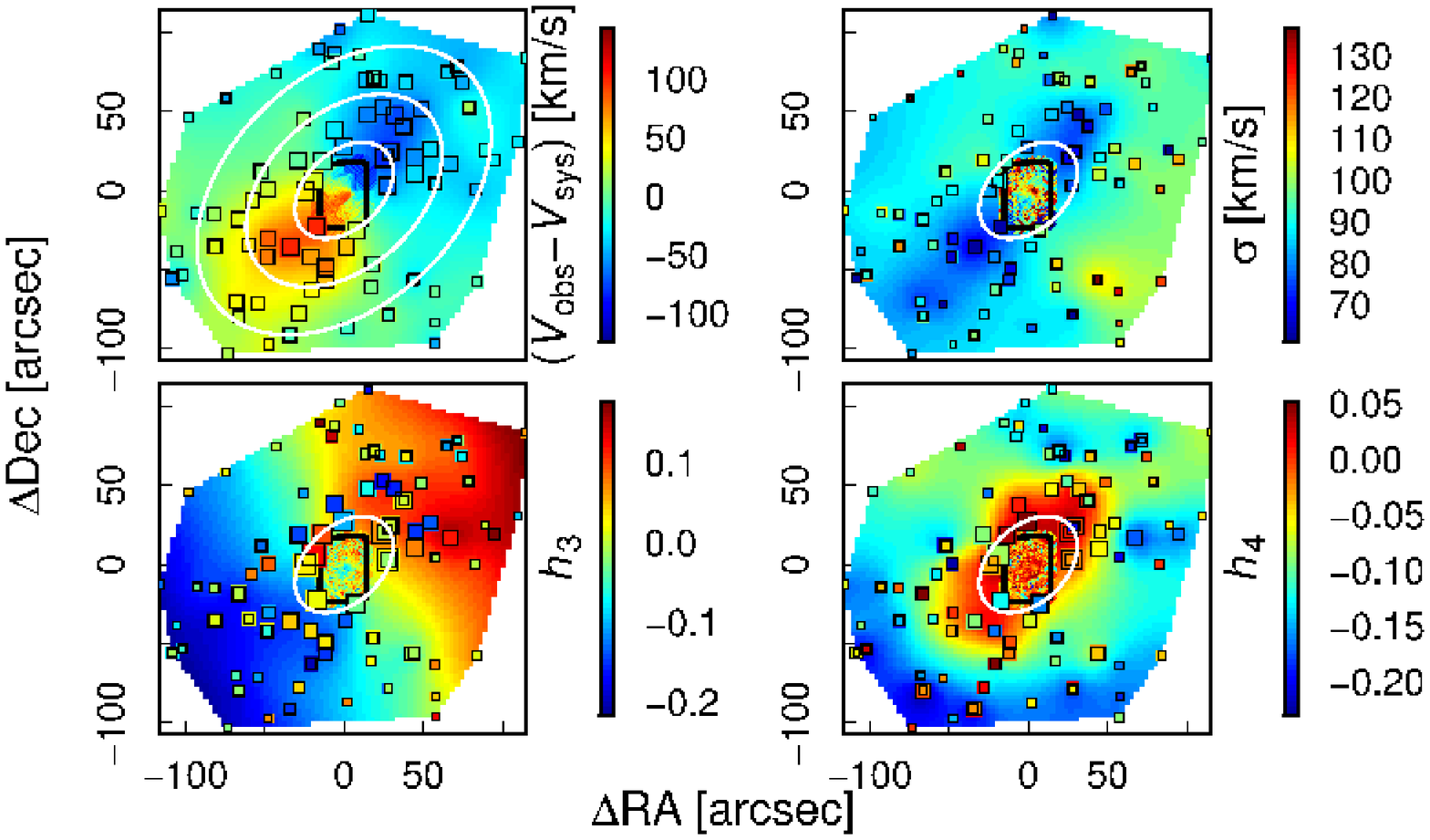}
        }
    \end{center}
    \caption{Continued. Kriging maps for NGC~3377.}
\end{figure*}

\begin{figure*} 
 \ContinuedFloat
     \begin{center}
        \subfigure{
           \hspace{-0.2in}\includegraphics[width=130mm]{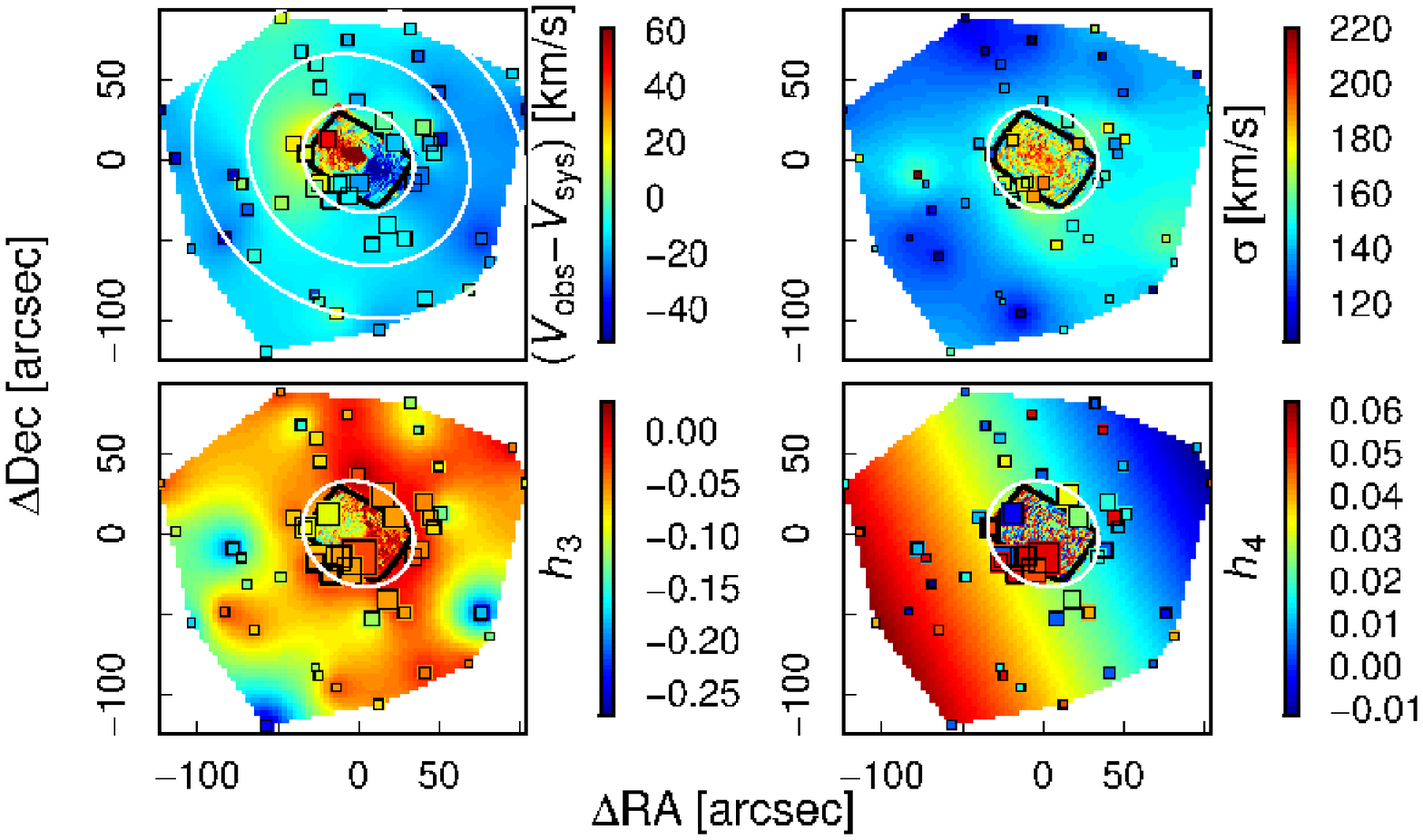}
        }
    \end{center}
    \caption{Continued. Kriging maps for NGC~3607.}
\end{figure*}

\begin{figure*} 
 \ContinuedFloat
     \begin{center}
        \subfigure{
            \hspace{-0.2in}\includegraphics[width=130mm]{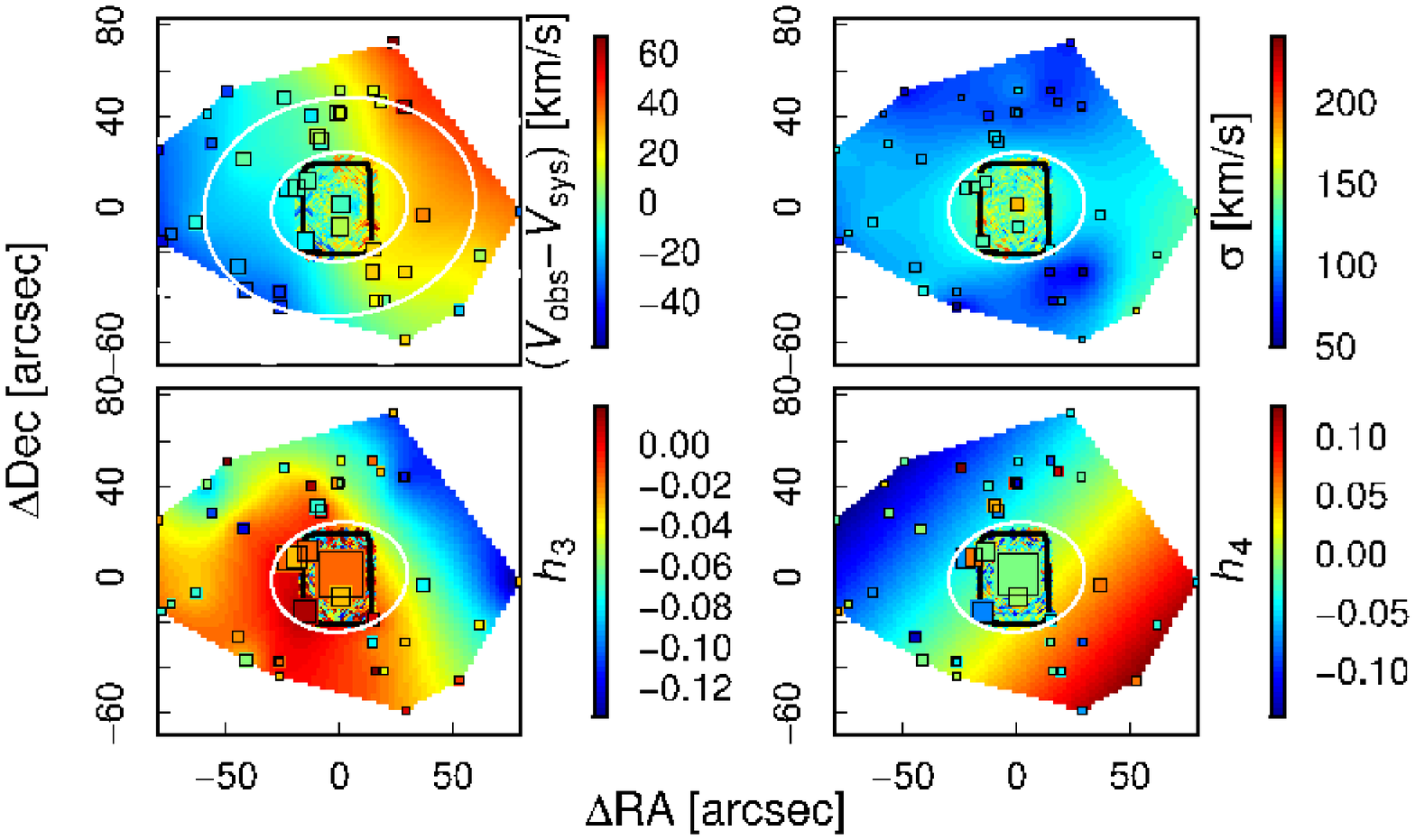}
        }    
    \end{center}
    \caption{Continued. Kriging maps for NGC~3608.}
\end{figure*}

\begin{figure*} 
 \ContinuedFloat
     \begin{center}
        \subfigure{
            \hspace{-0.2in}\includegraphics[width=130mm]{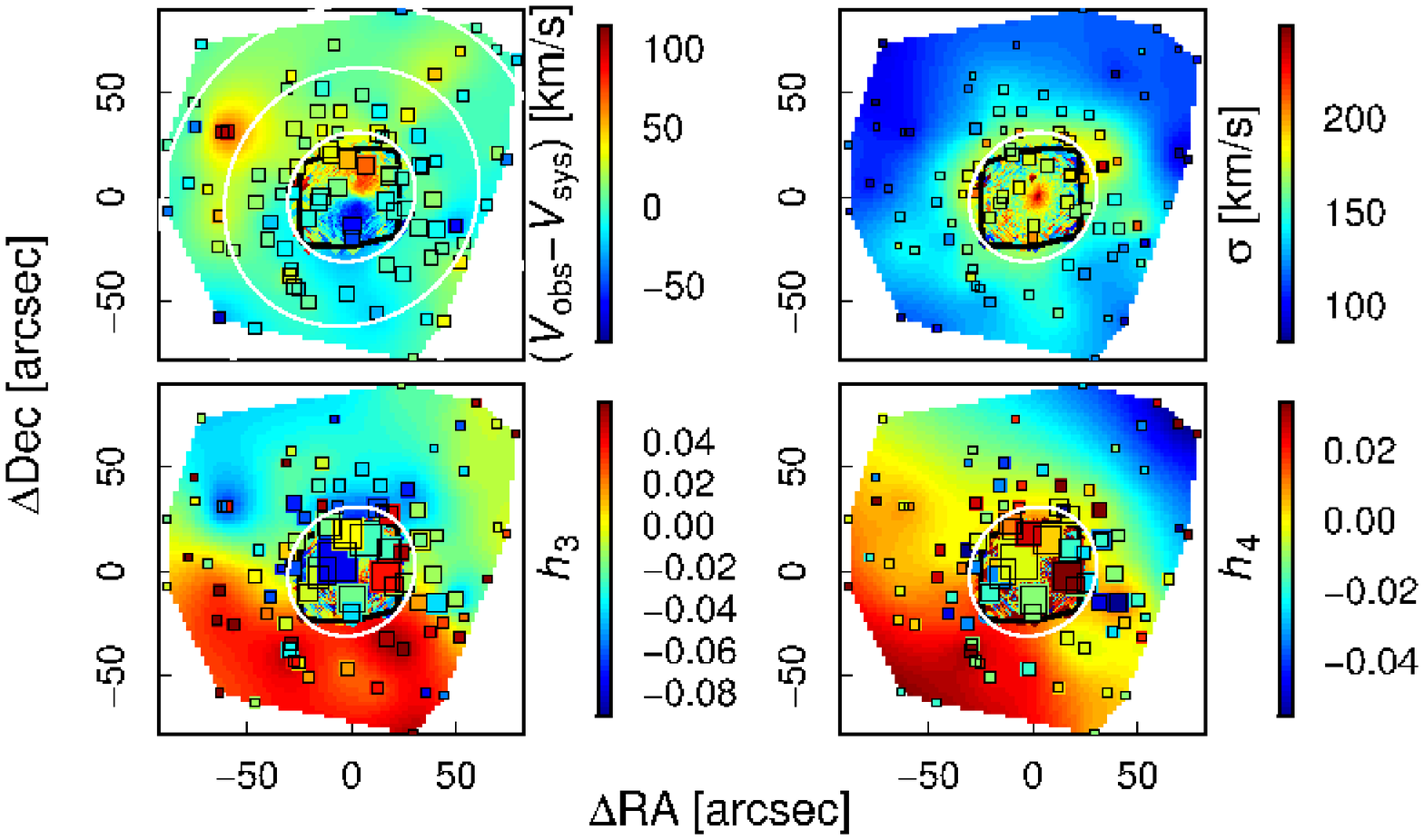}
        }
    \end{center}
    \caption{Continued. Kriging maps for NGC~4278.}
\end{figure*}

\begin{figure*} 
 \ContinuedFloat
     \begin{center}
        \subfigure{
           \hspace{-0.2in}\includegraphics[width=130mm]{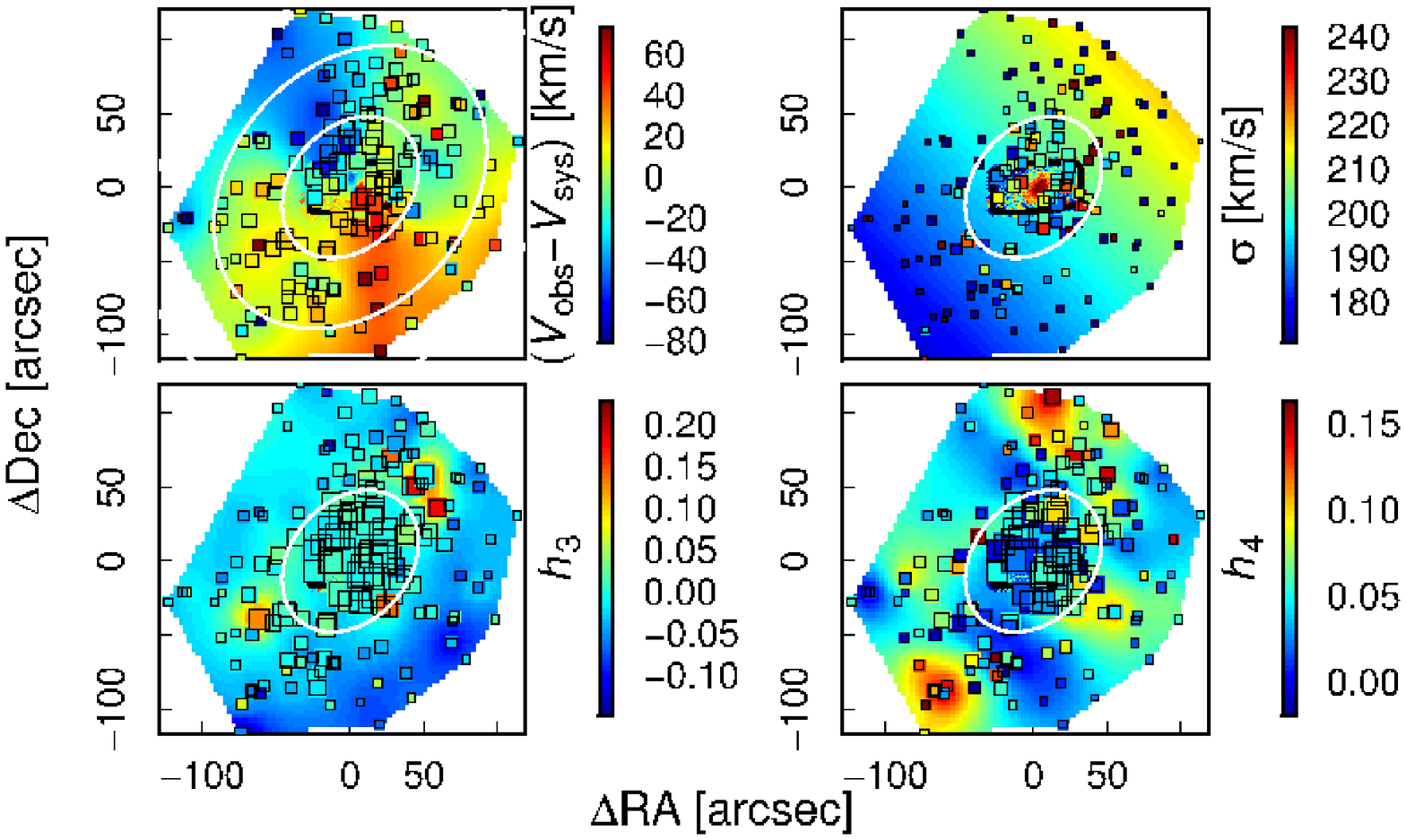}
        }
    \end{center}
    \caption{Continued. Kriging maps for NGC~4365.}
\end{figure*}

\begin{figure*} 
 \ContinuedFloat
     \begin{center}
        \subfigure{
            \hspace{-0.2in}\includegraphics[width=130mm]{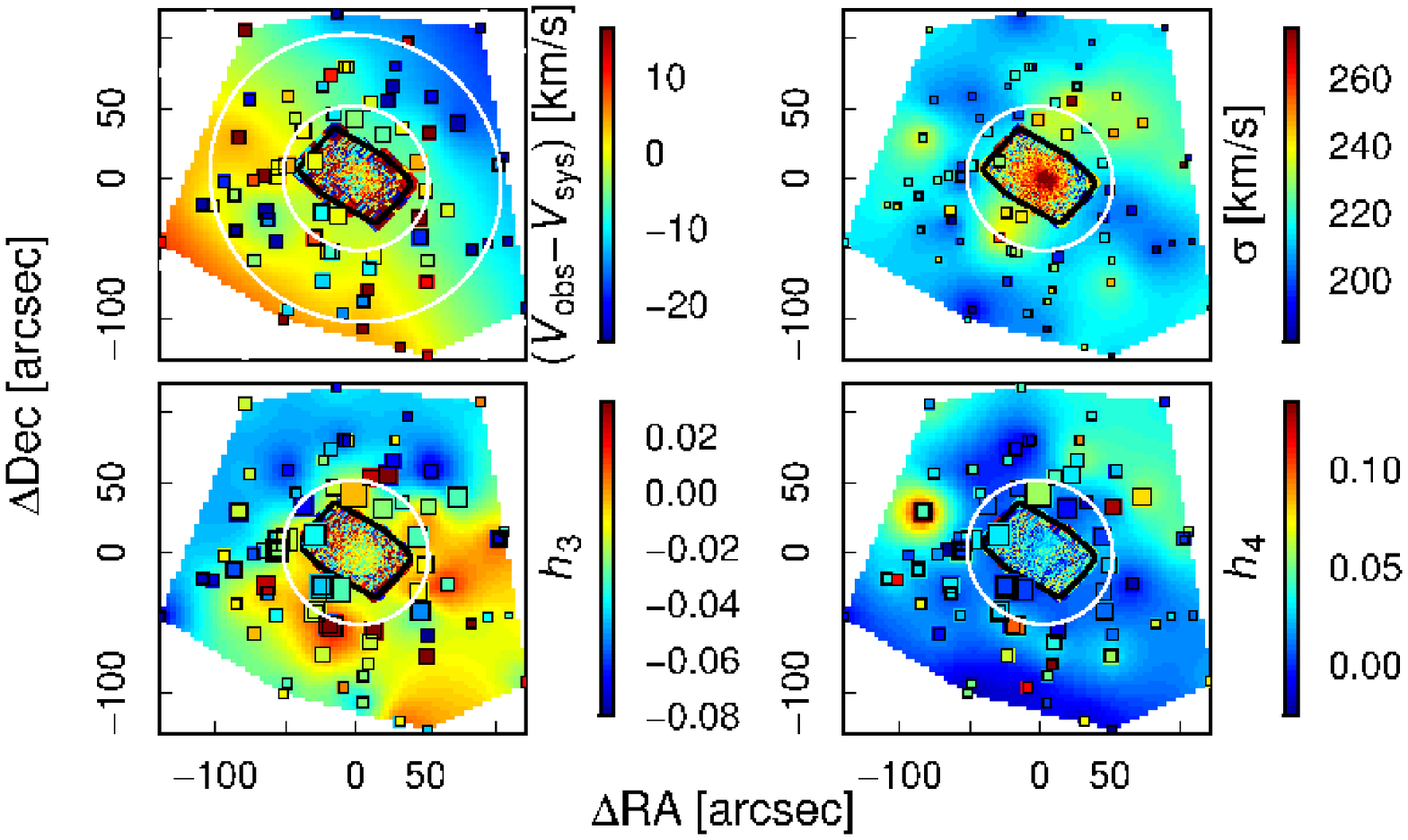}
        }
    \end{center}
    \caption{Continued. Kriging maps for NGC~4374 }
\end{figure*}

\begin{figure*} 
 \ContinuedFloat
     \begin{center}
        \subfigure{
            \hspace{-0.2in}\includegraphics[width=130mm]{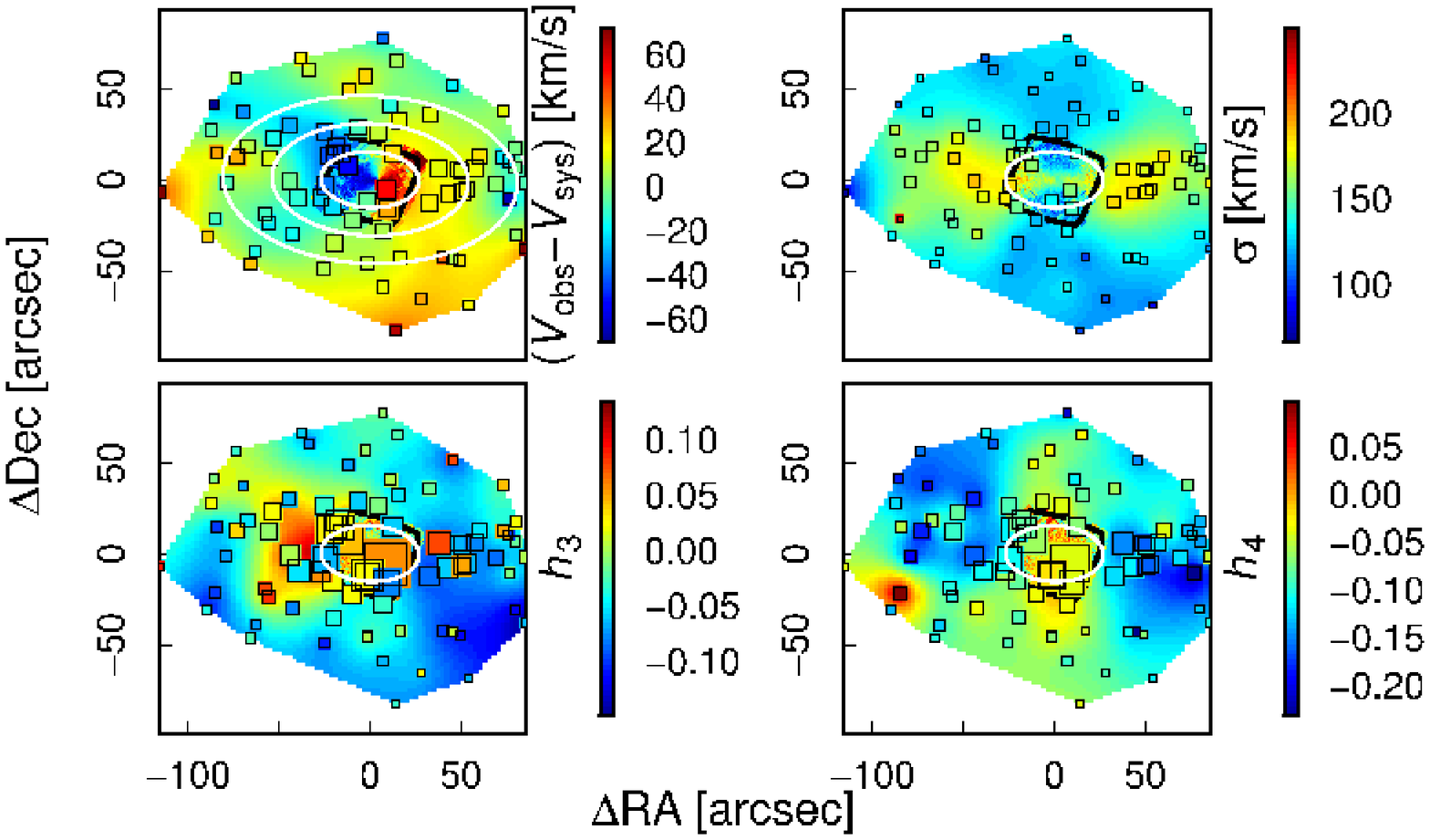}
        }
    \end{center}
    \caption{Continued. Kriging maps for NGC~4473.}
\end{figure*}

\clearpage

\begin{figure*} 
 \ContinuedFloat
     \begin{center}
        \subfigure{
            \hspace{-0.2in}\includegraphics[width=130mm]{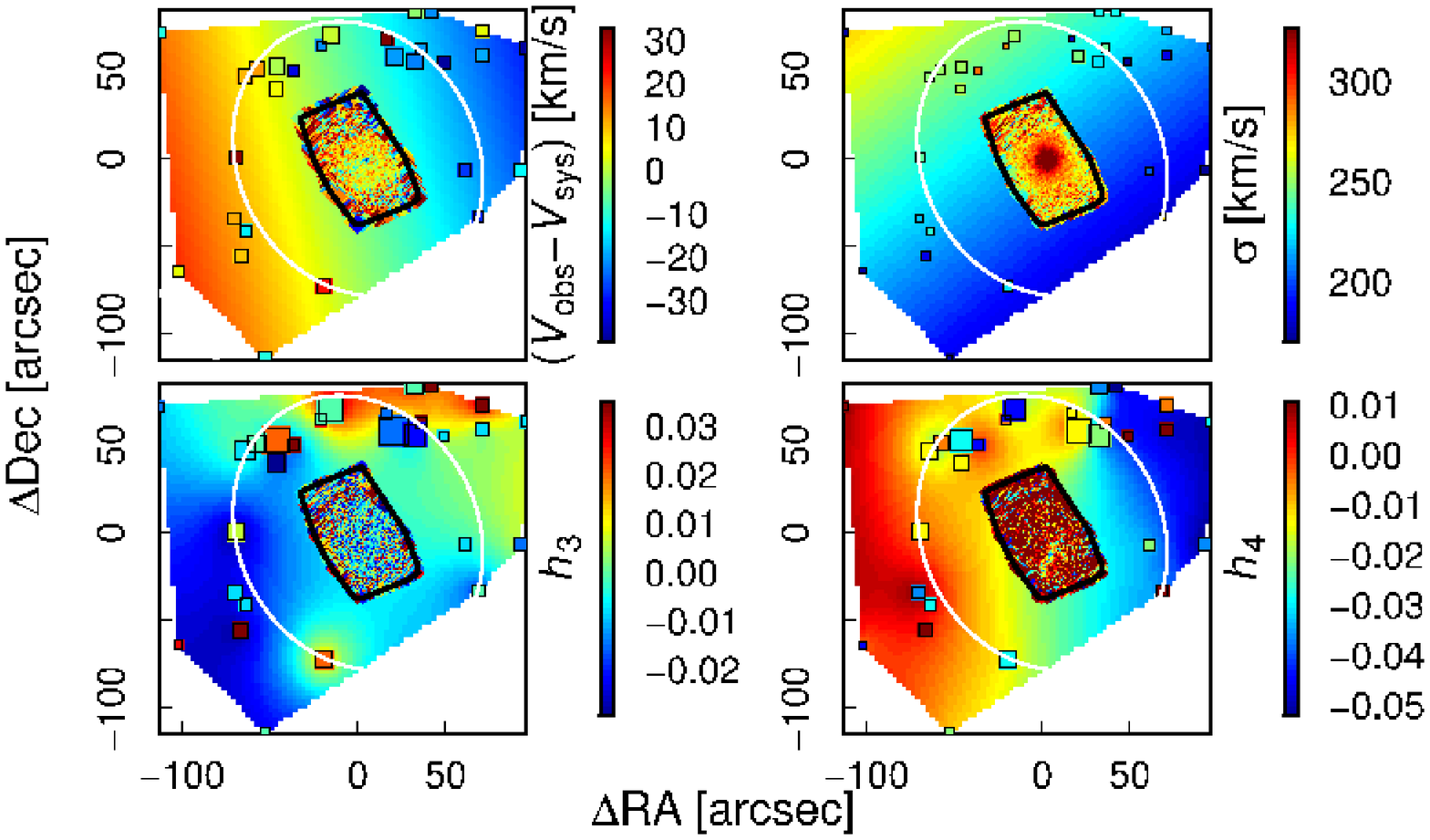}
        }
    \end{center}
    \caption{Continued. Kriging maps for NGC~4486.}
\end{figure*}

\begin{figure*} 
 \ContinuedFloat
     \begin{center}
        \subfigure{
           \hspace{-0.2in}\includegraphics[width=130mm]{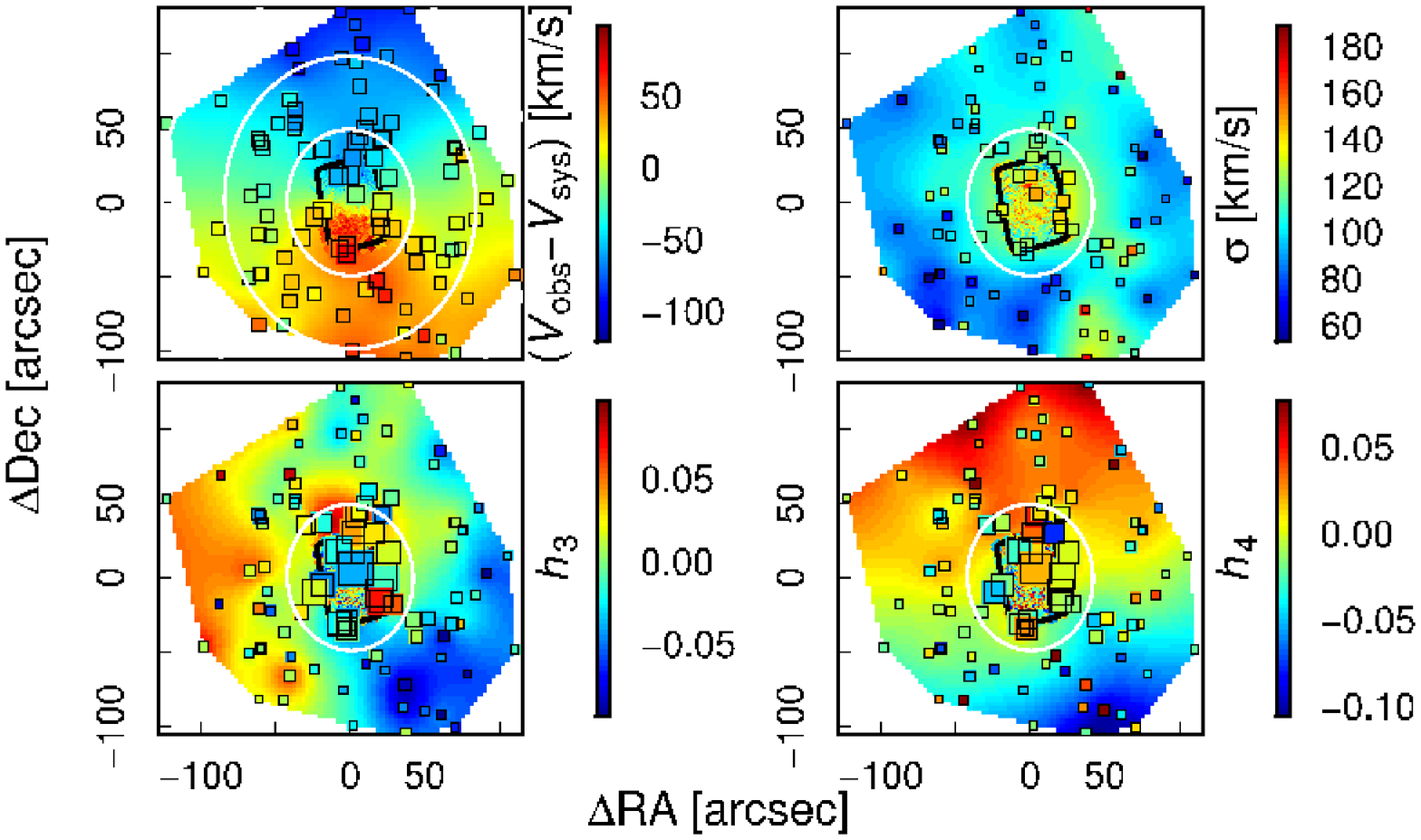}
        }
    \end{center}
    \caption{Continued. Kriging maps for NGC~4494.}
\end{figure*}

\begin{figure*} 
 \ContinuedFloat
     \begin{center}
        \subfigure{
            \hspace{-0.2in}\includegraphics[width=130mm]{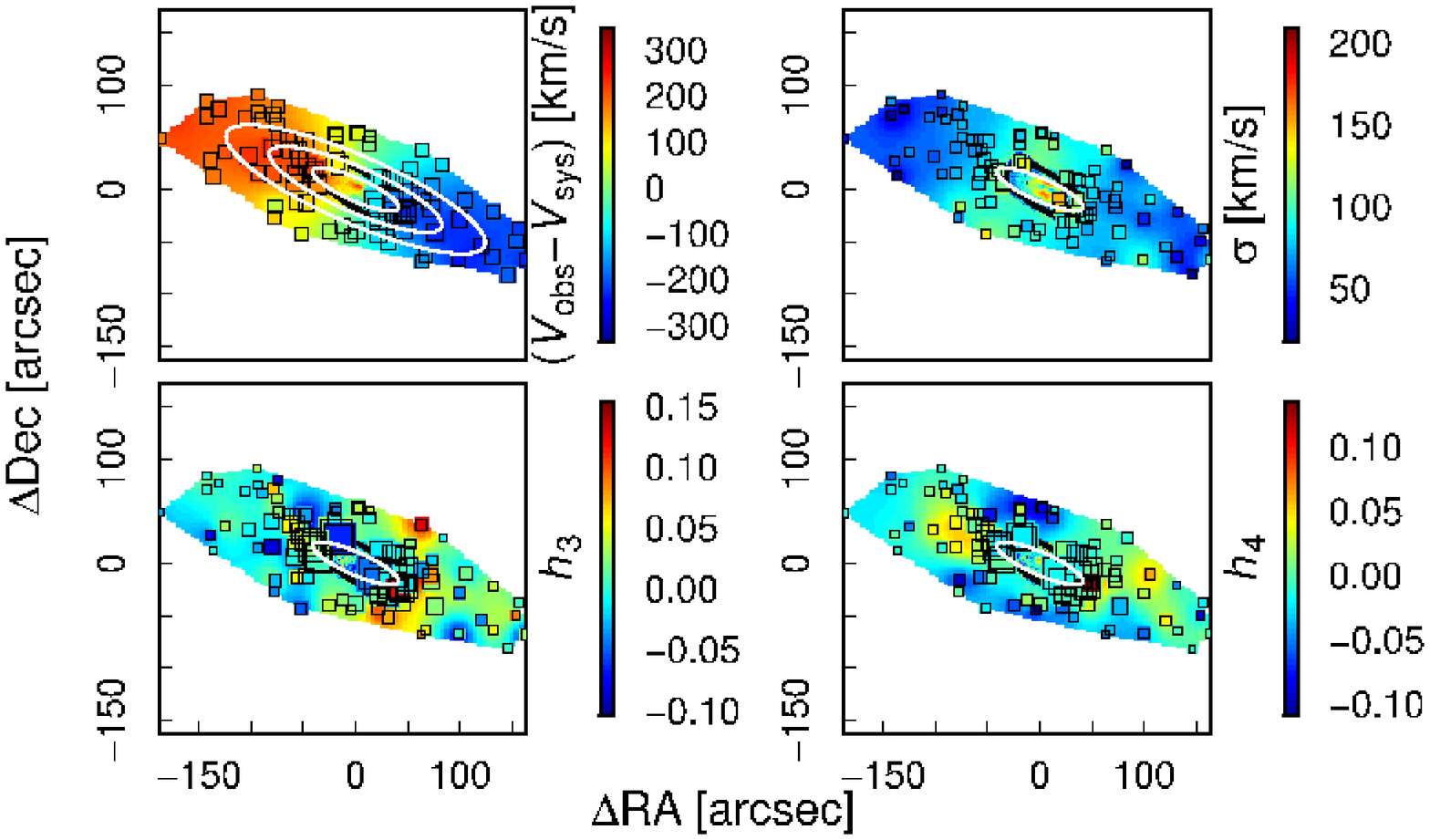}
        }
    \end{center}
    \caption{Continued. Kriging maps for NGC~4526.}
\end{figure*}

\begin{figure*} 
 \ContinuedFloat
     \begin{center}
        \subfigure{
            \hspace{-0.2in}\includegraphics[width=130mm]{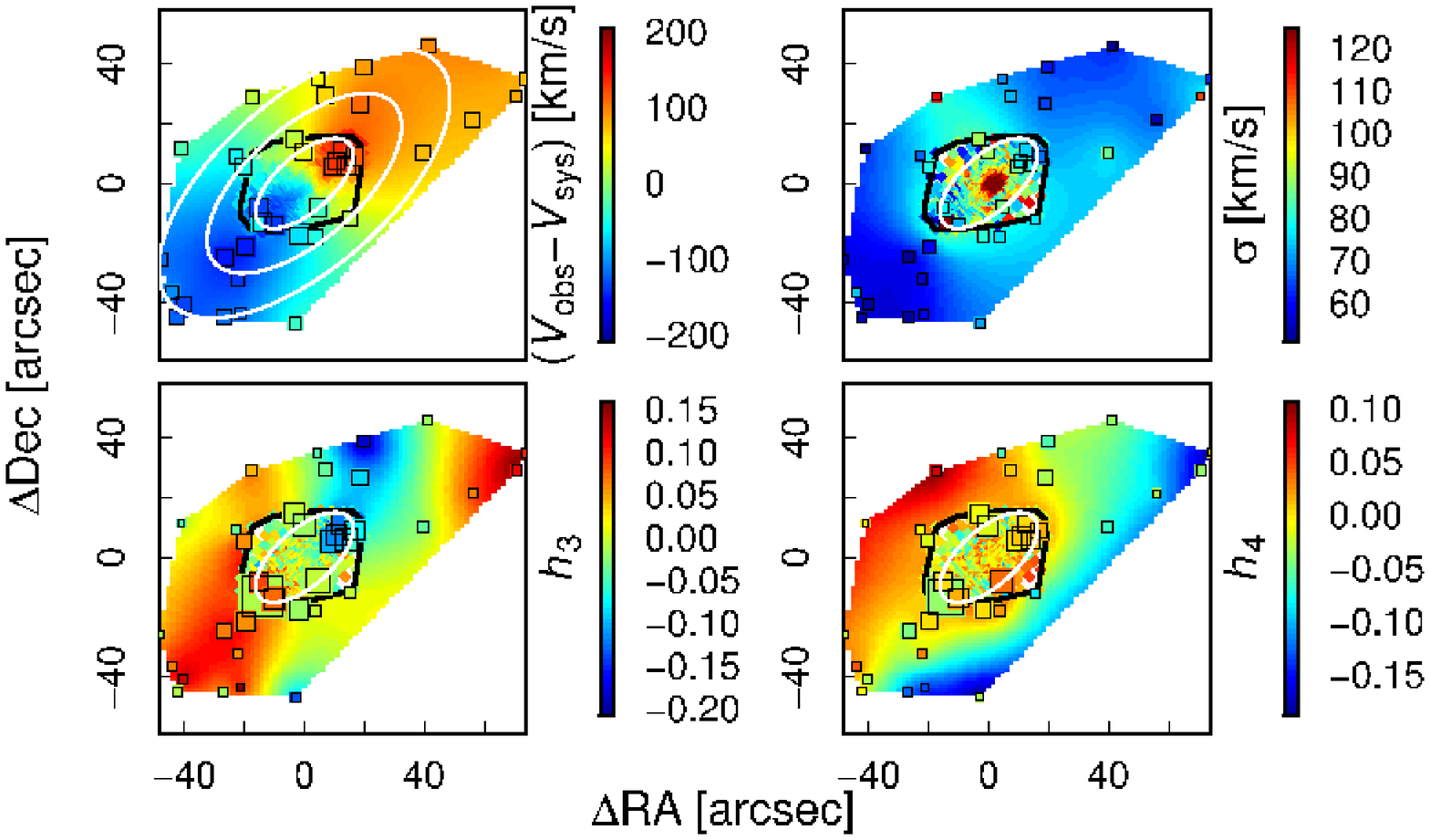}
        }
    \end{center}
    \caption{Continued. Kriging maps for NGC~4564.}
\end{figure*}

\begin{figure*} 
 \ContinuedFloat
     \begin{center}
        \subfigure{
            \hspace{-0.2in}\includegraphics[width=130mm]{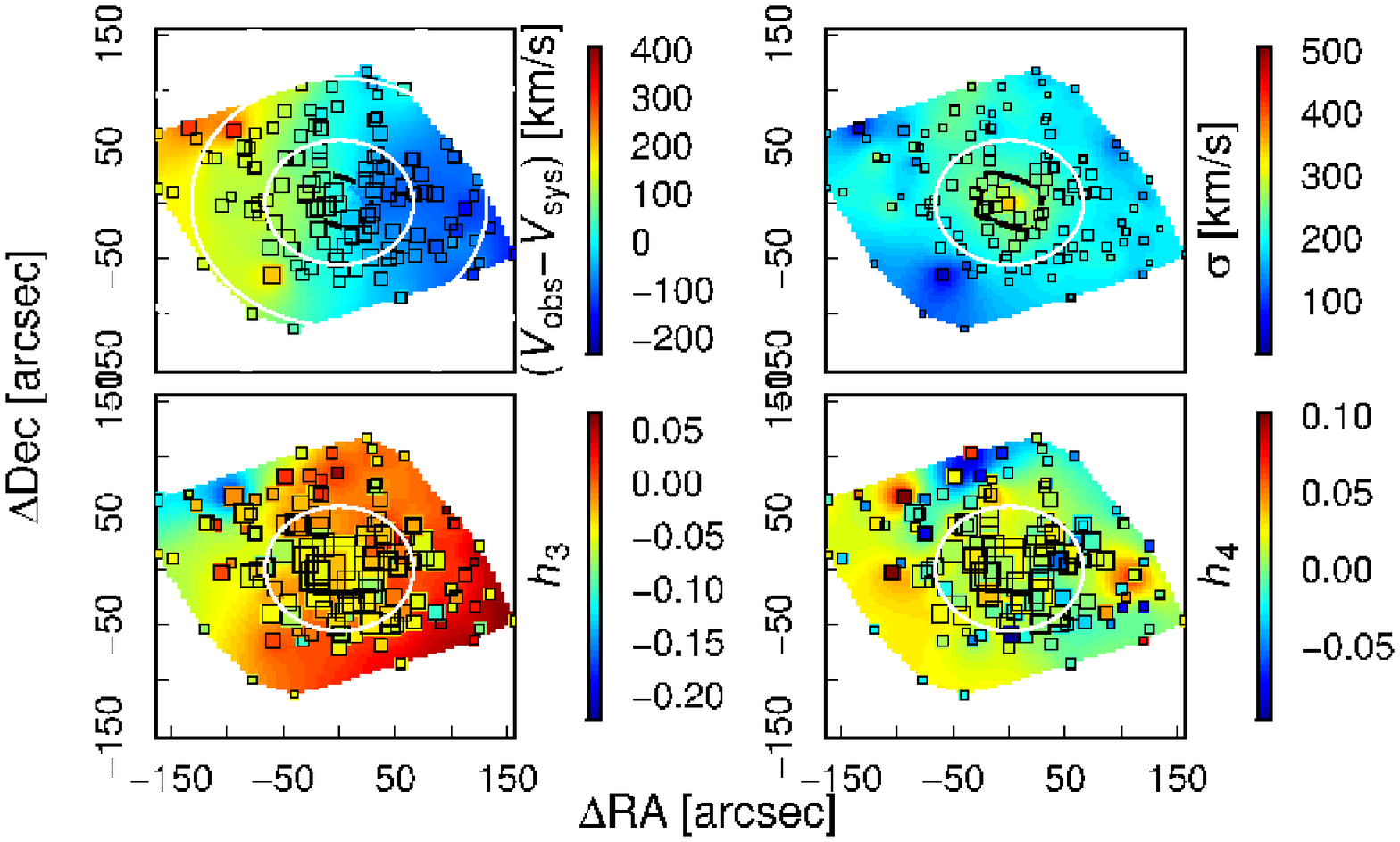}
        }
    \end{center}
    \caption{Continued. Kriging maps for NGC~4649.}
\end{figure*}

\begin{figure*} 
 \ContinuedFloat
     \begin{center}
        \subfigure{
            \hspace{-0.2in}\includegraphics[width=130mm]{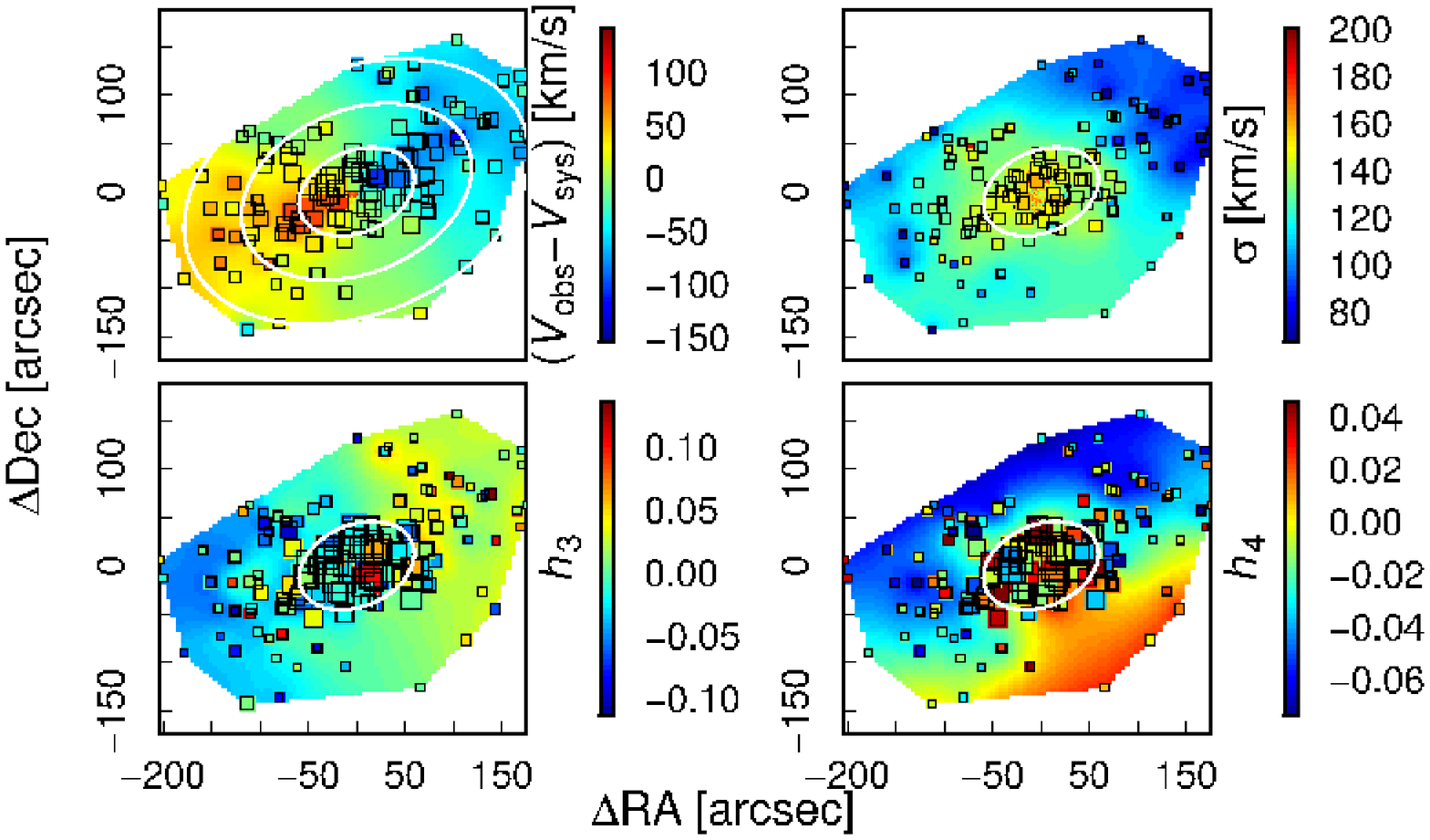}
        }
    \end{center}
    \caption{Continued. Kriging maps for NGC~4697     }
\end{figure*}

\begin{figure*} 
 \ContinuedFloat
     \begin{center}
        \subfigure{
           \hspace{-0.2in}\includegraphics[width=130mm]{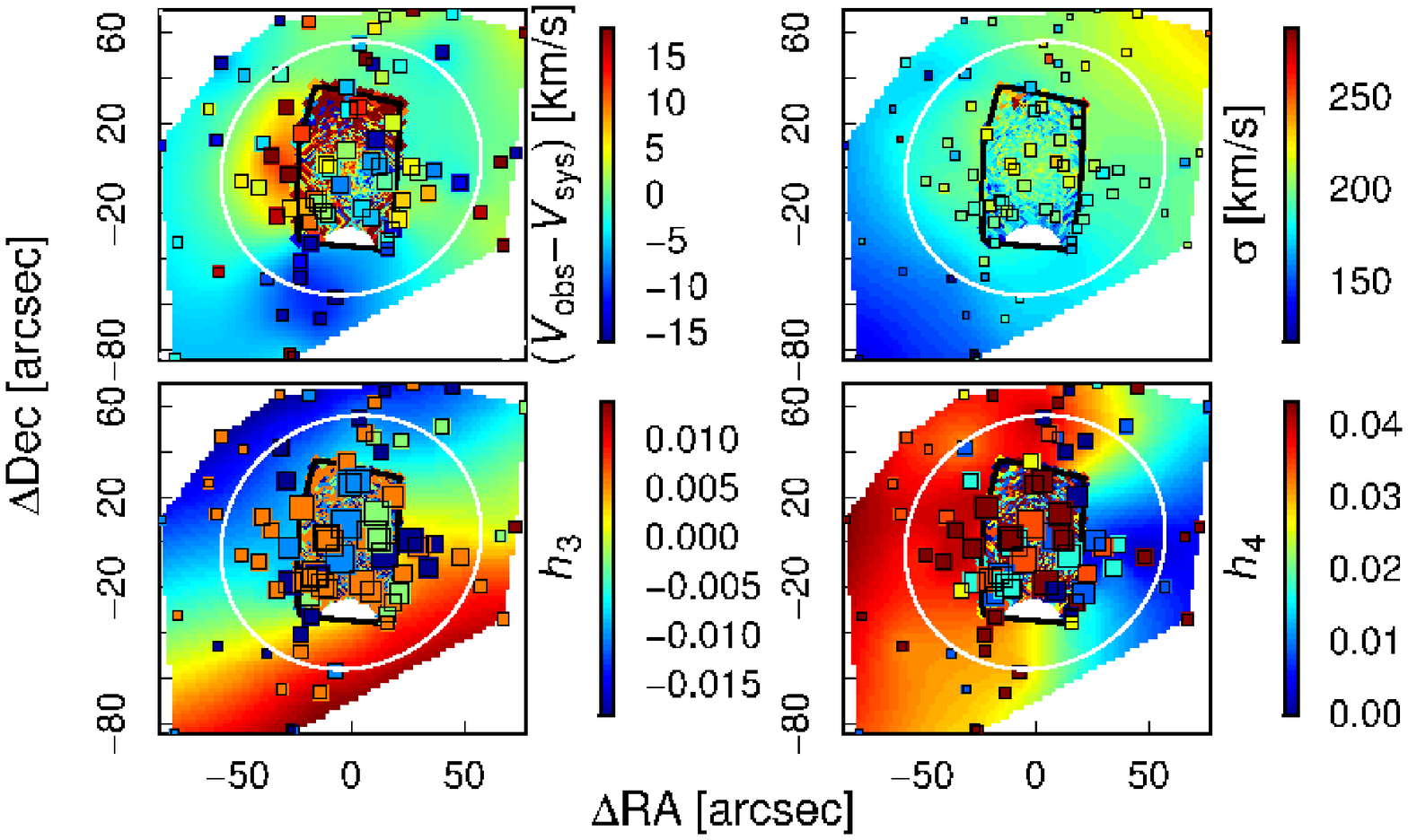}
        }
    \end{center}
    \caption{Continued. Kriging maps for NGC~5846. Slits associated with nearby galaxy NGC~5846a are removed.}
\end{figure*}

\begin{figure*} 
 \ContinuedFloat
     \begin{center}
        \subfigure{
           \hspace{-0.2in}\includegraphics[width=130mm]{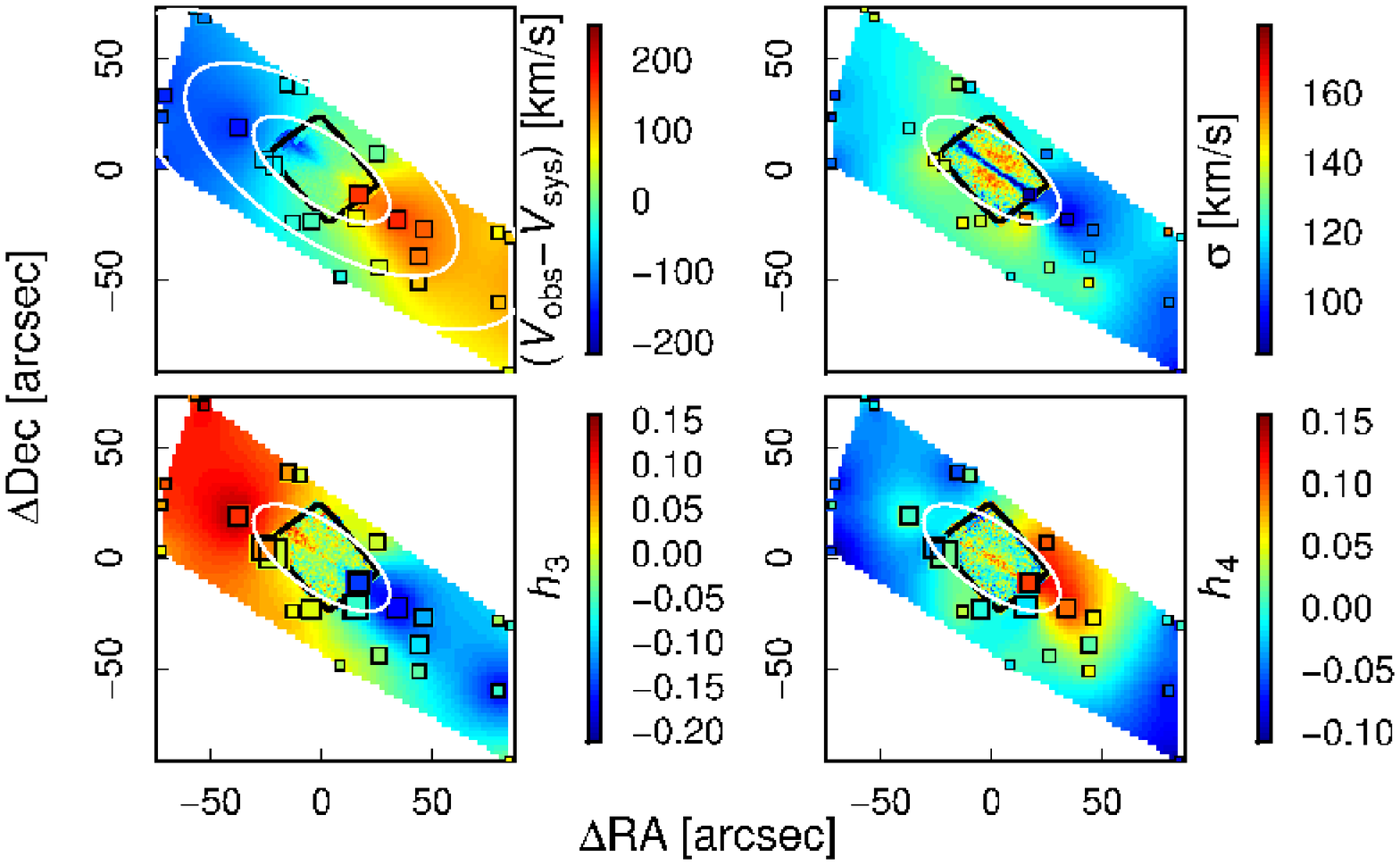}
        }
    \end{center}
    \caption{Continued. Kriging maps for NGC~5866.}
\end{figure*}

\begin{figure*} 
 \ContinuedFloat
     \begin{center}
        \subfigure{
            \hspace{-0.2in}\includegraphics[width=130mm]{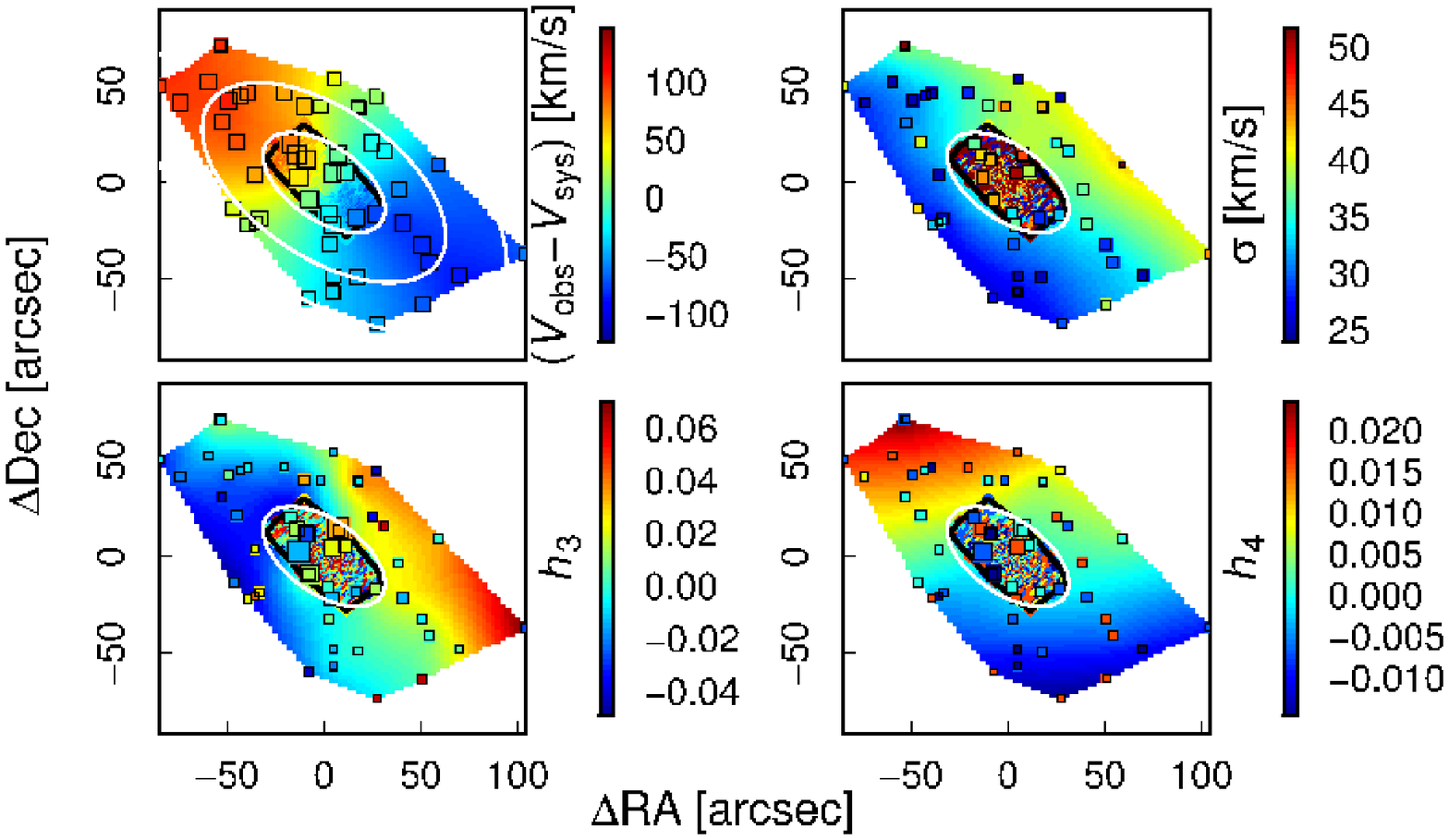}
        }
    \end{center}
    \caption{Continued. Kriging maps for NGC~7457.}
\end{figure*}

\clearpage

\section{Kinemetry figure}\label{section:kinemetryfigures}

\begin{figure}
        \subfigure{
            \hspace{-0.5cm}\includegraphics[width=90mm]{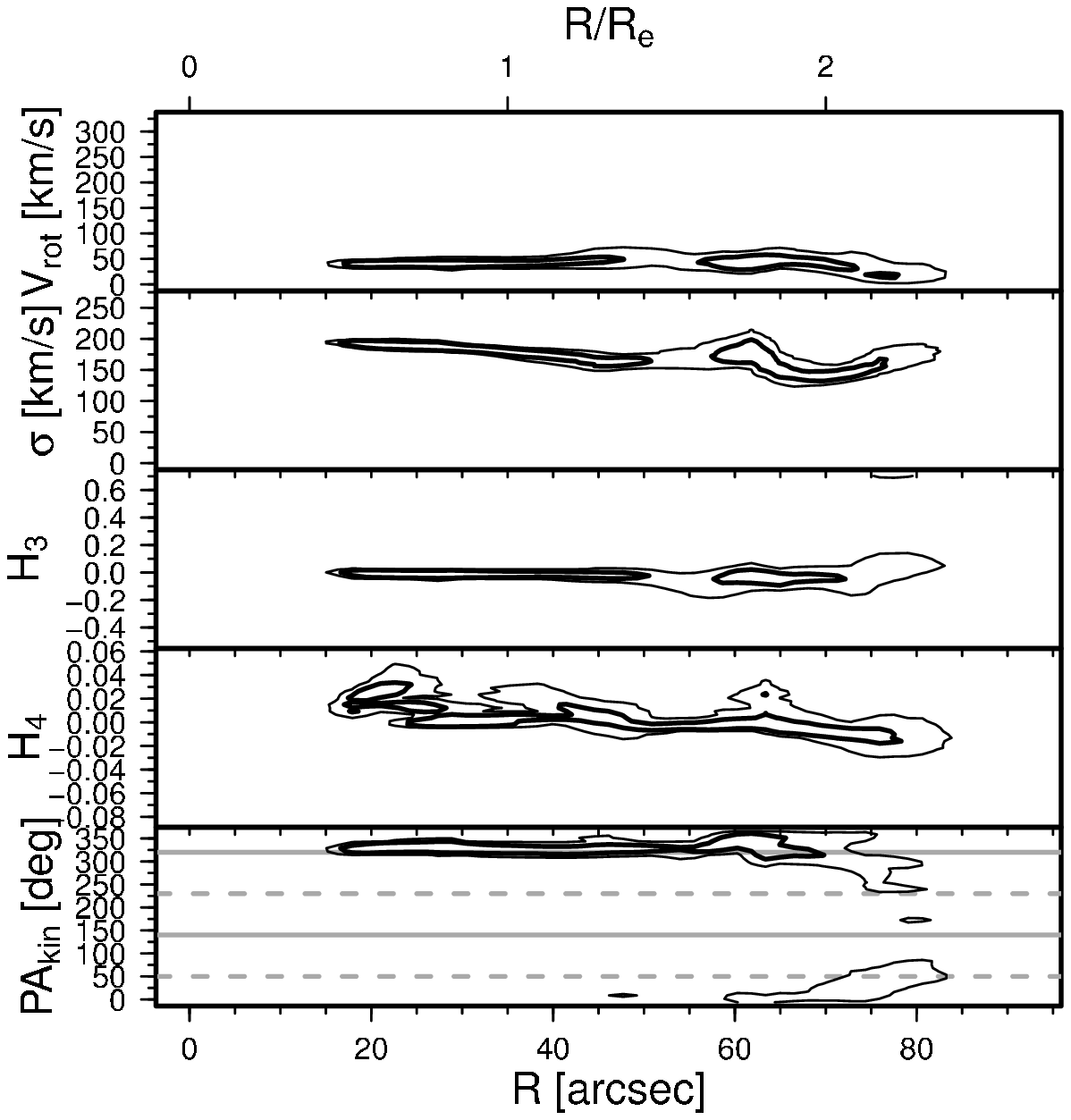}
        }
    \caption{Results of the radial (circularised radius) kinemetry for NGC~720. The rotational velocity ($V_{\rm rot}$), velocity dispersion ($\sigma$), amplitudes of $h_3$ and $h_4$ ($H_3$ and $H_4$, respectively) and kinematic position angle ($PA_{\rm kin}$) profiles are shown in their respective panels whenever they are fitted. Black thick and thin contours are the 1 and 2 $\sigma$ confidence intervals, respectively, for the SLUGGS data. Similarly, thick and thin red contours show the 1 and 2 $\sigma$ confidence intervals, respectively, of the ATLAS$^{\rm 3D}$ kinemetry when available. Green solid lines show our measured light profile when available. Grey solid and dashed lines show the literature photometric major axis/axis ratio and minor axis, respectively, as given in Table \ref{table:sample}.}
   \label{fig:kinemetry}
\end{figure}

\begin{figure} 
 \ContinuedFloat
        \subfigure{
            \hspace{-0.5cm}\includegraphics[width=90mm]{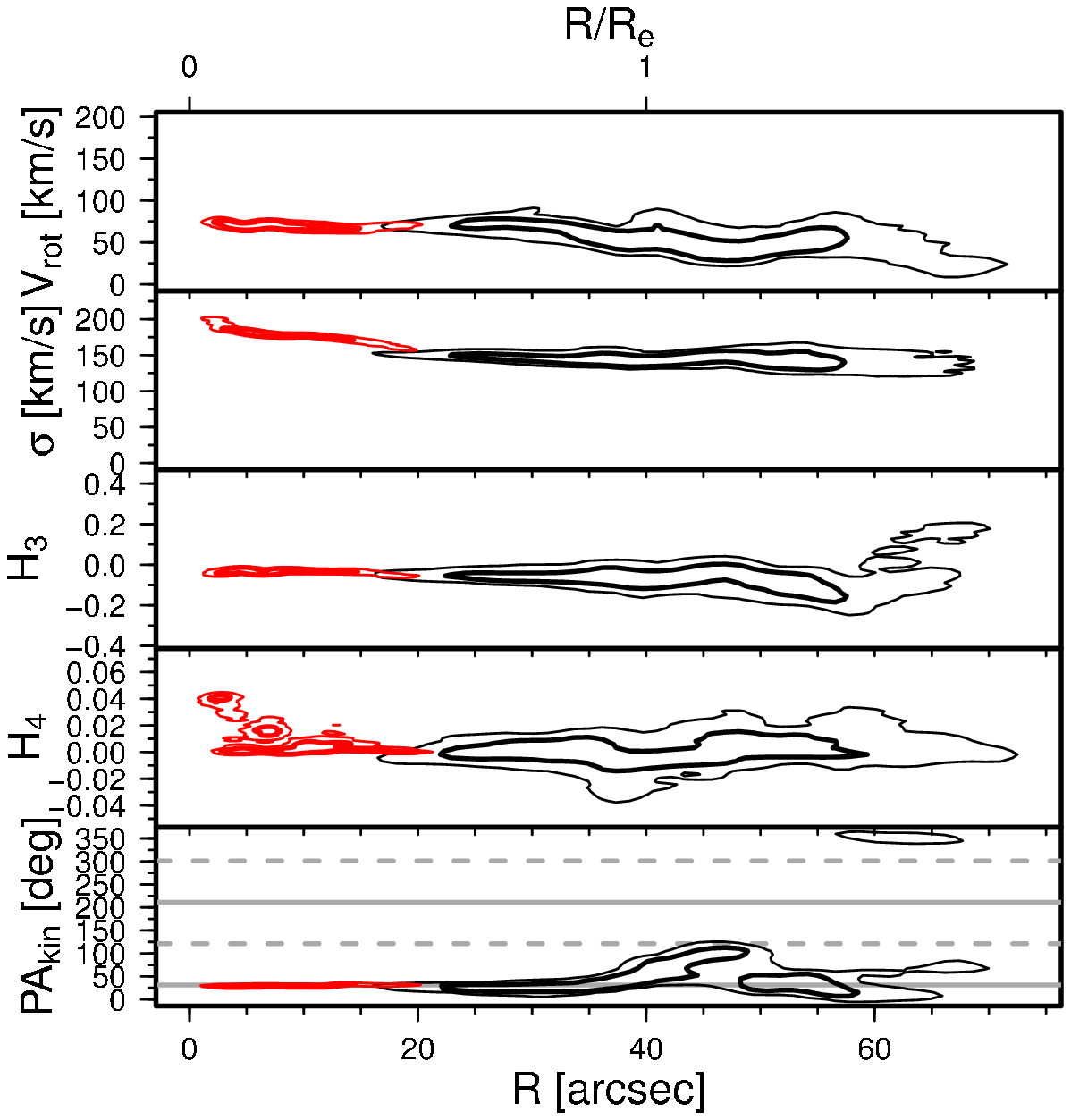}
        }
    \caption{Continued. Results of the radial kinemetry for NGC~821.}
\end{figure}

\begin{figure} 
 \ContinuedFloat
        \subfigure{
             \hspace{-0.5cm}\includegraphics[width=90mm]{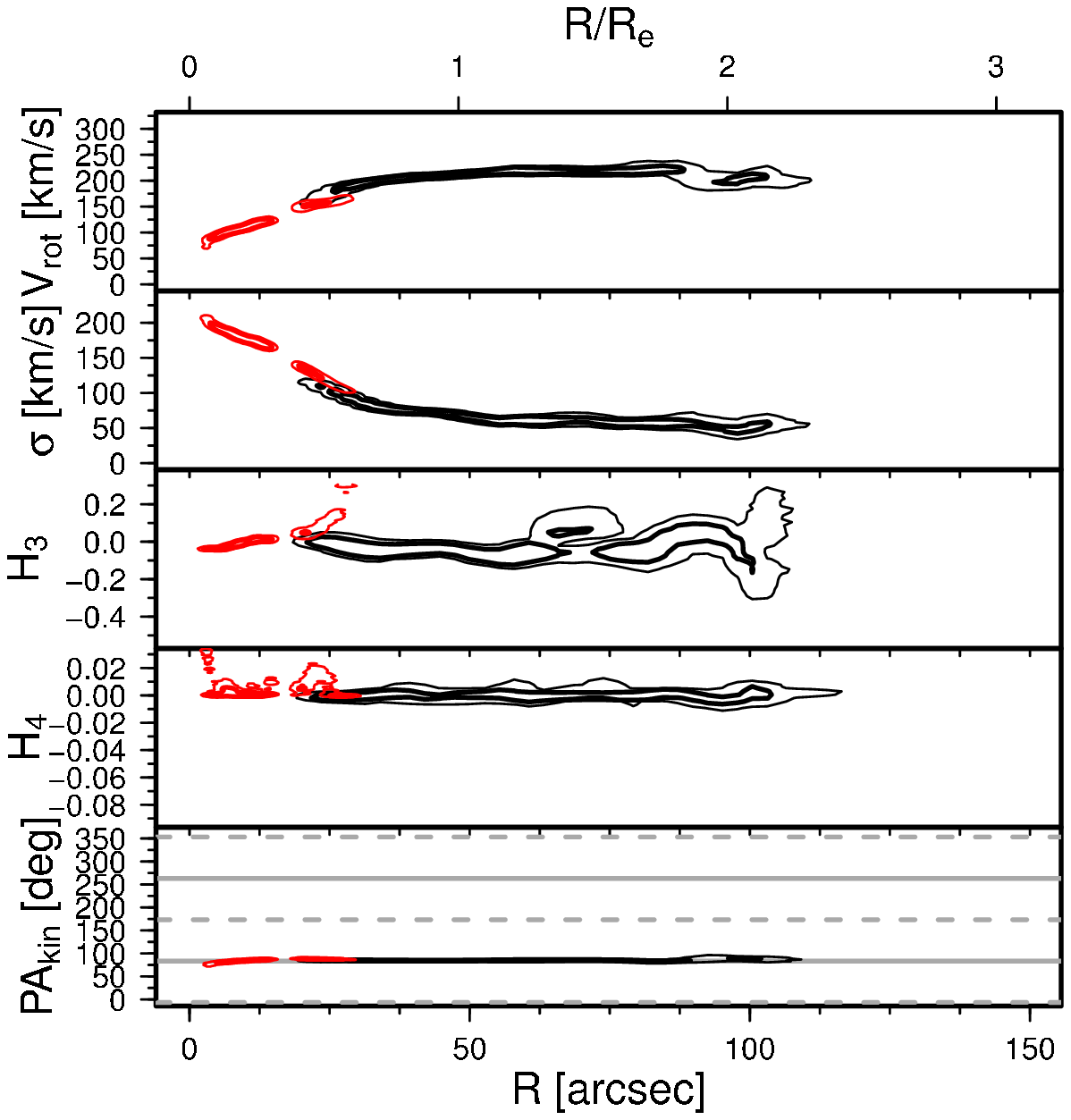}
        }
    \caption{Continued. Results of the radial kinemetry for NGC~1023.}
\end{figure}

\begin{figure} 
 \ContinuedFloat
        \subfigure{
             \hspace{-0.5cm}\includegraphics[width=90mm]{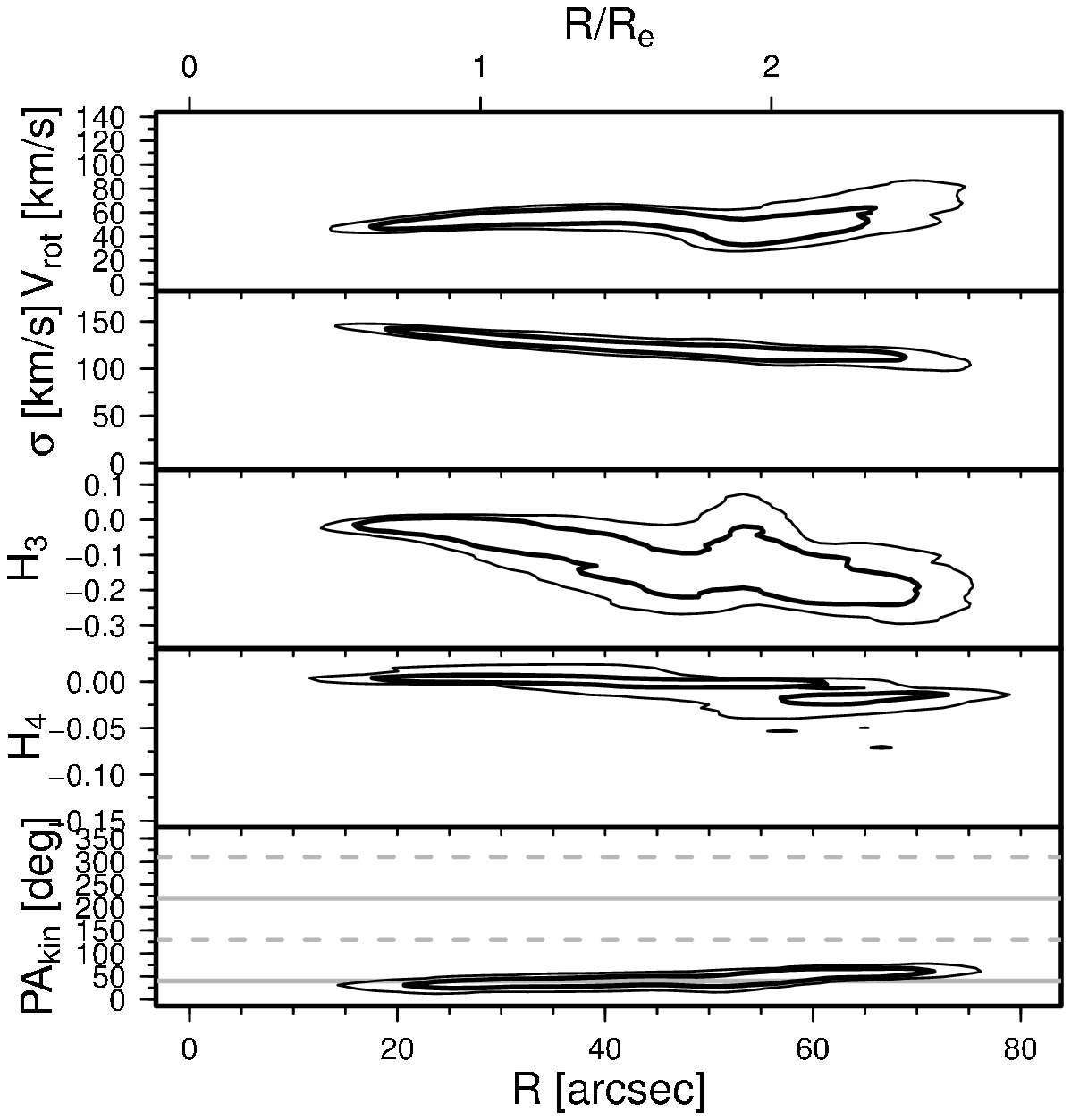}
        }
    \caption{Continued. Results of the radial kinemetry for NGC~1400.}
\end{figure}

\begin{figure} 
 \ContinuedFloat
        \subfigure{
             \hspace{-0.5cm}\includegraphics[width=90mm]{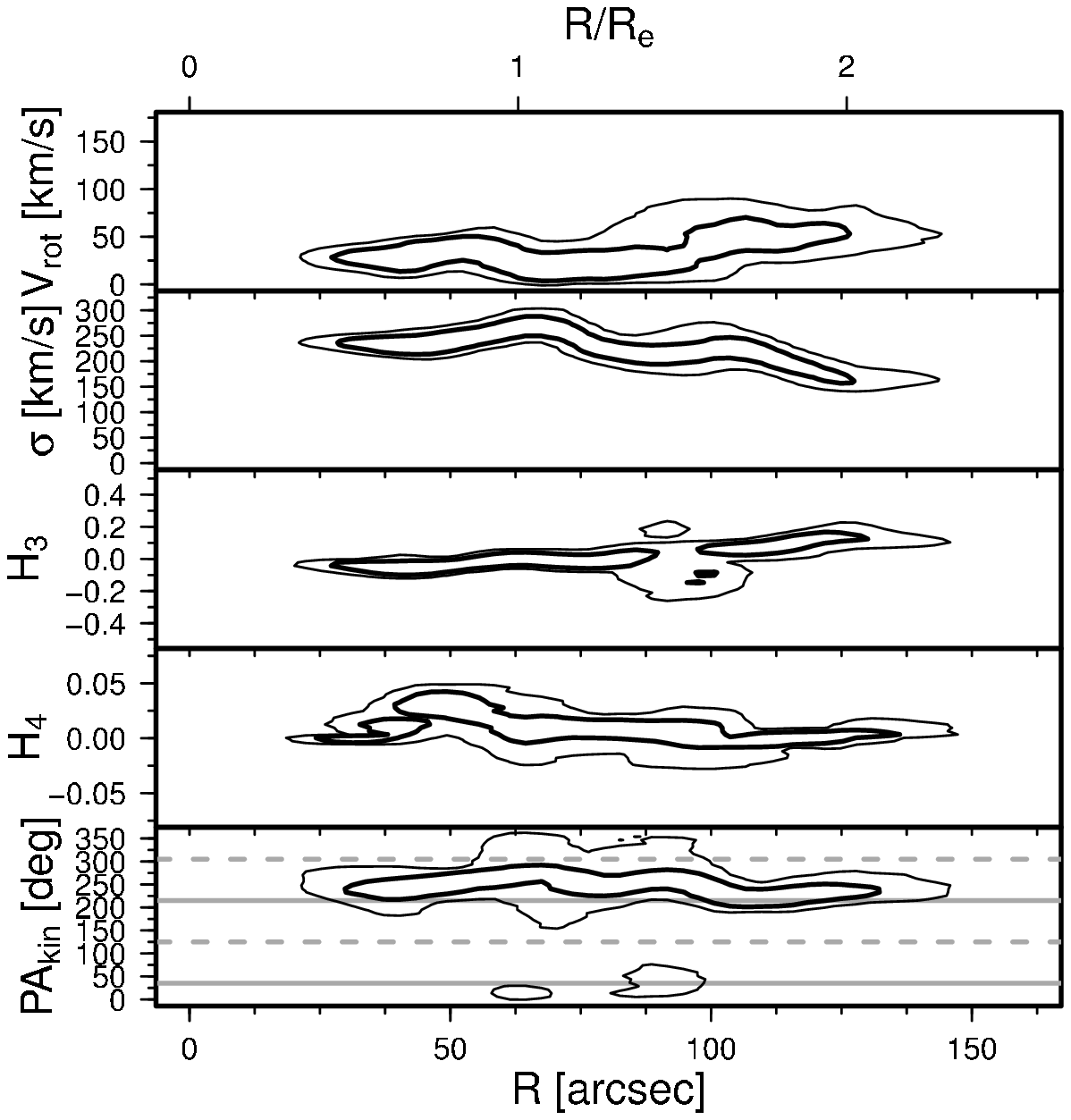}
        }
    \caption{Continued. Results of the radial kinemetry for NGC~1407.}
\end{figure}

\begin{figure} 
 \ContinuedFloat
        \subfigure{
            \hspace{-0.5cm}\includegraphics[width=90mm]{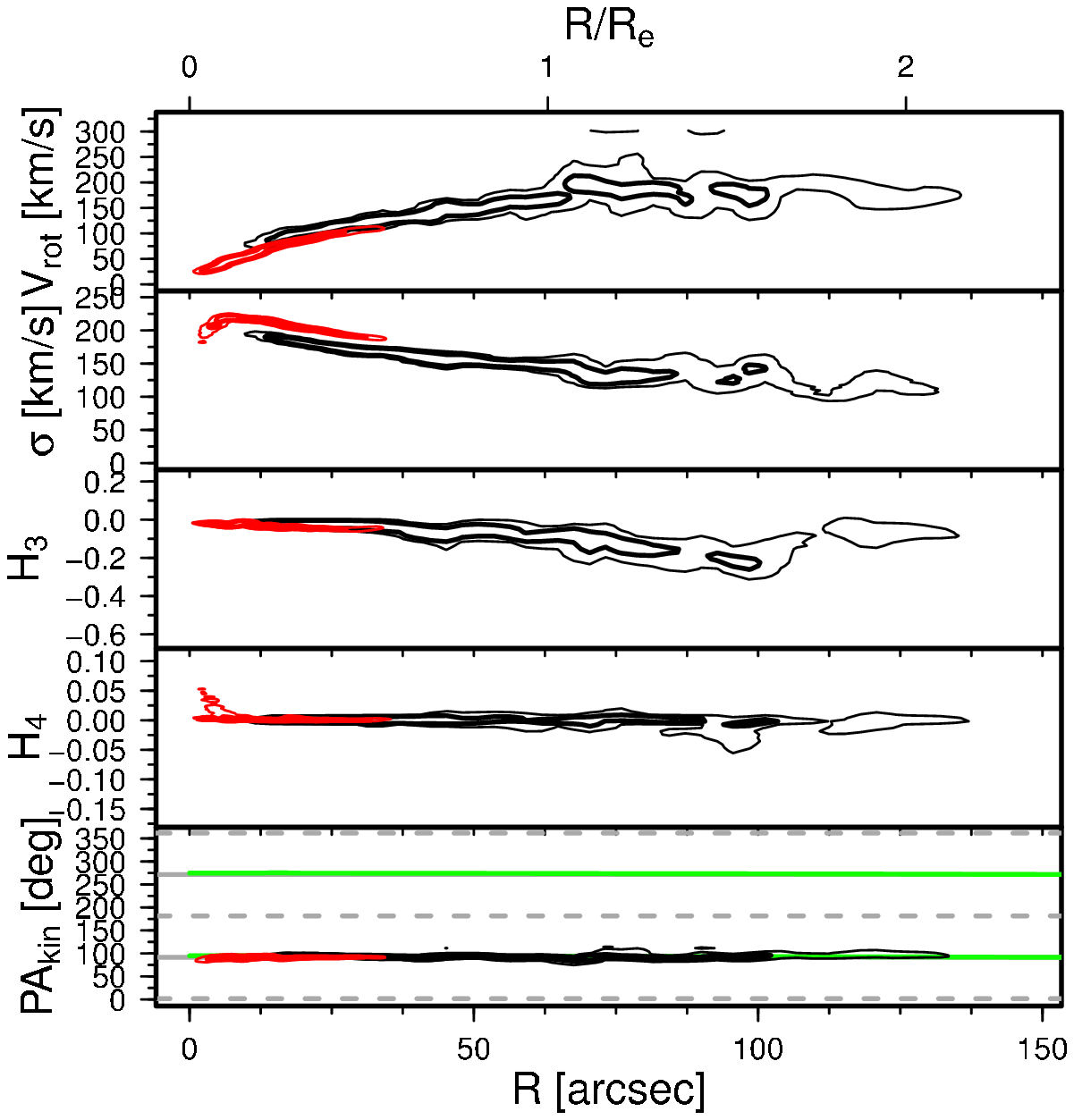}
        }
    \caption{Continued. Results of the radial kinemetry for NGC~2768.}
\end{figure}

\begin{figure} 
 \ContinuedFloat
        \subfigure{
             \hspace{-0.5cm}\includegraphics[width=90mm]{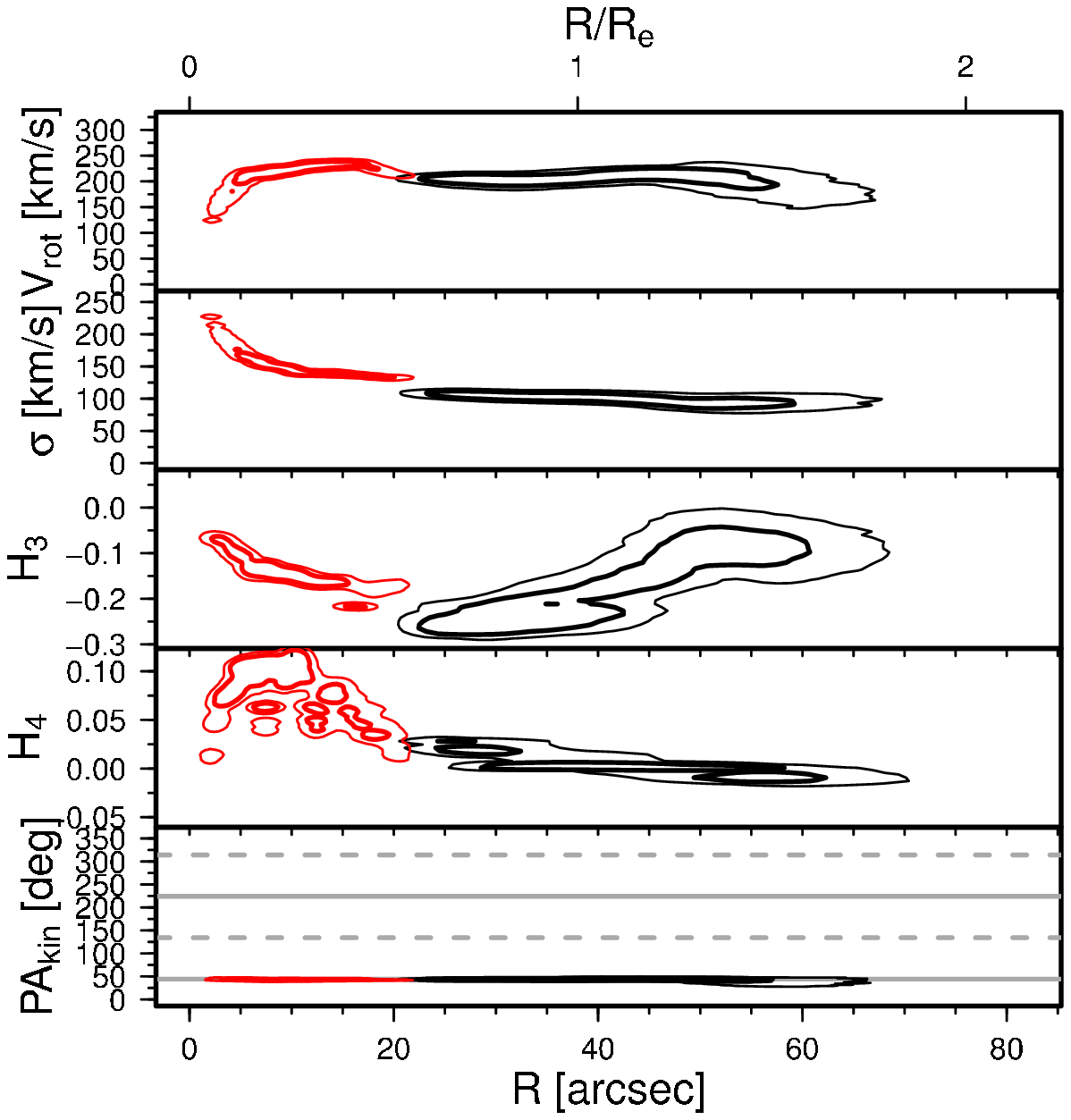}
        }
    \caption{Continued. Results of the radial kinemetry for NGC~2974.}
\end{figure}

\begin{figure} 
 \ContinuedFloat
        \subfigure{
             \hspace{-0.5cm}\includegraphics[width=90mm]{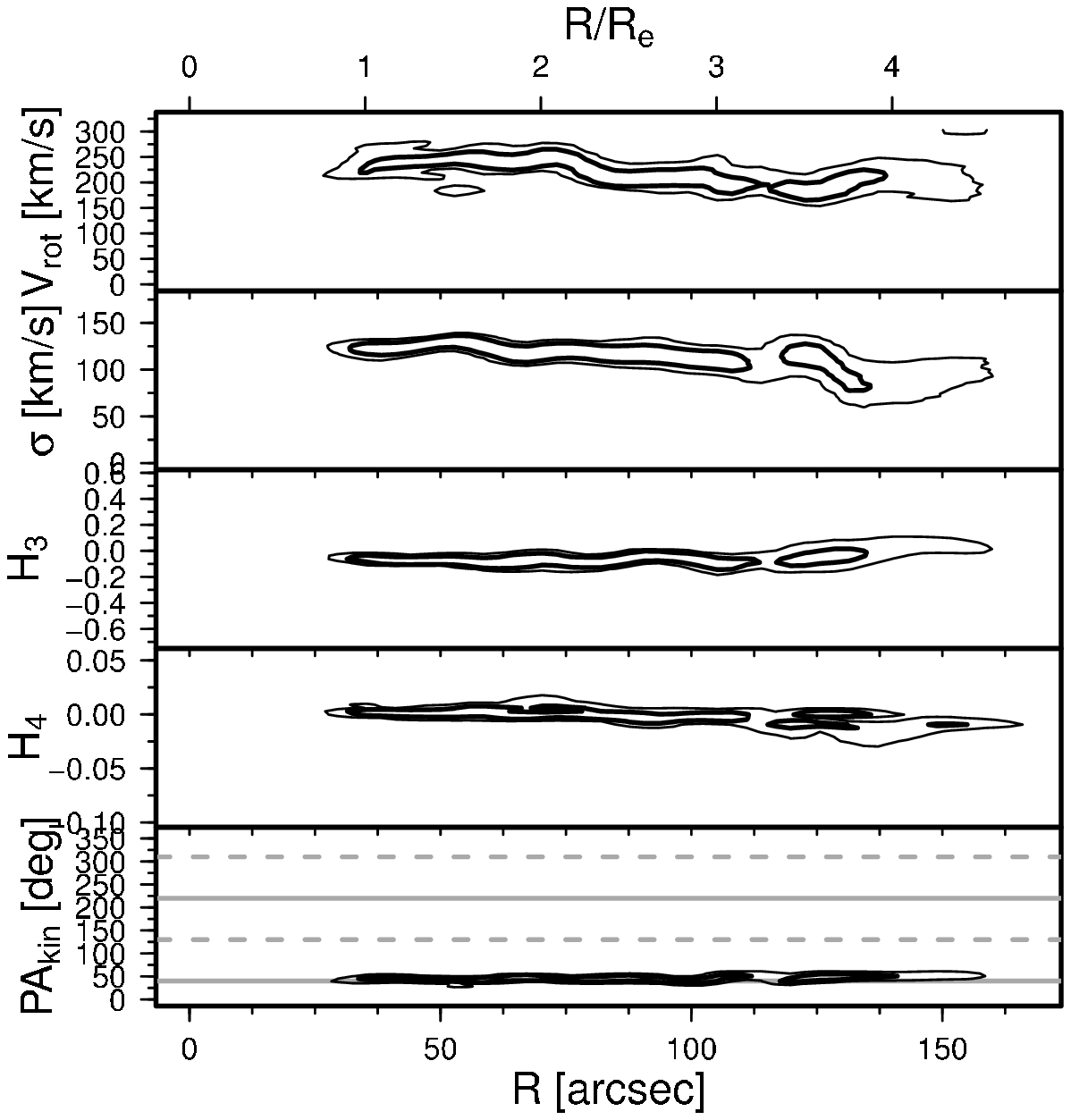}
        }
    \caption{Continued. Results of the radial kinemetry for NGC~3115.}
\end{figure}

\begin{figure} 
 \ContinuedFloat
        \subfigure{
             \hspace{-0.5cm}\includegraphics[width=90mm]{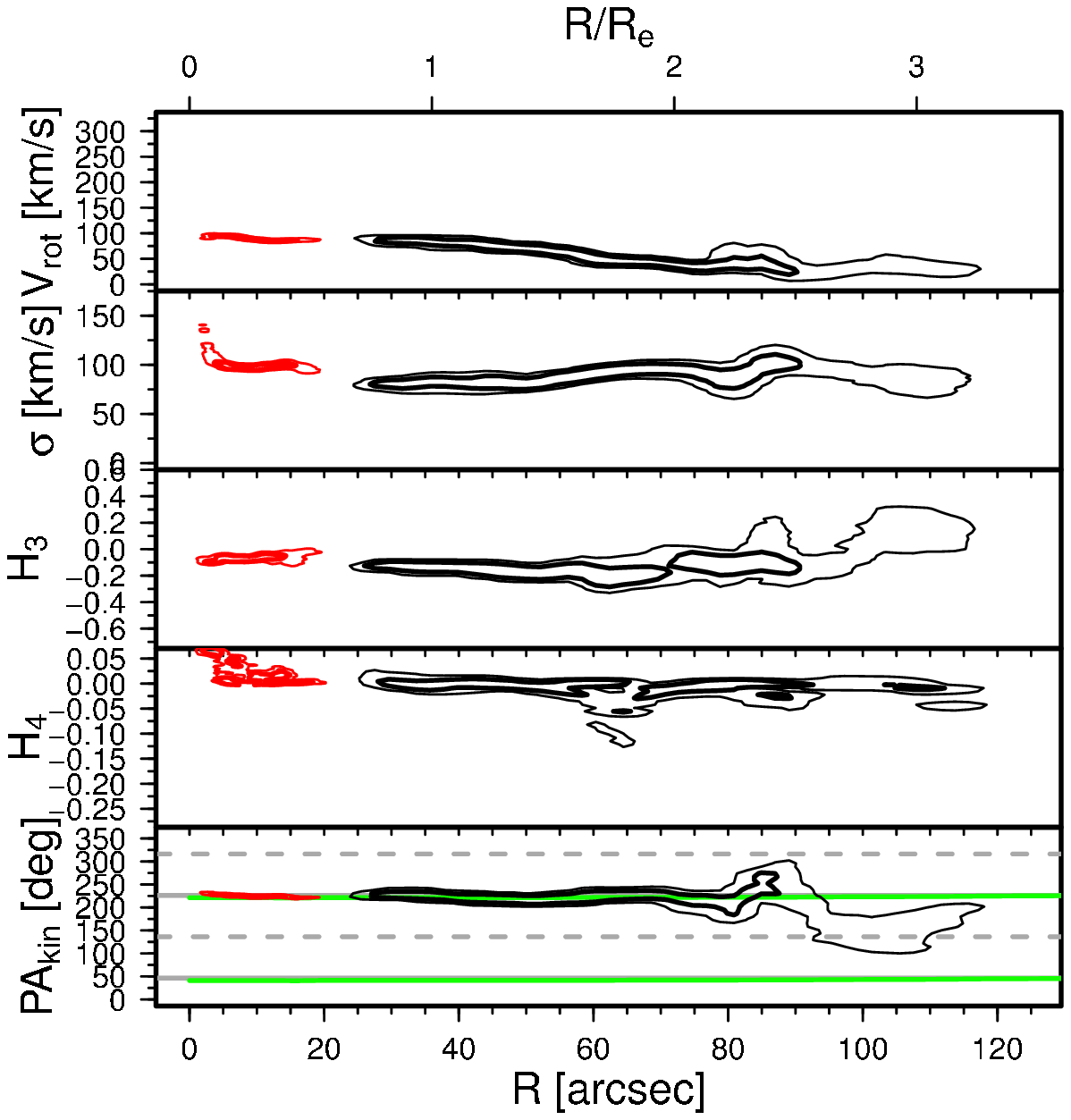}
        }
    \caption{Continued. Results of the radial kinemetry for NGC~3377.}
\end{figure}

\begin{figure} 
 \ContinuedFloat
        \subfigure{
            \hspace{-0.5cm}\includegraphics[width=90mm]{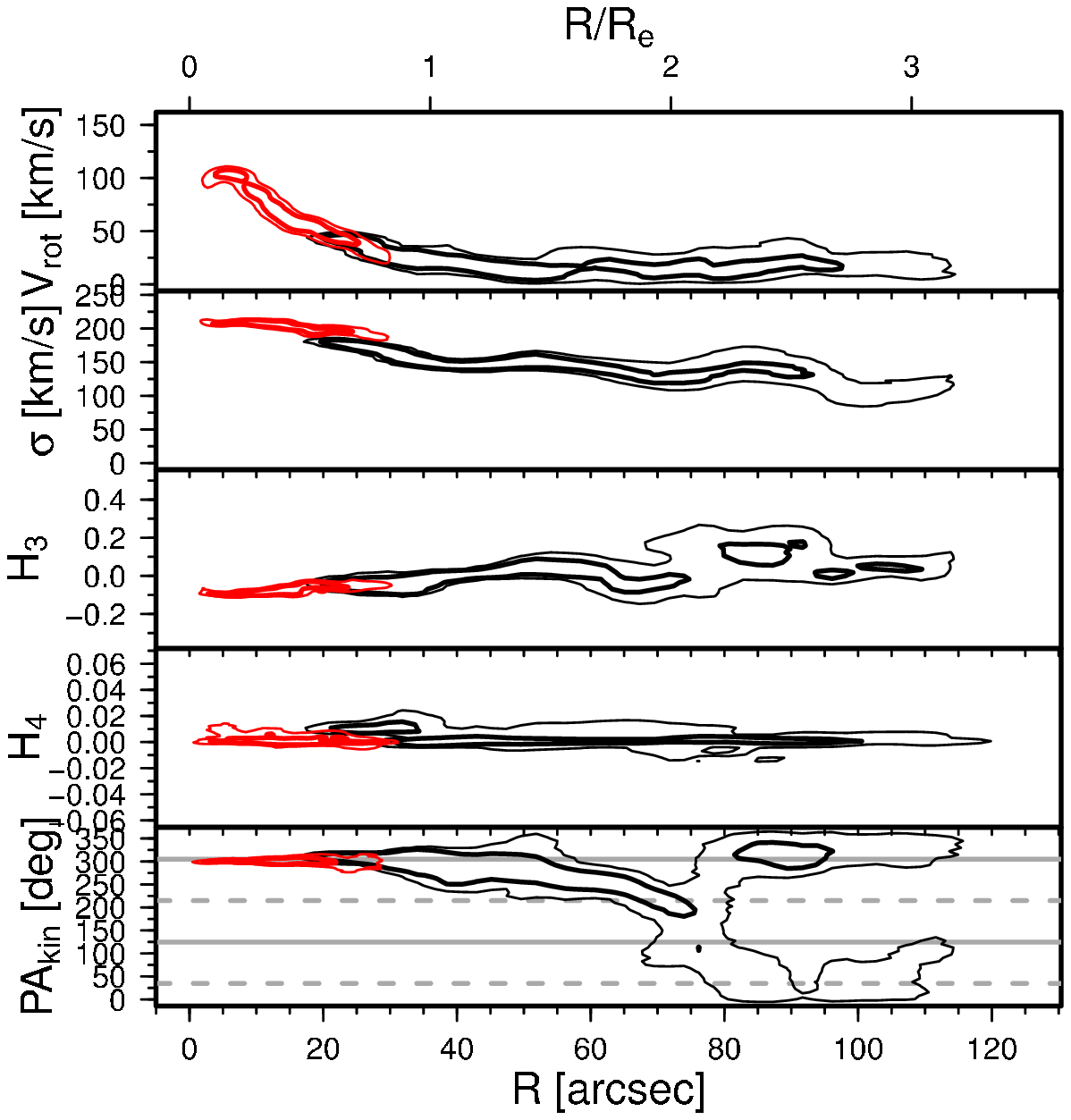}
        }
    \caption{Continued. Results of the radial kinemetry for NGC~3607.}
\end{figure}

\begin{figure} 
 \ContinuedFloat
        \subfigure{
             \hspace{-0.5cm}\includegraphics[width=84mm]{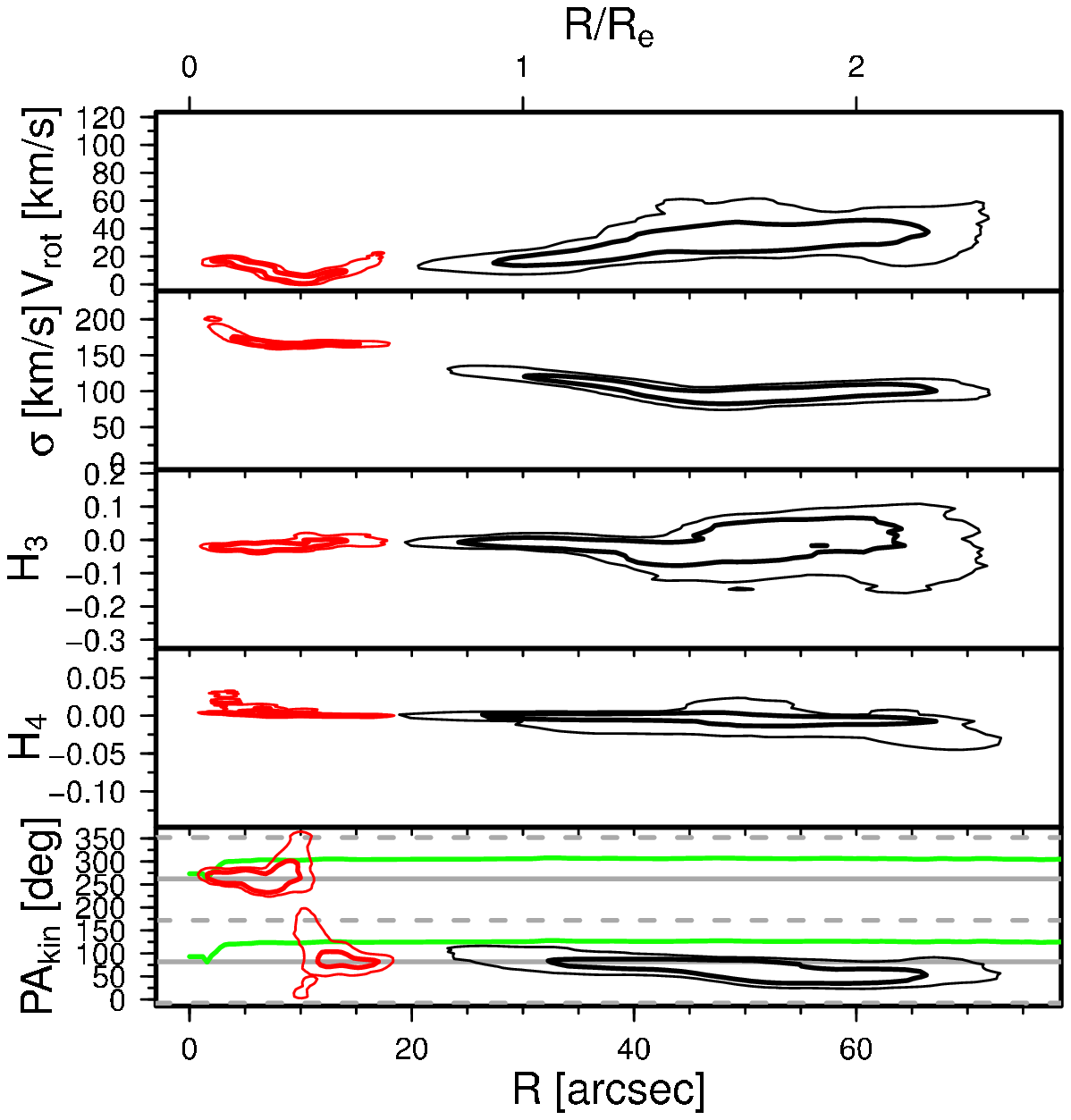}
        }
    \caption{Continued. Results of the radial kinemetry for NGC~3608.}
\end{figure}

\begin{figure} 
 \ContinuedFloat
        \subfigure{
             \hspace{-0.5cm}\includegraphics[width=90mm]{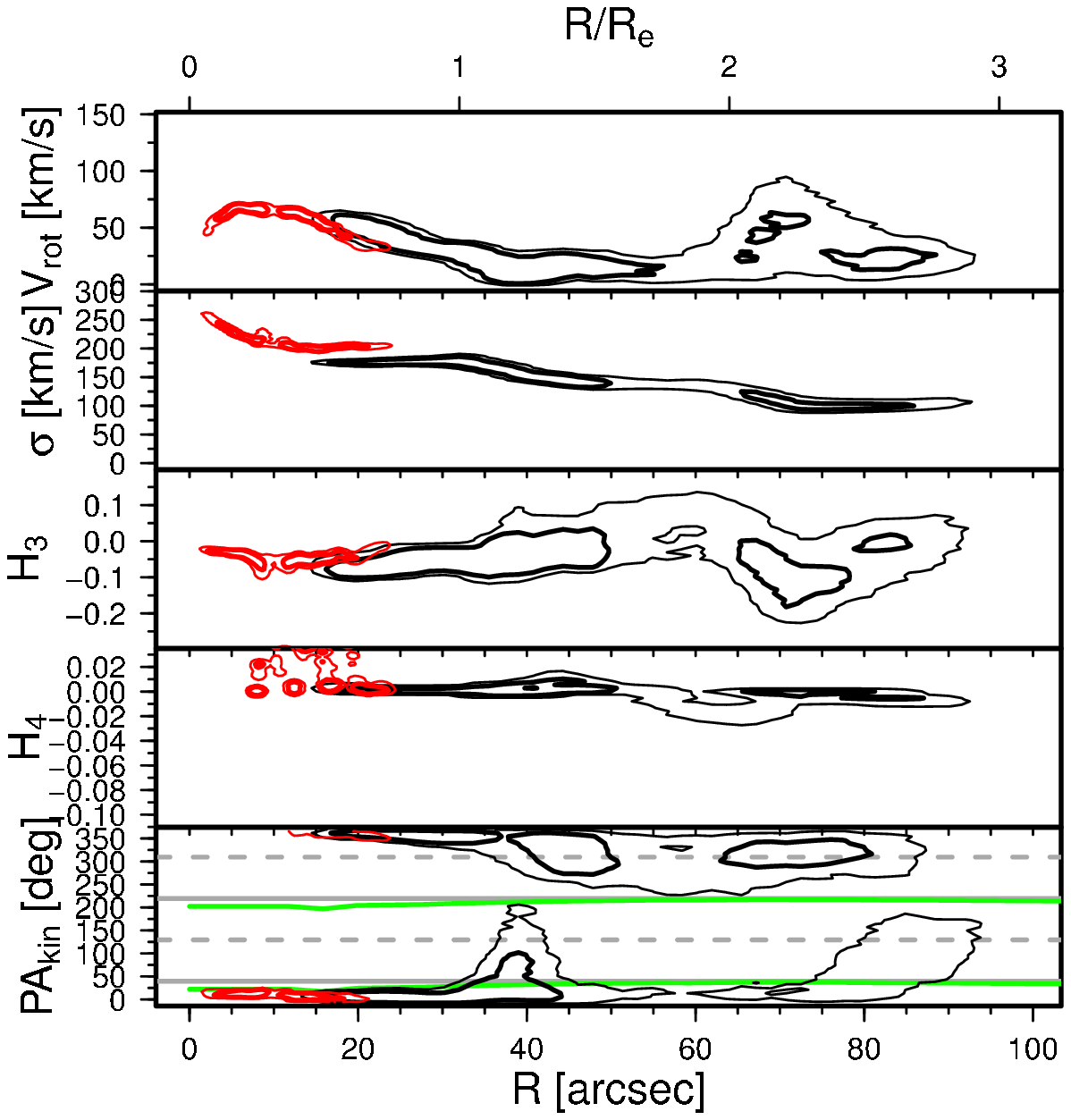}
        }
    \caption{Continued. Results of the radial kinemetry for NGC~4278.}
\end{figure}

\begin{figure} 
 \ContinuedFloat
        \subfigure{
            \hspace{-0.5cm}\includegraphics[width=90mm]{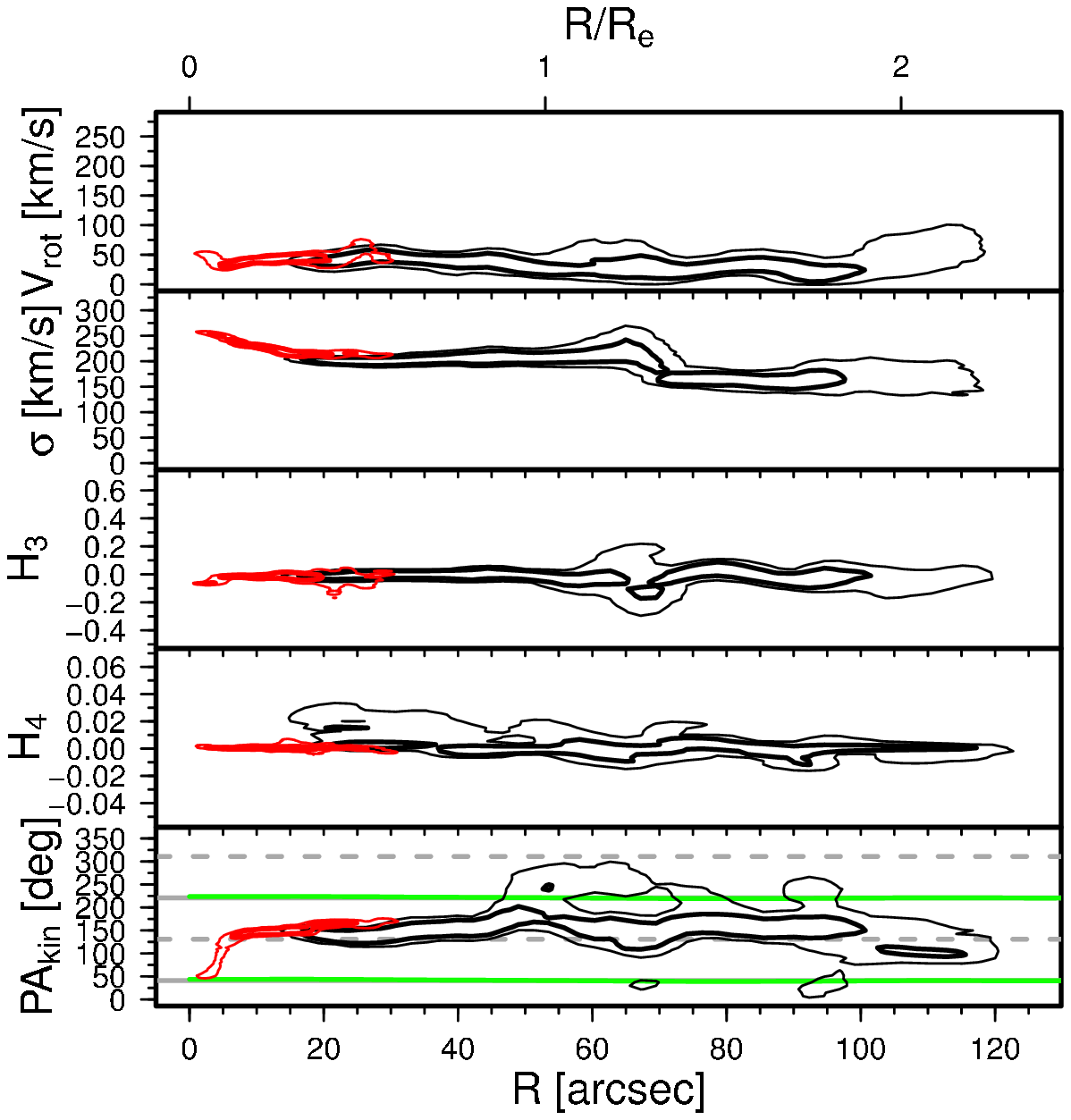}
        }
    \caption{Continued. Results of the radial kinemetry for NGC~4365.}
\end{figure}

\clearpage

\begin{figure} 
 \ContinuedFloat
        \subfigure{
            \hspace{-0.5cm} \includegraphics[width=90mm]{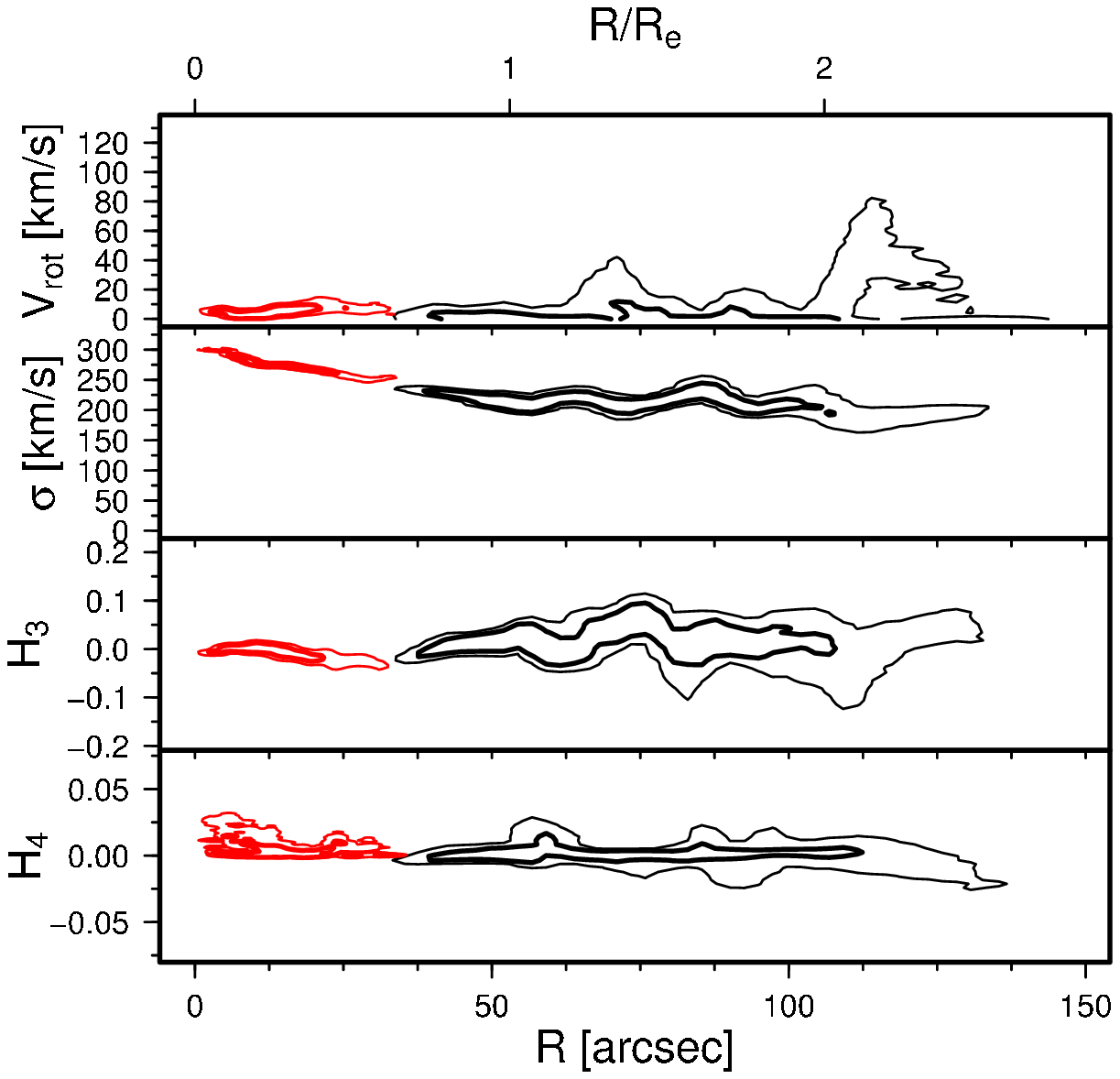}
        }
    \caption{Continued. Results of the radial kinemetry for NGC~4374.}
\end{figure}

\begin{figure} 
 \ContinuedFloat
        \subfigure{
             \hspace{-0.5cm}\includegraphics[width=90mm]{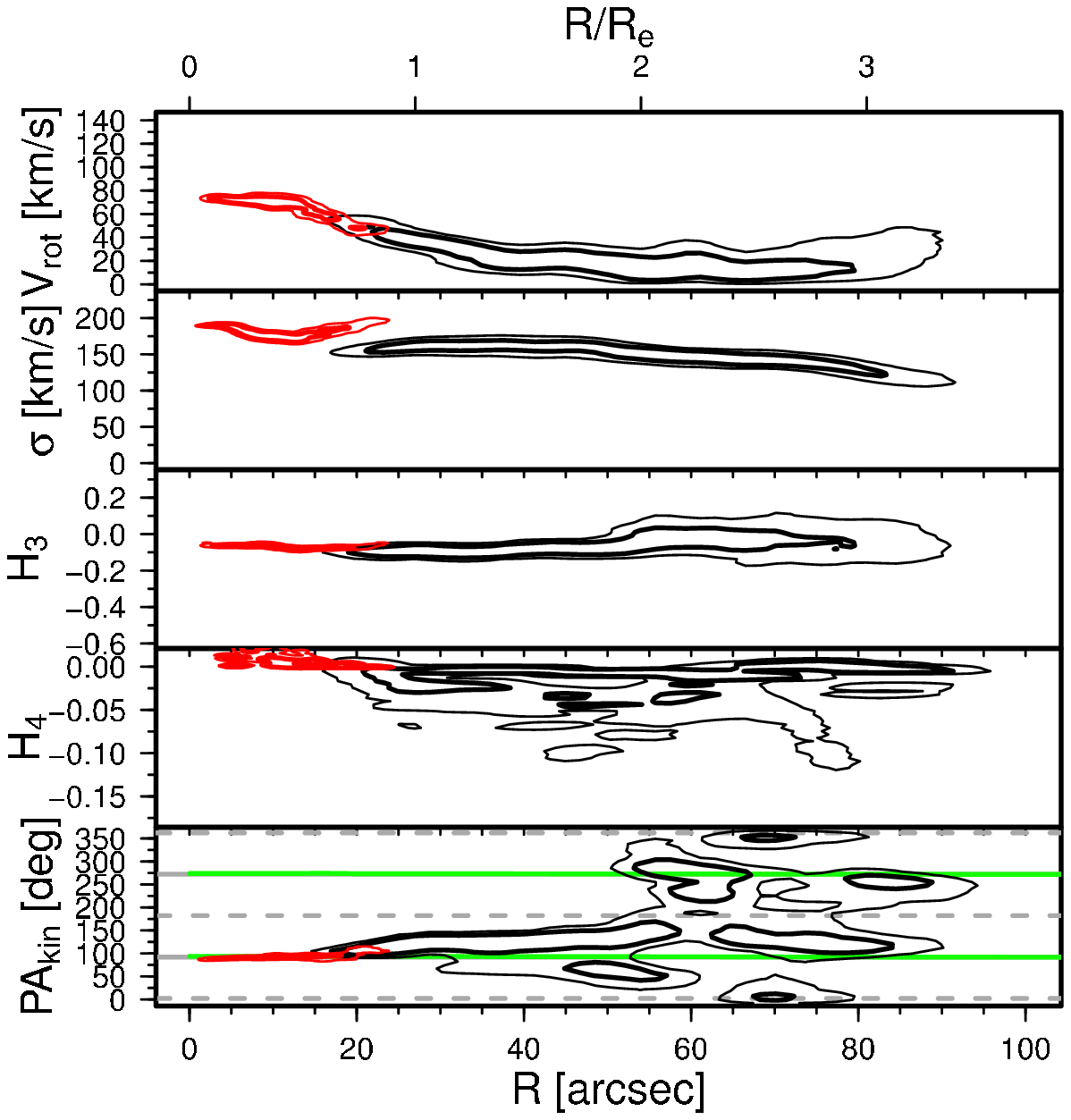}
        }
    \caption{Continued. Results of the radial kinemetry for NGC~4473.}
\end{figure}

\begin{figure} 
 \ContinuedFloat
        \subfigure{
             \hspace{-0.5cm}\includegraphics[width=90mm]{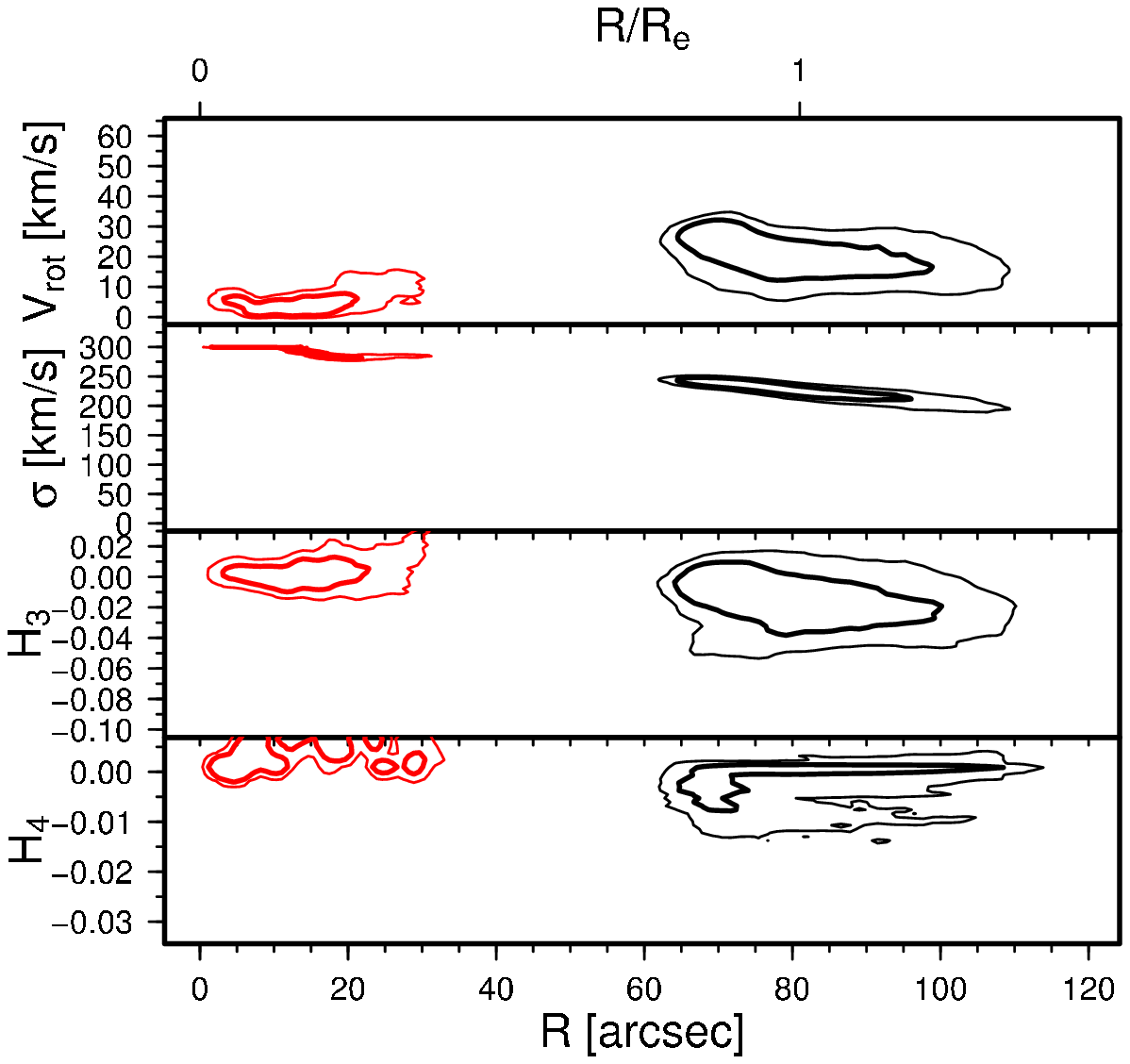}
        }
    \caption{Continued. Results of the radial kinemetry for NGC~4486.}
\end{figure}

\begin{figure} 
 \ContinuedFloat
        \subfigure{
            \hspace{-0.5cm}\includegraphics[width=90mm]{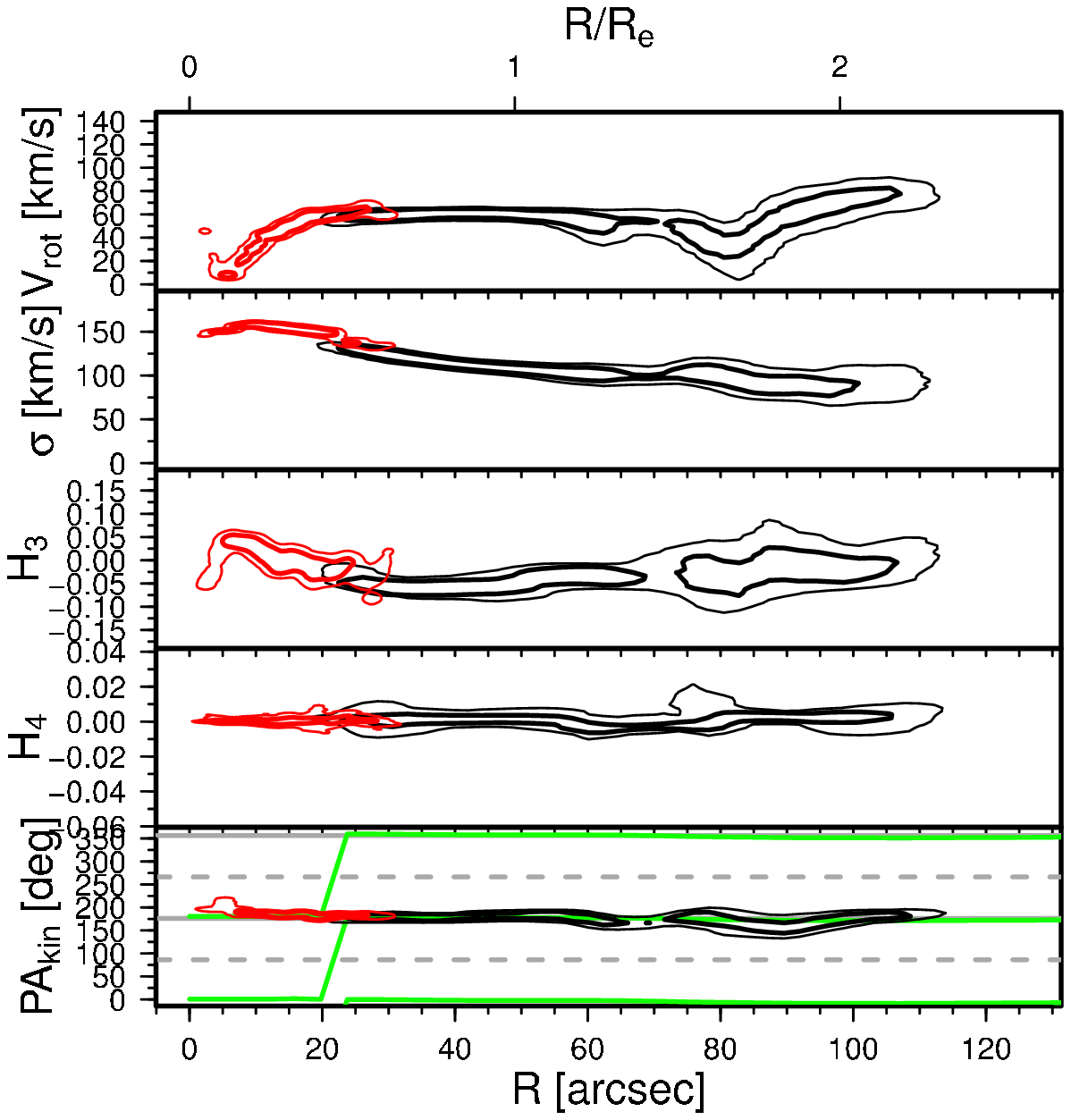}
        }
    \caption{Continued. Results of the radial kinemetry for NGC~4494.}
\end{figure}

\begin{figure} 
 \ContinuedFloat
        \subfigure{
            \hspace{-0.5cm} \includegraphics[width=90mm]{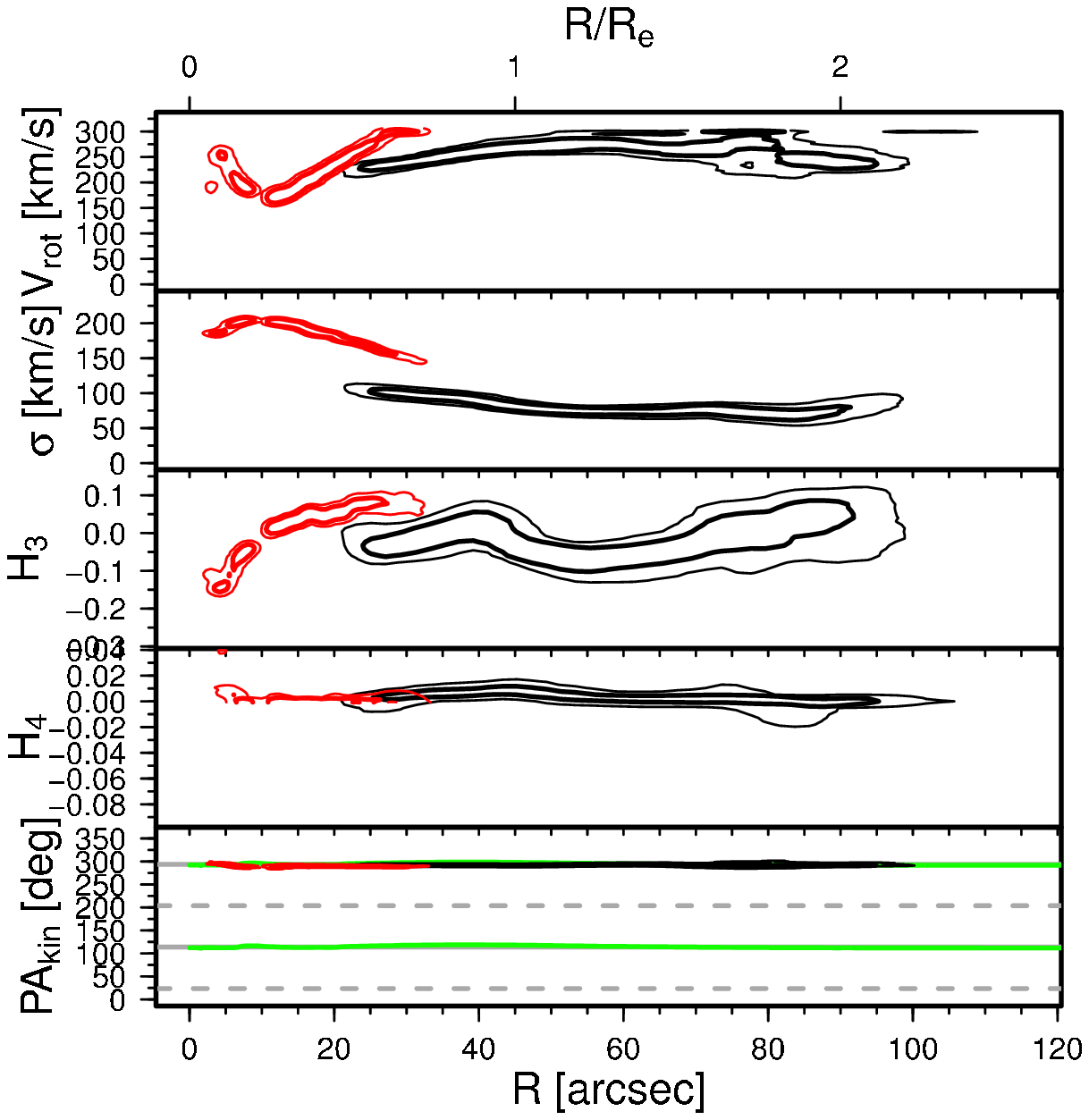}
         }
    \caption{Continued. Results of the radial kinemetry for NGC~4526.}
\end{figure}

\begin{figure} 
 \ContinuedFloat
        \subfigure{
             \hspace{-0.5cm}\includegraphics[width=90mm]{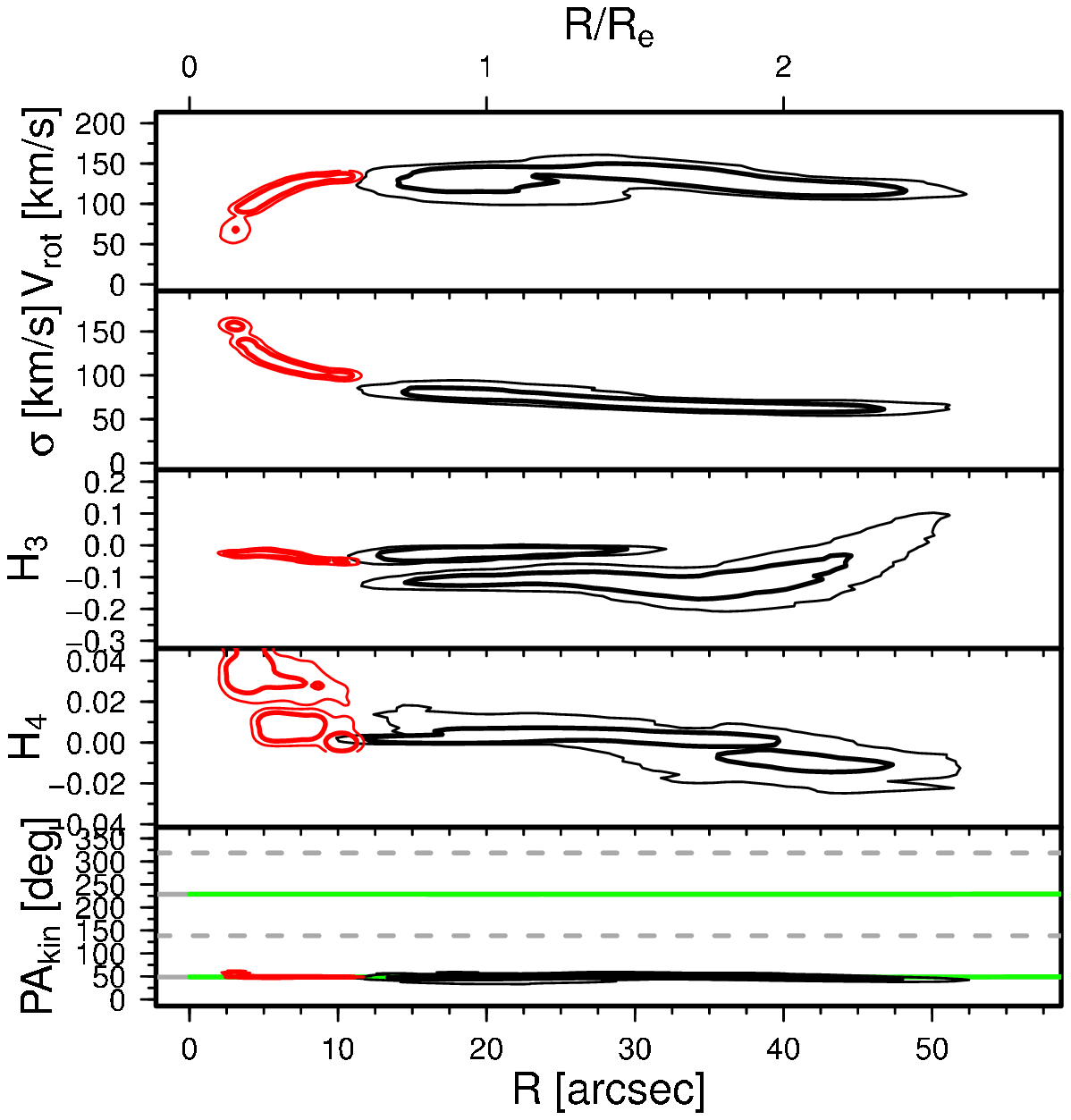}
        }
    \caption{Continued. Results of the radial kinemetry for NGC~4564.}
\end{figure}

\begin{figure} 
 \ContinuedFloat
        \subfigure{
             \hspace{-0.5cm}\includegraphics[width=90mm]{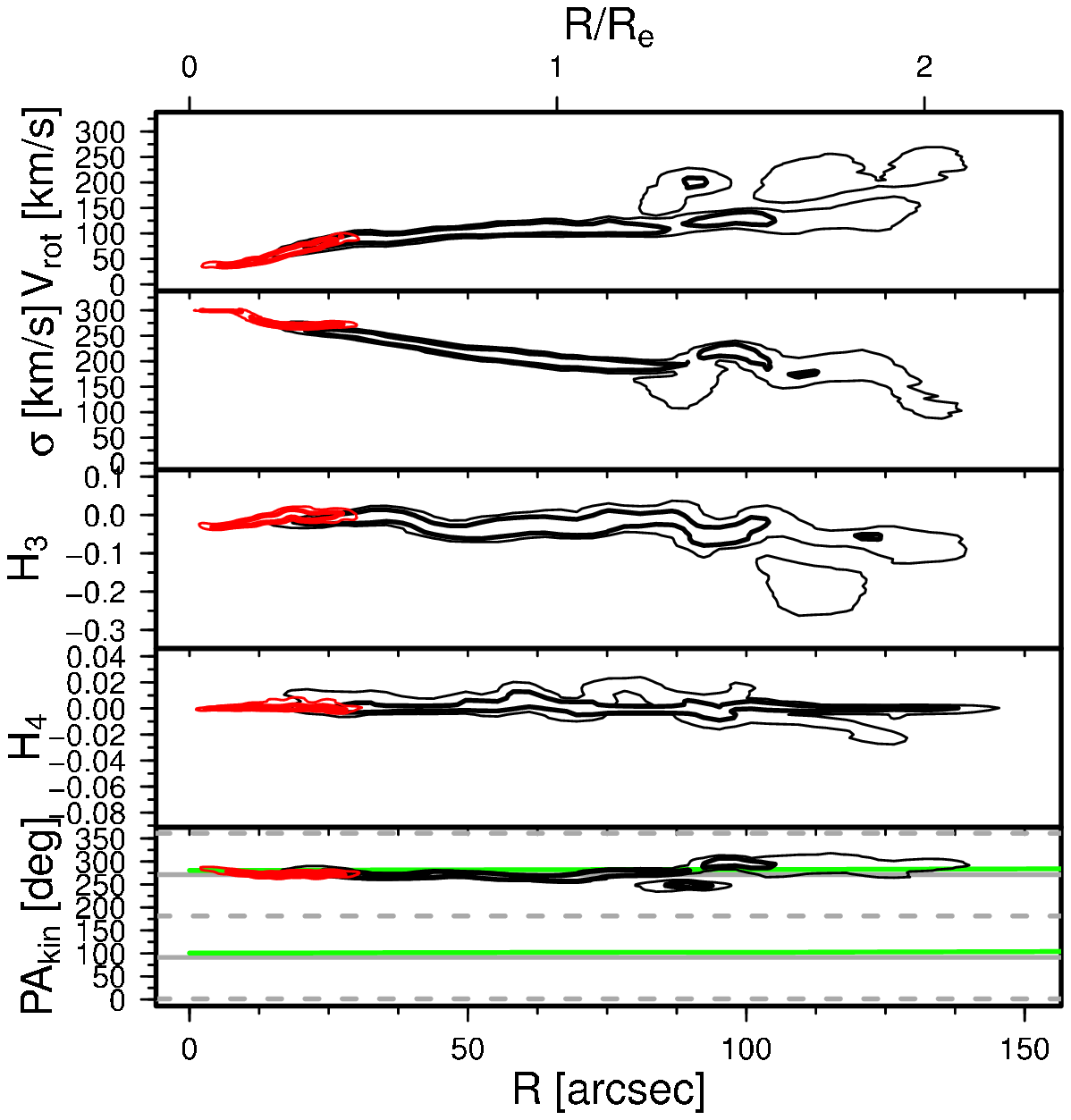}
        }
    \caption{Continued. Results of the radial kinemetry for NGC~4649.}
\end{figure}

\begin{figure} 
 \ContinuedFloat
        \subfigure{
            \hspace{-0.5cm} \includegraphics[width=90mm]{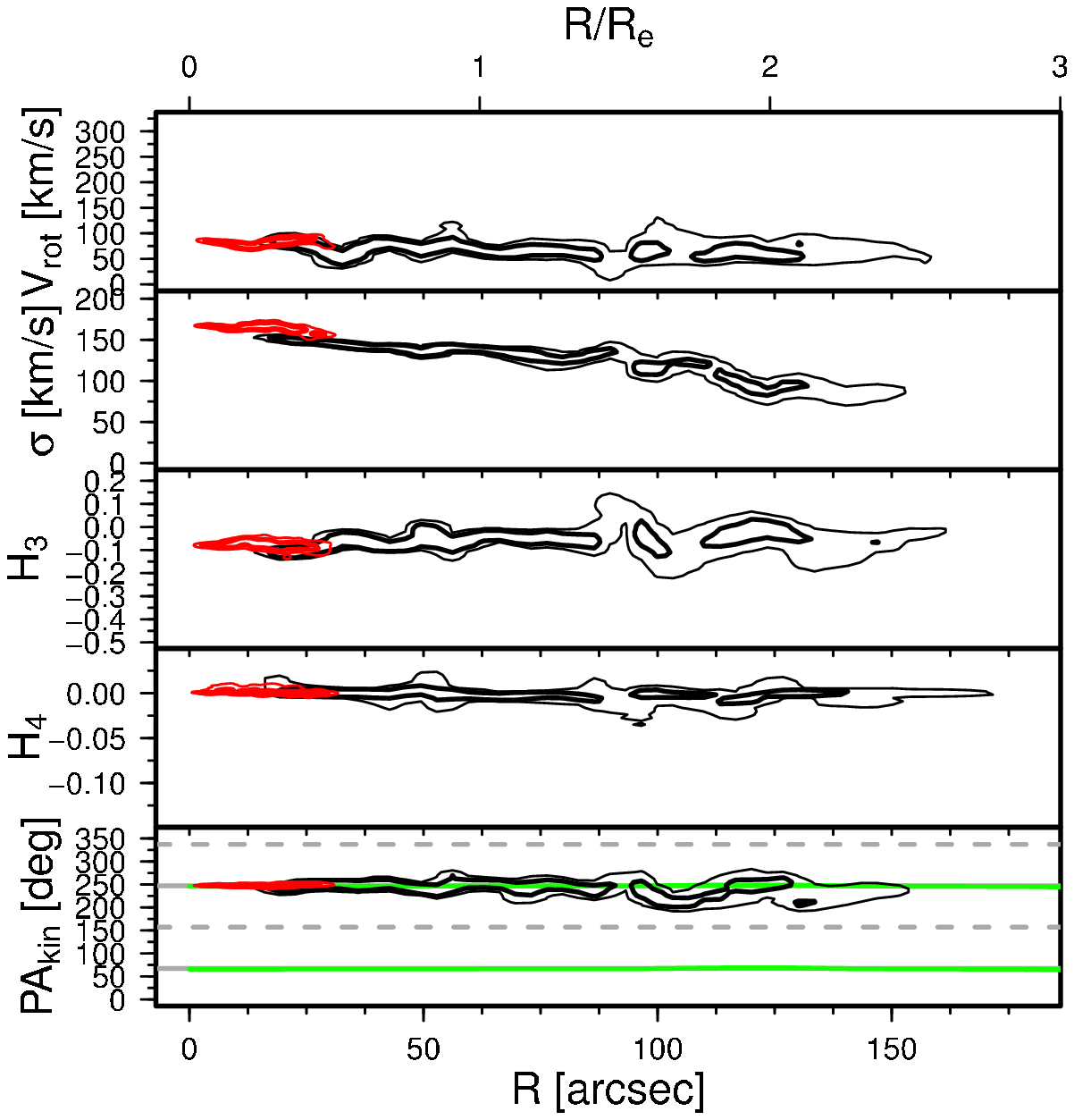}
        }
    \caption{Continued. Results of the radial kinemetry for NGC~4697.}
\end{figure}

\begin{figure} 
 \ContinuedFloat
        \subfigure{
            \hspace{-0.5cm}\includegraphics[width=90mm]{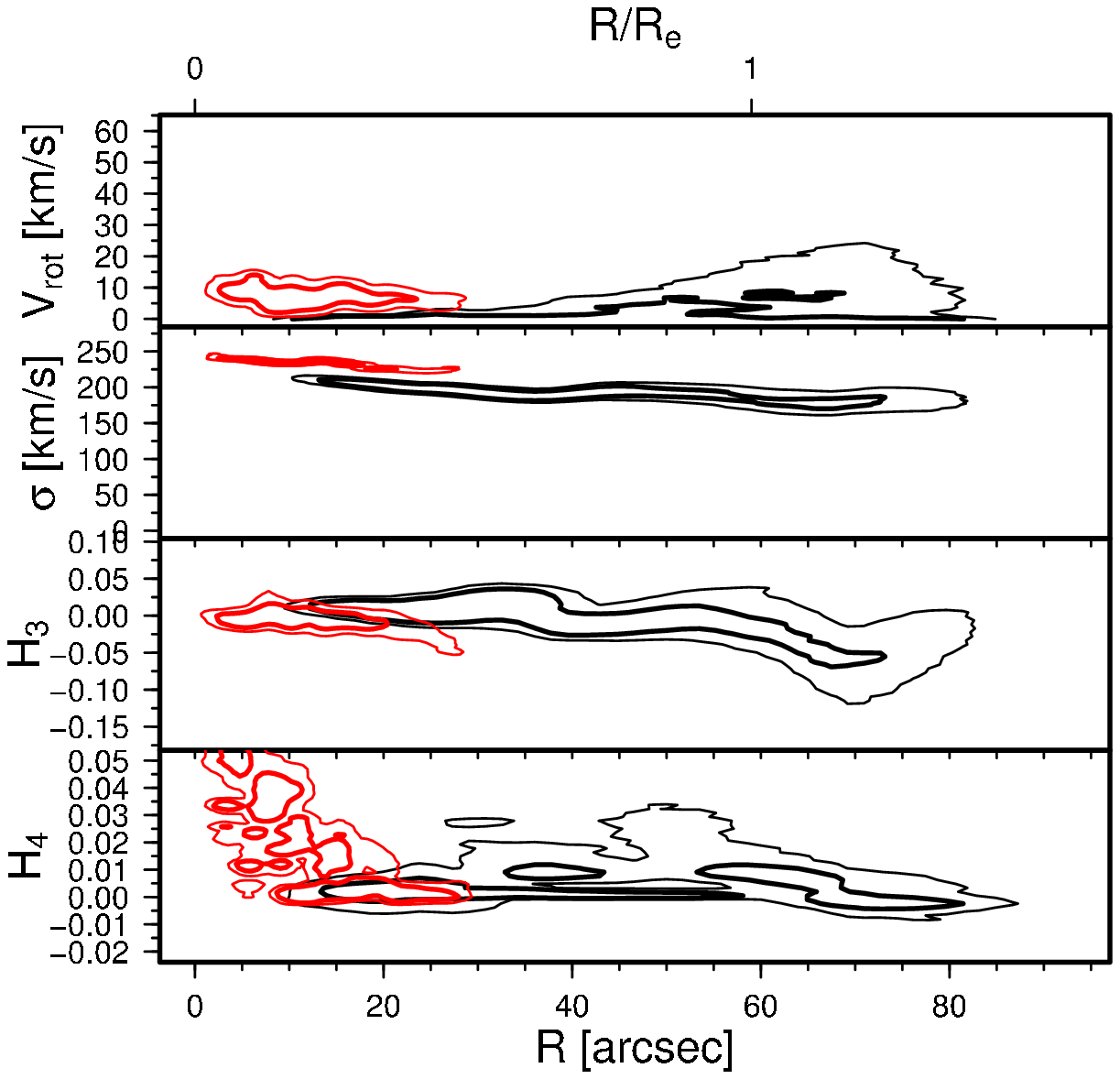}
        }
    \caption{Continued. Results of the radial kinemetry for NGC~5846. Data associated with nearby galaxy NGC~5846a are removed.}
\end{figure}

\begin{figure} 
 \ContinuedFloat
        \subfigure{
            \hspace{-0.5cm}\includegraphics[width=90mm]{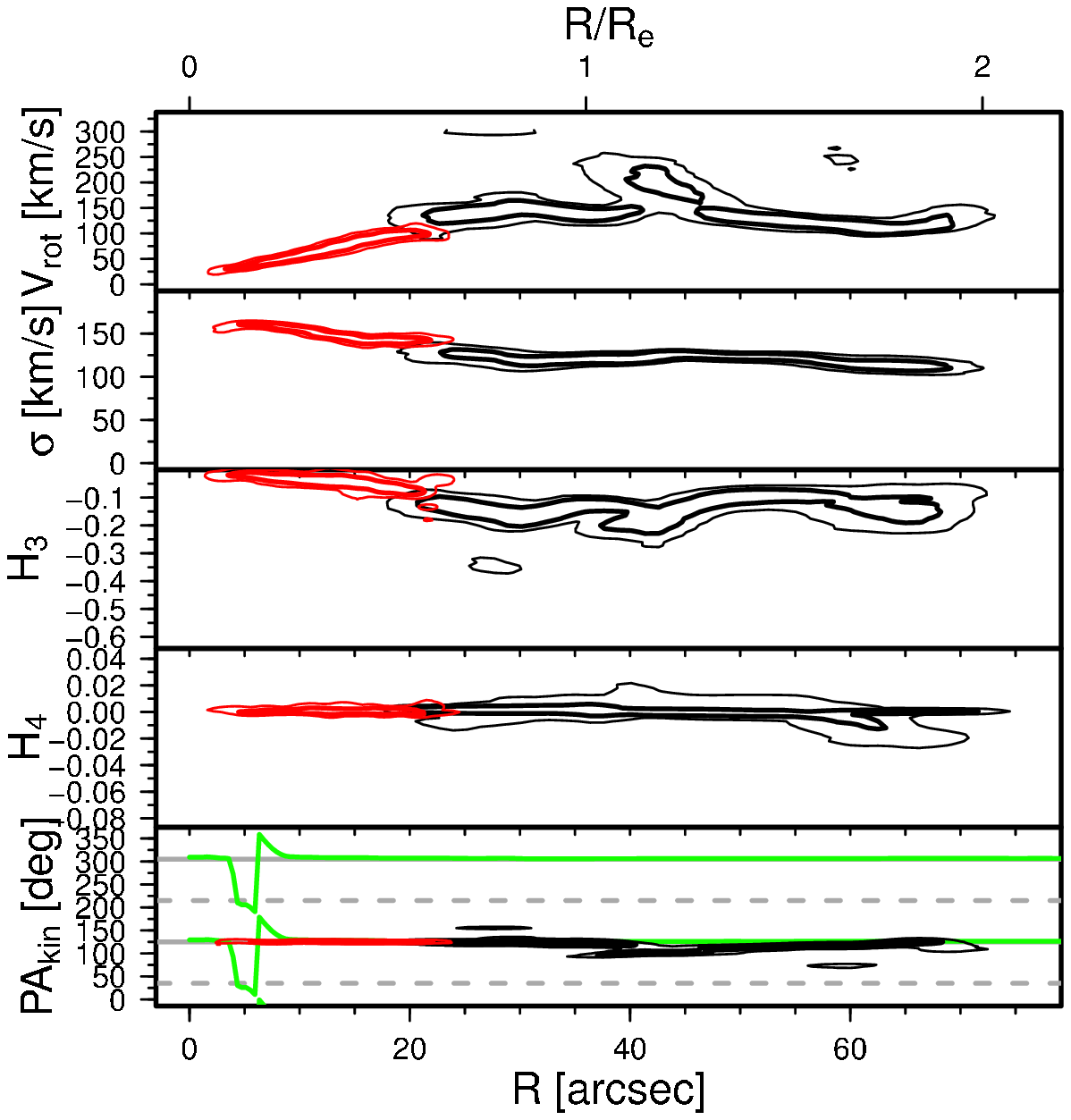}
        }
    \caption{Continued. Results of the radial kinemetry for NGC~5866.}
\end{figure}

\begin{figure} 
 \ContinuedFloat
        \subfigure{
             \hspace{-0.5cm}\includegraphics[width=90mm]{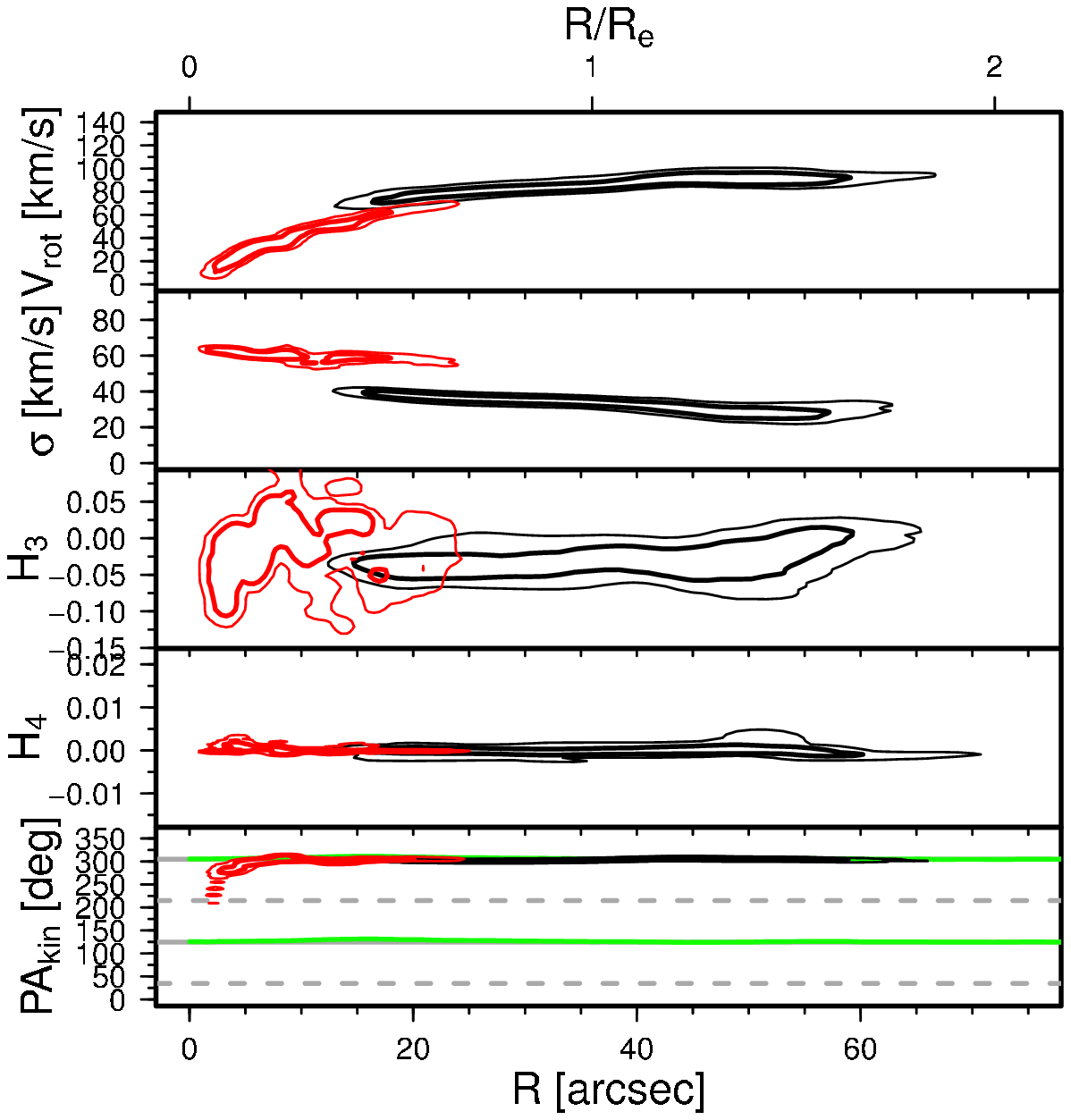}
        }
    \caption{Continued. Results of the radial kinemetry for NGC~7457.}
\end{figure}

\end{appendix}

\end{document}